\documentclass[longauth]{aa} 

\usepackage{graphicx}
\usepackage{txfonts}
\usepackage{xcolor}
\usepackage{amsmath} % for matrices

\usepackage{float}          % to prevent tables and images to float around
\usepackage{subcaption}     % enables subfigure package too

\usepackage{multirow}       % to enable multirow and multicolumn

\usepackage{array}          % to avoid table cluttering
\usepackage{longtable}

\usepackage{lscape}         % to rotate a single page table, example from A&A temp. Oct24

\usepackage{natbib}
\bibpunct{(}{)}{;}{a}{}{,} % to follow the A&A style

\usepackage{hyperref} 
\hypersetup{
    colorlinks,
    linkcolor={red!50!black},
    citecolor={blue!50!black},
    urlcolor={green!80!black}
} % to remove ugly link boxes of overleaf, but still show links as clickable

\begin{document} 

   \title{A comprehensive study on radial velocity signals using ESPRESSO: Pushing precision to the 10\,cm/s level\thanks{\tiny{Based on Guaranteed Time Observations collected at the European Southern Observatory under ESO PIds 106.21M2.001, 106.21M2.003, 106.21M2.004, 106.21M2.006, 108.2254.001, 108.2254.003, 108.2254.004, 108.2254.006, 110.24CD.001, 110.24CD.003, 110.24CD.009, 1102.C-0744, 1102.C-0958, 1104.C-0350 by the ESPRESSO Consortium and calibration data taken under 60.A-9128 and 60.A-9129.}}}

   \author{P. Figueira\inst{1,2},
            J. P. Faria\inst{1},
            A. M. Silva\inst{2,3},
            A. Castro-Gonz\'{a}lez\inst{4},
            J. Gomes da Silva\inst{2},
            S. G. Sousa\inst{2},
            D. Bossini\inst{5},
            M. R. Zapatero-Osorio\inst{4},
            O. Balsalobre-Ruza\inst{4},
            J. Lillo-Box\inst{4},
            H. M. Tabernero\inst{6},
            V. Adibekyan\inst{2},
            R. Allart\inst{7},
            S. Benatti\inst{8},
            F. Bouchy\inst{1},
            A. Cabral\inst{9,10},
            S. Cristiani\inst{11},
            X. Dumusque\inst{1},
            J. I. Gonz\'alez-Hern\'andez\inst{12,13},
            N. Hara\inst{14},
            G. Lo Curto\inst{15},
            C. Lovis\inst{1},
            A. Mehner\inst{15},
            P. Molaro\inst{11,16},
            F. Pepe\inst{1},
            N. C. Santos\inst{2,3},
            D. S\'{e}gransan\inst{1},
            D. Sosnowska\inst{1},
            R. Rebolo\inst{12,13},
            A.~Su{\'a}rez Mascare{\~n}o\inst{12,13},
            A. Sozzetti\inst{17},
            S. Udry\inst{1},
            B. Wehbe\inst{9,10}
          }

   \institute{Observatoire Astronomique de l’Universit\'{e} de Gen\`{e}ve, Chemin Pegasi 51b, 1290 Versoix, Switzerland \\ 
              \email{pedro.figueira@unige.ch}
        \and
        Instituto de Astrof\'{i}sica e Ci\^{e}ncias do Espa\c{c}o, Universidade do Porto, CAUP, Rua das Estrelas, 4150-762 Porto, Portugal 
        \and
        Departamento de F\'{i}sica e Astronomia, Faculdade de Ci\^{e}ncias, Universidade do Porto, Rua do Campo Alegre, 4169-007 Porto, Portugal
        \and
        Centro de Astrobiolog\'{i}a, CSIC-INTA, Camino Bajo del Castillo s/n, 28692 Villanueva de la Ca\~{n}ada, Madrid, Spain
        \and
        Department of Physics and Astronomy G. Galilei, University of Padova, Vicolo dell’Osservatorio 3, I-35122, Padova, Italy
        \and
        Departamento de F{\'i}sica de la Tierra y Astrof{\'i}sica \& IPARCOS-UCM (Instituto de F\'{i}sica de Part\'{i}culas y del Cosmos de la UCM), Facultad de Ciencias F{\'i}sicas, Universidad Complutense de Madrid, 28040 Madrid, Spain
        \and
        D\'{e}partement de Physique, Institut Trottier de Recherche sur les Exoplan\`{e}tes, Universit\'{e} de Montr\'{e}al, Montr\'{e}al, Qu\'{e}bec, H3T 1J4, Canada
        \and
        INAF - Osservatorio Astronomico di Palermo, Piazza del Parlamento, 1, 90134, Palermo, Italy
        \and
        Instituto de Astrofí+\'{i}sica e Ci\^{e}ncias do Espa\c{c}o, Universidade de Lisboa, Campo Grande, P-1749-016, Lisboa, Portugal
        \and
        Departamento de F\'{i}sica, Faculdade de Ci\^{e}ncias, Universidade de Lisboa, Campo Grande, P-1749-016, Lisboa, Portugal
        \and
        INAF – Osservatorio Astronomico di Trieste, via G. B. Tiepolo 11, I-34143, Trieste, Italy
        \and
        Instituto de Astrof\'{\i}sica de Canarias, c/ V\'ia L\'actea s/n, 38205 La Laguna, Tenerife, Spain
        \and 
        Departamento de Astrof\'{\i}sica, Universidad de La Laguna, 38206 La Laguna, Tenerife, Spain
        \and
        Aix Marseille Univ, CNRS, CNES, LAM, 13007 Marseille, France
        \and
        ESO - European Southern Observatory, Av. Alonso de Cordova 3107, Vitacura, Santiago, Chile
        \and
        Institute of Fundamental Physics of the Universe, IFPU, Via Beirut, 2, Trieste, 34151, Italy
        \and
        INAF – Osservatorio Astrofisico di Torino, Via Osservatorio 20, 10025 Pino Torinese, Italy
             }

   \date{Received January 23, 2025; accepted June 20, 2025}
 
  \abstract
   {}
   {We analyse ESPRESSO data for the stars HD\,10700 ($\tau$ Ceti), HD\,20794 ($e$ Eridani), HD\,102365, and HD\,304636 acquired via its Guaranteed Time Observations (GTO) programme. We characterise the stars' radial velocity (RV) signals down to a precision of 10 cm/s on timescales ranging from minutes to planetary periods falling within the host's habitable zone (HZ). We study the RV signature of pulsation, granulation, and stellar activity, inferring the potential presence of planets around these stars. Thus, we outline the population of planets that while undetectable remain compatible with the available data.}
   {We derived the stellar parameters through different methods for a complete characterisation of the star. We used these parameters to model the effects of stellar pulsations on intra-night RV variations and of stellar activity on nightly averaged values. The RVs were derived both with the cross-correlation method and template matching, as well as over the blue and red ESPRESSO detectors independently to identify colour-dependent parasitic effects of an instrumental or stellar nature. The study of RVs was complemented by an investigation of stellar activity indicators using photospheric information and chromospheric indexes.}
   {A simple model of stellar pulsations successfully reproduced the intra-night RV scatter of HD\,10700 down to a few cm/s. For HD\,102365 and HD\,20794, an additional source of scatter at the level of several 10\,cm/s remains necessary to explain the data. 
   
   A \texttt{kima} analysis was used to evaluate the number of planets supported by the nightly averaged time series of each of these three stars, under the assumption that a quasi-periodic Gaussian process (GP) regression is able to model the activity signal. While a frequency analysis of HD\,10700 RVs is able to identify a periodic signal at 20\,d, when it is modelled along with the activity signal the signal is formally non-significant. Moreover, its physical origin remains uncertain due to the similarity with the first harmonic of the stellar rotation. ESPRESSO data on their own do not provide conclusive evidence to support the existence of planets around HD\,10700, HD\,102365, or HD\,304636. In addition, the comparison of RVs with the contemporaneous indicators displays a strong correlation for HD\,102365. The direct interpretation is that half of the RV variance on this star is directly attributed to activity.
   }
   {ESPRESSO is shown to reach an on-sky RV precision of better than 10 cm/s on short timescales ($<$1\,h) and of 40\,cm/s over 3.5\,yr. A subdivision of the datasets showcases a precision reaching 20-30\,cm/s over one year. These results impose stringent constraints on the impact of granulation mechanisms on RV. In spite of no detections, our analysis of HD\,10700 RVs demonstrates a sensitivity to planets with a mass of 1.7\,M$_{\oplus}$ for periods of up to 100\,d, and a mass of 2-5M$_{\oplus}$ for the star's HZ.}


   \titlerunning{ESPRESSO's RVs down to the 10 cm/s level}
   \authorrunning{Figueira et al. (2025)}
   \maketitle
%
%-------------------------------------------------------------------

\section{Introduction}

The search for exoplanets around nearby stars is now mainstream science, with roughly 6000 planets known to date\footnote{As listed in \href{http://exoplanet.eu/}{exoplanet.eu} as of June 2025, when filtering for planets with mass lower than 13\,M$_{Jup}$.}. Within this sphere, the interest for the lowest-mass exoplanets was recently spurred by the capabilities of the ESPRESSO spectrograph \citep{2021A&A...645A..96P}. 
Designed with the objective of detecting an Earth-mass planet within the habitable zone (HZ) of a Sun-like star, the instrument was built to reach a radial velocity (RV) precision of down to 10\,cm/s over timescales of up to several years. The first studies in the field, such as that of \cite{2020A&A...635A..13F} and \cite{2020A&A...642A.121L}, led to very detailed planetary system characterisations. The data obtained in the context of the consortium's Guaranteed Time of Observations (GTO) enabled results such as those reported in \cite{2020A&A...639A..77S} and \cite{2022A&A...658A.115F}, whereby a system of two planets was detected around Proxima\,Cen and the M star's rotational-modulated activity was modelled and corrected down to a level of 30\,cm/s, close to the photon noise of the measurements. 

As discussed in \cite{2021A&A...645A..96P}, the GTO team, the associated observations, and their analysis are organised into working groups (WG). WG1 is dedicated to the detection and characterisation of Earth-like planets via blind RV searches. Proxima was one of $\sim$25 stars observed intensively within ESPRESSO GTO WG1 to cover periodicities up to the stars' HZs. For the preparatory work and original list of stars, we refer to \cite{2019A&A...629A..80H}.

In this work, we selected four ESPRESSO GTO WG1 stars for a detailed analysis: HD\,10700 ($\tau$ Ceti), HD\,20794 ($e$ Eridani), HD\,102365, and HD\,304636. These specific stars were selected for having enough data points for an RV characterisation, while also being representative of our blind RV campaign. 
We first used ESPRESSO and ancillary data to characterise each star and then followed up with RV time series calculations and their analysis. We study the presence and characteristics of signals from the shortest to the longest timescales: pulsations, granulation, stellar activity, and, finally, the planetary signals up to the HZ of the host stars. We discuss the currently achievable level of radial velocity precision, our analysis choices, and the current limitations, along with potential avenues to explore. 

In Section\,\ref{Sec:Targets}, we present a summary of the literature on the targets. In Section\,\ref{Sec:Host}, we perform our own stellar characterisation using a range of methods. In Section\,\ref{Sec:Data}, we describe the ESPRESSO data acquisition and RV derivation. We discuss the intra-night RV variations in Section\,\ref{Sec:RVintra}, along with the night-to-night and longer-term RVs and their associated quantities variations in Section\,\ref{Sec:RVnight-to-night}. In Section\,\ref{Sec:RVkima}, we characterise the long-term variations with a planetary and stellar activity model. We discuss our results in Section\,\ref{Sec:Disc} and wrap up our conclusions in Section\,\ref{Sec:Conc}.

%--------------------------------------------------------------------

\section{Targets and previous works}\label{Sec:Targets}

\subsection{HD\,10700 ($\tau$ Ceti)}

The star HD\,10700, commonly know as $\tau$ Ceti, is a very bright (V\,=\,3.50) star located only at 3.65\,pc from the Sun. It is stable photometrically and has been considered to be representative of chromosphericaly inactive stars, as underlined by \cite{1994ApJ...427.1042G}. The same work reports no evidence for a rotational modulation on temperature, granulation or chromospheric activity, along with only a weak magnetic cycle at a period of roughly 11\,yr. 

The star was target of an RV asteroseismology campaign with HARPS, providing accurate values for stellar mass, radius, and luminosity \citep{2009A&A...494..237T}. Recently, \cite{2023AJ....166..123K} used long-baseline optical interferometric to provide additional constraints on stellar parameters, deriving a rotation period, P$_{\mathrm{rot}}$, of 46$\pm$4\,d and a stellar axis inclination $i$ of 7$\pm$7$^{\circ}$. A brief summary of the parameters available in the literature is given in Table\,\ref{Tab:starliteraturedata}. 

\setlength{\extrarowheight}{4pt}

\begin{table*} 
\caption{Literature parameters for the stars discussed in the paper.}\label{Tab:starliteraturedata}

\centering
\small
\begin{tabular}{ l | c c | c c | c c | c c }
 \hline \hline
 Parameter  &  \multicolumn{2}{c|}{HD\,10700} & \multicolumn{2}{c|}{HD\,20794} & \multicolumn{2}{c|}{HD\,102365}  & \multicolumn{2}{c}{HD\,304636} \\
  \hline
\multirow{2}{*}{Spectral type} & G8.5V & \tiny{[1]} &  G8V  & \tiny{[4]} & G5  & \tiny{[9]} & M0 & \tiny{[14]/[15]} \\
&  K0 & \tiny{[2]} & K0 & \tiny{[2]} & G8 & \tiny{[2]} & M1 & \tiny{[2]/[16]} \\

\multirow{2}{*}{T$_{\mathrm{eff}}$ \tiny{[K]}} & \multirow{2}{*}{5339$\pm$19} & \multirow{2}{*}{\tiny{[2]}} & \multirow{2}{*}{5388$\pm$14} & \multirow{2}{*}{\tiny{[2]}} & 5652$\pm$18 & \tiny{[2]} & 3820$\pm$92 & \tiny{[2]} \\ 
 & & & & & 5618$\pm$14 & \tiny{[10]} & 3770$\pm$40 & \tiny{[16]} \\

\multirow{2}{*}{[Fe/H]} & \multirow{2}{*}{-0.51$\pm$0.01} & \multirow{2}{*}{\tiny{[2]}} & \multirow{2}{*}{-0.385$\pm$0.011} & \multirow{2}{*}{\tiny{[2]}} & -0.3$\pm$0.01 & \tiny{[2]} & -0.16$\pm$0.05 & \tiny{[2]} \\
 & & & & & -0.31$\pm$0.02 & \tiny{[11]} & -0.16$\pm$0.09 & \tiny{[15]} \\

log g \tiny{[cm.s$^{-2}$]} & 4.45$\pm$0.06 & \tiny{[2]} & 4.379$\pm$0.036 & \tiny{[2]} & 4.4$\pm$0.05 & \tiny{[2]} & 4.72$\pm$0.02 & \tiny{[16]} \\
 
$\xi_{t}$\tiny{[km/s]} & 0.67$\pm$0.04 & \tiny{[2]} & 0.568$\pm$0.040 & \tiny{[2]} & 0.98$\pm$0.03 & \tiny{[2]} & -- & -- \\

\multirow{2}{*}{v.sin(i) \tiny{[km/s]}} & \multirow{2}{*}{$\sim$1} & \multirow{2}{*}{\tiny{[3]}} & \multirow{2}{*}{$<$3} & \multirow{2}{*}{\tiny{[4]}} & 0.7 & \tiny{[9]} & \multirow{2}{*}{2.32$\pm$0.65} & \multirow{2}{*}{\tiny{[2]}} \\
& & & & & $<$2 & \tiny{[2]} & & \\

\multirow{2}{*}{log($R'_{HK}$)} & -4.96$\pm$0.003 & \tiny{[4]} & -4.976$\pm$0.003 & \tiny{[4]} & -4.99 & \tiny{[9]} & \multirow{2}{*}{-4.94$\pm$0.01} & \multirow{2}{*}{\tiny{[2]}} \\
 & -4.90$\pm$0.01 & \tiny{[2]} & -5.00$\pm$0.01 & \tiny{[2]} & -4.88$\pm$0.01 & \tiny{[2]} & & \\

\multirow{2}{*}{P$_{\mathrm{rot}}$ \tiny{[d]}} & 34 & \tiny{[5]} & \multirow{2}{*}{33.19$\pm$3.61} & \multirow{2}{*}{\tiny{[4]}} & \multirow{2}{*}{40} & \multirow{2}{*}{\tiny{[12]}} & \multirow{2}{*}{31$\pm$5} & \multirow{2}{*}{\tiny{[2]}}\\
& 46$\pm$4 & \tiny{[6]} & & & & & &  \\

$\pi$ \tiny{[mas]} & 273.8097$\pm$0.1701 & \tiny{[7]} & 165.5242$\pm$0.0784 & \tiny{[7]} & 107.3024$\pm$0.0873 & \tiny{[7]} & 95.6982$\pm$0.0145 & \tiny{[7]} \\

d \tiny{[pc]} & 3.65 & \tiny{[7]} & 6.04 & \tiny{[7]} & 9.32 & \tiny{[7]}  & 10.45 & \tiny{[7]}\\

\multirow{3}{*}{M$_{\mathrm{star}}$\tiny{[M$_{\odot}$]}} & \multirow{3}{*}{0.783$\pm$0.012} & \multirow{3}{*}{\tiny{[8]}} & \multirow{3}{*}{0.70} & \multirow{3}{*}{\tiny{[4]}} & \multirow{3}{*}{0.88$^{+0.02}_{-0.03}$} & \multirow{3}{*}{\tiny{[13]}} & 0.55$\pm$0.05 & \tiny{[17]}\\
& & & & & & & 0.53$\pm$0.05 & \tiny{[15]}  \\
& & & & & & & 0.48$\pm$0.03 & \tiny{[16]} \\
R$_{\mathrm{star}}$\tiny{[R$_{\odot}$]} & 0.793$\pm$0.004 & \tiny{[8]} & -- & -- & -- & --  & 0.5066$^{+0.0169}_{-0.0172}$ & \tiny{[16]}  \\
L$_{\mathrm{star}}$\tiny{[L$_{\odot}$]} & 0.488$\pm$0.010 & \tiny{[8]} & 0.656$\pm$0.003 & \tiny{[4]} & -- & --  & 0.046$\pm$0.0015 & \tiny{[16]} \\[2pt]
\hline
\end{tabular}
\footnotesize
\vspace{0.15cm}

References: 
[1]: \cite{2006AJ....132..161G}, 
[2]: \cite{2019A&A...629A..80H}, 
[3]: \cite{1994ApJ...427.1042G}, 
[4]: \cite{2011A&A...534A..58P}, 
[5]: \cite{1996ApJ...457L..99B}, 
[6]: \cite{2023AJ....166..123K}, 
[7]: \cite{2023A&A...674A...1G}, 
[8]: \cite{2009A&A...494..237T}, 
[9]: \cite{2011ApJ...727..103T}, 
[10]: \cite{2022A&A...658A..57M}, 
[11]: \cite{2022A&A...663A...4S}, 
[12]: \cite{1984ApJ...287..769N}, 
[13]: \cite{2018A&A...614A..55A}, 
[14]: \cite{2010MNRAS.403.1949K}, 
[15]: \cite{2019A&A...624A..94M}, 
[16]: \cite{2025ApJ...979..214D},
[17]: \cite{2013A&A...551A..36N}
\end{table*}

\setlength{\extrarowheight}{2pt}

The star's brightness and perceived stability made it a prime target for RV studies. The work of \cite{2011A&A...534A..58P}, using HARPS \citep{2003Msngr.114...20M} data accumulated over more than 7 years, reported an RV dispersion of only 1.5\,m/s for intra-night measurements and a long-term scatter of 92\,cm/s when nightly averaged RVs are considered. This study reported that several signals seemed to be present in the data but none was confidently detected. 

The work of \cite{2013A&A...551A..79T} put together HARPS together with AAPS and HIRES data to find evidence for five planets at periods of 13.9, 35.4, 94, 168, and 640\,d with a minimum mass of 2.0, 3.1, 3.6, 4.3, and 6.6\,M$_{\oplus}$, respectively, and amplitudes between 58 and 75\,cm/s. The same team published an updated analysis in \cite{2017AJ....154..135F} reporting the detection of only four planets, with periods of 20.0, 49.4, 162, and 636\,d and a minimum mass of 1.75, 1.83, 3.93, and 3.93\,M$_{\oplus}$, having amplitudes between 35 and 55\,cm/s.\footnote{In Ursula K. Le Guin's book \textit{The Dispossessed}, the star is reported to be orbited by the twin planets Urras and Anarres. It is one of the few works of fiction to win all three \textit{Hugo}, \textit{Locus}, and \textit{Nebula} awards for best novel, and if you haven't read it you must absolutely do so.}

\subsection{HD\,20794 ($e$ Eridani)}

The bright (V\,=\,4.26) star HD\,20794, also known as $e$ Eridani, has been intensively studied in exoplanet searches. It was subject to an intense RV monitoring with HARPS and had three super-Earths announced in \cite{2011A&A...534A..58P}. The data were later revisited by \cite{2017A&A...605A.103F}, who confirmed two of the previously presented planets and reported a different characterisation of the third one. The stellar characterisation available is summarised in Table \ref{Tab:starliteraturedata}. A recent analysis of ESPRESSO data on HD\,20794 is presented in \cite{2025A&A...693A.297N}. One particular study that is especially pertinent to our analysis is the commissioning work on HARPS, presented in \cite{2003Msngr.114...20M}, which reported a root-mean-square (rms) RV dispersion of HD\,20794 of 80\,cm/s, ascribed to p-modes.

\subsection{HD\,102365}

HD\,102365 is a bright (V\,=\,4.89) star that stood out among ESPRESSO GTO WG1 targets for its brightness, low rotational velocity, and low activity level (as measured by its log($R'_{HK}$) index). The main stellar properties are compiled in Table\,\ref{Tab:starliteraturedata}.

\cite{2011ApJ...727..103T} detected a planet around the star using 149 UCLES \citep{1990SPIE.1235..562D} RVs. A period inspection was done with Lomb-Scargle (LS) tools and a significant peak with a false alarm probability (FAP) of 0.1\% was found at 122\,d. The best-fit parameters led to a planetary orbital period of 122.1$\pm$0.3\,d, an amplitude of 2.40$\pm$0.35\,m/s, and an eccentricity of 0.34$\pm0$.14; these values corresponded to a lower mass limit $m.\sin{i}$ of 16.0$\pm$2.6 M$_{\oplus}$. 

\cite{2023AJ....165..176L} performed an archival RV analysis on several stars, covering HD\,102365. The authors updated the orbital parameters using additional observations; the previously published planetary period and eccentricity are recovered, but the semi-amplitude is revised down to 1.38$\pm$0.23\,m/s, a value 58\% lower than announced by \cite{2011ApJ...727..103T}. Finally, \cite{2023AJ....165...34H} consider HD\,102365\,b as the most favourable HZ non-transiting exoplanets for a spectroscopic characterisation, underlining the proposed planet's importance. 

\subsection{HD\,304636}

HD\,304636 is listed as an M0 from \textit{Hipparcos} colours \citep{2010MNRAS.403.1949K} that is comparatively bright for its assigned spectral type (V\,=\,9.465). 
Its parameters were derived in \cite{2019A&A...624A..94M} using ratios of pseudo-equivalent widths of spectral features sensitive to the effective temperature and the stellar metallicity. The literature values on stellar characterisation are presented in Table\,\ref{Tab:starliteraturedata}.

The TESS transit survey \citep{2014SPIE.9143E..20R} identified a planet candidate around HD\,304636 with a periodicity of 7.57\,d and a transit depth of 0.29 parts per thousand (ppt), corresponding to a radius of $\sim$0.87\,R$_{\oplus}$. The star was then listed as a TESS Object of Interest, TOI-741. In a recent work \cite{2025ApJ...979..214D} used HARPS RV observations to attempt a detection of the candidate. While a detection was not possible, the authors put an upper limits of 4.0 M$_{\oplus}$ (at 3$\sigma$) on the mass of the planet (if indeed present).

%--------------------------------------------------------------------

\section{Stellar host characterisation}\label{Sec:Host}

The modelling of stellar phenomena capable of creating RV signals requires an accurate determination of the stellar parameters. We use different methodologies and ancillary data to do so. After critically assessing and comparing the derived values, we conclude by summarising the best estimation for each star's parameters.

\subsection{Spectroscopic analysis}\label{Sec:specanalysis}

We used ESPRESSO spectra to derive the stellar atmospheric parameters $T_{\mathrm{eff}}$, $\log g$, microturbulence $\xi_{t}$, and [Fe/H]. The individual spectra were co-added to create a master spectrum with S/N up to 2500 for each star. In this regime, the uncertainty on the derived parameters is not limited by the spectra S/N, but by the intrinsic precision and accuracy of the method instead. 

For the first three stars, we used the methodology described in \cite{2021A&A...656A..53S} and references therein. In brief, we used \texttt{ARES}\footnote{\url{https://github.com/sousasag/ARES}} \citep{2007A&A...469..783S, 2015A&A...577A..67S} to measure the equivalent widths (EW) of iron lines on the combined spectra followed by a minimisation process to find ionisation and excitation equilibrium and converge to the best set of spectroscopic parameters. This process makes use of a grid of Kurucz model atmospheres \citep{1993sssp.book.....K} and the radiative transfer code \texttt{MOOG} \citep{1973PhDT.......180S}. 

For M-stars like HD\,304636 the stellar parameters cannot be derived with the same technique. We used the \texttt{ODUSSEAS} code \citep{2020A&A...636A...9A}\footnote{\url{https://github.com/AlexandrosAntoniadis/ODUSSEAS}} that employs a machine learning approach on pseudo-EW of more than 4000 stellar absorption lines to predict the stellar effective temperature and metallicity. We applied the calibration presented in \cite{2024A&A...690A..58A}, which was demonstrated to provide more accurate and consistent results than its previous iteration. In parallel, we used the Bayesian tool \texttt{SteParSyn} described in \cite{2022A&A...657A..66T}.   

% SteParSyn for HD 304636
%
%Teff = 3799 +- 10 K
%log(g) = 4.87 +- 0.03 dex
%[Fe/H] = -0.25 +- 0.02 dex

We used the derived $T_{\mathrm{eff}}$ and GAIA DR3 \citep{2023A&A...674A...1G} data and followed the procedure described in \cite{2021A&A...656A..53S} to calculate the trigonometric surface gravity $\log g_\mathrm{trig}$. The work of \cite{2013ApJS..208....9P} was used to assign a spectral type to the $T_{\mathrm{eff}}$ of each star. The spectroscopic parameters $T_{\mathrm{eff}}$, $\log g$, microturbulence $\xi_{t}$, and [Fe/H], plus the associated $\log g_\mathrm{trig}$ and spectral type are presented in Table\,\ref{Tab:specESP}. 

% NOTE: taken from:
% https://www.pas.rochester.edu/~emamajek/EEM_dwarf_UBVIJHK_colors_Teff.txt

\begin{table*} 
\centering
\caption{Stellar parameters derived from ESPRESSO spectra, plus their associated quantities.}
\label{Tab:specESP}
\begin{tabular}{ l c | c c c c | c | c }
 \hline \hline
 \multirow{2}{*}{Star} & \multirow{2}{*}{Method} & T$_\mathrm{eff}$ & logg & \multirow{2}{*}{[Fe/H]}  & \multirow{2}{*}{$\xi_{t}$} &logg$_\mathrm{trig}$  & \multirow{2}{*}{\small{Spectral type}} \\ 
 & & \tiny{[K]} & \tiny{[cm.s$^{-2}$]} & & & \tiny{[cm.s$^{-2}$]} & \\
  \hline
HD\,10700 & \multirow{3}{*}{\texttt{ARES}+\texttt{MOOG}} & 5278$\pm$66 & 4.40$\pm$0.11 & -0.55$\pm$0.04 & 0.45$\pm$0.10 & 4.46$\pm$0.03 & K0V \\ 
HD\,20794 &  &5368$\pm$64& 4.32$\pm$0.11& -0.42$\pm$0.04& 0.56$\pm$0.05 & 4.39$\pm$0.03 & G9V \\ 

HD\,102365 & &5592$\pm$62 & 4.37$\pm$0.10 & -0.34$\pm$0.04 & 0.81$\pm$0.04 & 4.38$\pm$0.03 & G6V\\ 

\multirow{2}{*}{HD\,304636} &  \texttt{ODUSSEAS} & 3747$\pm$64 & --- & -0.22$\pm$0.11 & --- & 4.62$\pm$0.06 & \multirow{2}{*}{M0.5V} \\
& \texttt{SteParSyn} & 3799$\pm$10 & --- & -0.25$\pm$0.02 & --- & 4.87$\pm$0.03 & \\

\hline
\end{tabular}
\end{table*}

\subsection{Spectral energy distribution analysis}\label{Sec:SED}

We used the \texttt{ARIADNE}\footnote{\url{https://github.com/jvines/astroARIADNE}} package described in \cite{2022MNRAS.513.2719V} to perform a spectral energy distribution (SED) fitting of stellar models and with it determine stellar parameters from photometric data. 
The program cross-matches source coordinates over a wide range of catalogues to retrieve photometry and associated uncertainties; it then fits the SED across the available photometric bands using several stellar atmosphere models and employing Bayesian model averaging (BMA). With this approach, the final posterior is created by drawing from the posterior of the different models, weighted by their evidence. The parameters posteriors are then wider due to the systematic differences between the models, and error bars drawn from confidence intervals are a better representation of the method accuracy.
Model averaging is done over PHOENIX \citep{2013A&A...553A...6H}, BT-Settl \citep{2012RSPTA.370.2765A}, ATLAS9 \citep{2003IAUS..210P.A20C}, and SYNTHE Kurucz \citep{1993sssp.book.....K} models. \texttt{ARIADNE} then queries Gaia DR3 for parameters, and provides the best estimate for several other stellar parameters. Since our aim is to obtain an independent stellar characterisation, we used the default (weakly informative) priors on the parameters to fit. 

The input photometry and model priors are detailed in Appendix\,\ref{App:ARIADNE}; the best-fit stellar parameters $T_{\mathrm{eff}}$; logg [Fe/H]; distance, d; luminosity, $L$; radius, $R;$ and mass, $M$ are presented in Table\,\ref{Tab:ARIADNEoutput}.  

\setlength{\extrarowheight}{3pt}

\begin{table}[H]
\centering
\caption{\texttt{ARIADNE} Bayesian model averaging output.}
\label{Tab:ARIADNEoutput}
\resizebox{\columnwidth}{!}{%
\begin{tabular}{ l c c c c }
 \hline \hline
 Parameter & HD\,10700 & HD\,20794 & HD\,102365 & HD304636 \\  
  \hline
T$_\mathrm{eff}$\,\tiny{[K]}  & 5296.8$^{+131.3}_{-185.8}$ & 5393.7$^{+90.2}_{-83.4}$ & 5677.1$^{+82.0}_{-71.7}$ & 3694.8$^{+32.7}_{-31.2}$ \\[4pt]
logg\,\tiny{[cm.s$^{-2}$]} & 4.34$^{+0.77}_{-0.63}$ & 4.15$^{+0.70}_{-0.37}$ & 4.49$^{+0.49}_{-0.54}$ & 5.25$^{+0.53}_{-0.55}$ \\[4pt]

[Fe/H] & -0.23$^{+0.24}_{-0.26}$ & -0.17$^{+0.18}_{-0.18}$ & -0.17$^{+0.20}_{-0.19}$ & -0.11$^{+0.21}_{-0.20}$ \\[4pt]

d\,\tiny{[pc]} & 3.657$^{+0.007}_{-0.003}$ & 6.047$^{+0.006}_{-0.003}$ & 9.332$^{+0.013}_{-0.008}$ & 10.452$ ^{+0.004}_ {-0.003}$ \\[4pt]

L$_{\mathrm{star}}$\,\tiny{[L$_{\odot}$]} & 0.49$^{+0.11}_{-0.08}$ & 0.66$^{+0.07}_{-0.07}$ & 0.85$^{+0.08}_{-0.07}$ & 0.050$^{+0.003}_{-0.003}$ \\[4pt]

M$_{\mathrm{star}}$\,\tiny{[M$_{\odot}$]} & 0.82$^{+0.14}_{-0.02}$ & 0.87$^{+0.08}_{-0.05}$ & 1.00$^{+0.04}_{-0.16}$ & 0.54$^{+0.03}_{-0.02}$ \\[4pt]

R$_{\mathrm{star}}$\,\tiny{[R$_{\odot}$]} & 0.84$^{+0.08}_{-0.05}$ & 0.93$^{+0.04}_{-0.04}$ & 0.95$^{+0.03}_{-0.03}$ & 0.54$^{+0.01}_ {-0.01}$ \\[4pt]

\hline
\end{tabular}
}
\end{table}

\setlength{\extrarowheight}{2pt}

\subsection{Mass and radius computations from evolutionary models}\label{Sec:EvolMod}

We derived the stellar mass and radius using the Bayesian code \texttt{PARAM}\footnote{\url{http://stev.oapd.inaf.it/cgi-bin/param}} \citep{2006A&A...458..609D, 2014MNRAS.445.2758R, 2017MNRAS.467.1433R}. A  grid of stellar evolutionary tracks was matched to the derived quantities of $T_{\rm eff}$, [Fe/H] from spectroscopic analysis and luminosity from \texttt{ARIADNE} as presented in the previous subsection. The grid of stellar evolutionary isochrones was taken from \texttt{PARSEC}\footnote{\url{http://stev.oapd.inaf.it/cgi-bin/cmd}} v2.1 (the \underline{PA}dova and T\underline{R}ieste \underline{S}tellar \underline{E}volution \underline{C}ode; \cite{2012MNRAS.427..127B}. 
The median and the 68\% confidence intervals (C.I.$_\mathrm{68\%}$) of the posterior distribution are used to derive the most probable values for mass and radius, presented in Table\,\ref{Tab:isoc}.

\setlength{\extrarowheight}{3pt}

\begin{table}[H]
\centering
\caption{Mass and radius derived using \texttt{PARAM} with \texttt{PARSEC} isochrones.}
\label{Tab:isoc}
\begin{tabular}{ l c c }
 \hline \hline
 Star & M [M$_{\odot}$]  & R$_{\odot}$ [R$_{\odot}$]  \\ 
  \hline
HD\,10700 & 0.69$^{+0.02}_{-0.03}$ & 0.77$^{+0.05}_{-0.06}$  \\
HD\,20794 & 0.72$\pm$0.02 & 0.90$\pm$0.04 \\

HD\,102365 & 0.79$^{+0.03}_{-0.02}$ & 0.97$\pm$0.05 \\
HD\,304636 & 0.51$^{+0.02}_{-0.01}$ & 0.52$^{+0.02}_{-0.01}$ \\[2pt] 

\hline
\end{tabular}
\end{table}

\setlength{\extrarowheight}{2pt}

\subsection{Rotational periods from photometry and chromospheric activity indexes}\label{Sec:Prot}

On top of the fundamental stellar parameters listed, an estimation of the rotational period is key for the characterisation of RV activity signals. We analysed the public ASAS-SN \citep{2014ApJ...788...48S} and TESS photometry \citep{2014SPIE.9143E..20R} to search for periodic signals that could represent the stellar rotation modulation of the four targets. We used the \texttt{tpfplotter}\footnote{\url{https://github.com/jlillo/tpfplotter}} \citep{2020A&A...635A.128A} and \texttt{TESS-cont}\footnote{\url{https://github.com/castro-gzlz/TESS-cont}} \citep{2024A&A...691A.233C} to ensure that most of the photometric flux comes from the target stars. The TESS Simple Aperture Photometry (SAP) and the systematics-corrected Presearch Data Conditioned Simple Aperture Photometry (PDCSAP) were downloaded from the Mikulski Archive for Space Telescopes (MAST)\footnote{\url{https://mast.stsci.edu/portal/Mashup/Clients/Mast/Portal.html}}. 

The ASAS-SN data were downloaded through the Sky Patrol web interface\footnote{\url{https://asas-sn.osu.edu/}}. The stars in our sample have relatively large proper motions (pm) of 1-3"/yr, and the instrumental FWHMs are comparable to the ASAS-SN photometric apertures (i.e. $\sim$2 pixels, with a pixel scale or 8"). To avoid flux losses, we downloaded the image subtraction (IS) photometry in short chunks based on the pm-corrected coordinates, similarly to \citet{2023A&A...675A..52C}. Given the brightness of the stars, we also downloaded the ASAS-SN photometry extracted through the novel Machine Learning (ML) approach, which is aimed at improving the photometric quality of saturated sources \citep[][]{2024ApJ...971...61W}. We ran the \texttt{GLS} periodogram on the four data sets (TESS/SAP, TESS/PDCSAP, ASAS-SN/IS, ASAS-SN/ML) for the four stars, and found no significant activity signals in any of them. This result is consistent with the low chromospheric activity previously reported in the literature (Sect.~\ref{Sec:Targets}). The null results on HD\,304636 are in line with the recent work of \cite{2025ApJ...979..214D}.

\vspace{0.3cm}

The rotational period of a star can be calibrated by the Ca II H+K chromospheric emission as measured by the log($R'_{HK}$) index. This relation was explored in the original study of \cite{1984ApJ...287..769N} and an updated version was presented in \cite{2008ApJ...687.1264M} that gained significant traction in the community. This study calibrates the Rossby number, $R_0$, as function of the activity index, and associates it to the rotational period, $P_{rot}$, via the relation $R_0\,=\,P_{rot}/\tau_c$, where $\tau_c$ is the convective turnover of the star. Since $\tau_c$ is a function of effective temperature, we then have P$_{\mathrm{rot}}$(\,log($R'_{HK}$)\,, T$_{\mathrm{eff}}$\,).

\cite{2016A&A...595A..12S} calibrated photometrically determined rotational periods against log($R'_{HK}$), considering 125 stars either as a monolithic set or as subsets of G, K, and M dwarfs. The authors estimated the accuracy of their calibration from the achieved dispersion, reaching values of the order of 20\% for low-activity stars (\,log($R'_{HK}$)\,$<$\,4.5\,). The spectral type of our stars are well defined from the study in Section\,\ref{Sec:specanalysis} and Section\,\ref{Sec:SED}. For HD\,10700 and HD\,102365, we used the calibration for K dwarfs, for HD\,20794 the calibration for G dwarfs, and for HD\,304636 the calibration for M dwarfs. HD\,102365 was paired with the K dwarfs due to its late spectral type within the G and due to the low-number statistics of the \cite{2016A&A...595A..12S} G dataset. For log($R'_{HK}$) we use the value derived by \cite{2019A&A...629A..80H} and available for the four stars. We considered an uncertainty 20\% for the calculated rotational period, as used by these authors. Thus, we achieved a value of 37.7$\pm$7.5\,d for HD\,10700, 27.7$\pm$5.5\,d for HD\,20794, 36.4$\pm$7.3\,d for HD\,102365, and 30.3$\pm$6.1\,d for HD\,304636. 
We caution, however, that this determination depends strongly on the assumed spectral type, as explained in Appendix\,\ref{App:Prot}.

\subsection{Final stellar parameters and their accuracy}

The stellar parameters derived from ESPRESSO spectra are compatible with those available in the literature (Table\,\ref{Tab:starliteraturedata}) and in particular with the ESPRESSO GTO preparatory work done in \cite{2019A&A...629A..80H}. The higher resolution of ESPRESSO and higher S/N of the spectra of this work allowed for slightly more accurate determinations. We note the match between \texttt{ODUSSEAS} and \texttt{SteParSyn} results for HD\,304636; the difference in formal uncertainty between the two methods is due to \texttt{ODUSSEAS} representing accuracy variance while \texttt{SteparSyn} representing only the formal error. Since we are interested in accuracy we use the values and uncertainties delivered by \texttt{ODUSSEAS}.

A comparison between the spectral parameters derivation and the SED fitting results shows agreement (within formal uncertainties) for the effective temperature and metallicity. A comparison between SED fitting and isochrones show agreement in radii but not in mass. The luminosity values derived in our work for HD\,10700 and HD\,102365 also match the values derived by \cite{2009A&A...494..237T} and \cite{2011A&A...534A..58P} very well, and are well within the formal error bars.

For the analysis in the remainder of the paper, if a parameter can be determined via multiple methods, we selected the one that provides the highest accuracy.
For the effective temperature, we chose  values derived by spectroscopic analysis, for the luminosity the values from \texttt{ARIADNE} SED, and for the mass and radius the values derived using \texttt{PARAM} and \texttt{PARSEC}. We note, however, that the different quantities are determined at a different accuracy level, which, in turn, depends on the method chosen. In \cite{2022ApJ...927...31T} the authors compared fundamental parameters obtained through different methods and concluded on the existence of a systematic uncertainty floor of approximately 2\% in effective temperature, 2\% in luminosity, 4\% in radius, and 5\% in mass. The systematic floor level on luminosity is smaller than the ones derived in our work and, thus, it is not expected to be a limitation. However, for effective temperature it is comparable or larger than the ones we derived; for mass and radii they are appreciably larger than derived by several of our methods and, thus, we might be limited by systematic errors. We only find evidence of such an issue with respect to mass.

For the stellar rotation period we consider the values calculated in Sect.\,\ref{Sec:Prot}. To our knowledge, there is no prior study on the accuracy of activity-indexes calibration in delivering rotational periods. The relative uncertainty of 20\% on the rotational period used by \cite{2016A&A...595A..12S} is conservative when compared with other derivations and is obtained directly from the scatter of the calibration. It should, nonetheless, be used with care as a wrong spectral type or variable activity index can lead to different rotational periods in a way that cannot be included in the uncertainty value.  

Using the stellar effective temperature, luminosity and stellar mass above, we followed \cite{2013ApJ...765..131K} and \cite{2014ApJ...787L..29K}\footnote{Using the interface hosted on \url{https://personal.ems.psu.edu/~ruk15/planets/}.}, to compute the conservative HZ limits for a planet with 1\,M$_{\oplus}$.
The values calculated using these methods are summarised in Table\,\ref{Tab:FinalParams} and used in the remainder of the paper. 

\begin{table*}
\centering
\caption{Selected stellar parameters for each star and associated HZ limits.}
\label{Tab:FinalParams}
\begin{tabular}{ l l | c c c c | c}
 \hline \hline
 \multicolumn{2}{l|}{Parameter} & HD\,10700 & HD\,20794 & HD\,102365 & HD\,304636 & method / reference\\  
  \hline
T$_\mathrm{eff}$ & \tiny{[K]} & 5278$\pm$66 & 5368$\pm$64 & 5592$\pm$62 & 3747$\pm$64 & spectroscopy (Sect\,\ref{Sec:specanalysis})\\[4pt]

L$_{\mathrm{star}}$ & \tiny{[L$_{\odot}$]} & 0.49$^{+0.11}_{-0.08}$ & 0.66$^{+0.07}_{-0.07}$ & 0.85$^{+0.08}_{-0.07}$ & 0.050$^{+0.003}_{-0.003}$ & SED fitting (Sect\,\ref{Sec:SED})\\[3pt]

M$_{\mathrm{star}}$ & \tiny{[M$_{\odot}$]} & 0.69$^{+0.02}_{-0.03}$ & 0.72$\pm$0.02 & 0.79$^{+0.03}_{-0.02}$ & 0.51$^{+0.02}_{-0.01}$  & evol. model (Sect\,\ref{Sec:EvolMod})  \\[3pt]

R$_{\mathrm{star}}$ & \tiny{[R$_{\odot}$]} & 0.77$^{+0.05}_{-0.06}$ & 0.90$\pm$0.04 & 0.97$\pm$0.05 & 0.52$^{+0.02}_{-0.01}$ & evol. model (Sect\,\ref{Sec:EvolMod})  \\[3pt]

P$_{\mathrm{rot}}$ & \tiny{[d]} & 37.7$\pm$7.5 & 27.7$\pm$5.5 & 36.4$\pm$7.3 & 30.3$\pm$6.1 & activity calib. (Sect\,\ref{Sec:Prot}) \\[1pt]

\multirow{2}{*}{HZ$_{1\,M_{\oplus}}$} & \tiny{[AU]} & (0.685,\,1.227) & (0.791,\,1.413) & (0.887,\,1.573) & (0.231,\,0.445) & \multirow{2}{*}{Kopparapu et al. (2013, 2014)}  \\[1pt]

 & \tiny{[d]} & (249,\,598) & (302,\,723) & (343,\,811) & (56,\,150) & \\[1pt]

\hline

\end{tabular}

\end{table*}

%--------------------------------------------------------------------

\section{ESPRESSO data, radial velocities and indicators}\label{Sec:Data}

ESPRESSO is a high-resolution echelle spectrograph stabilised in temperature and pressure, and with mechanically fixed optics. The instrument operates on the ESO VLT and is physically located in the Combined Coud\'{e} lab. Light feeding is done from any of the four UT telescopes (or any combination of these) with a Coud\'{e} train and enters the spectrograph via two parallel fibres; the first points to the centre of the telescope field-of-view and the second points either to the neighbouring sky or to a simultaneous calibration source. The spectra from the two fibres are recorded simultaneously in interweaving orders. 
A pupil slicer splits the interference orders into two parallel slices; the spectra are imaged onto and collected by two detectors, located inside two independent cryostats. The complete wavelength coverage goes from 370 to 788\,nm. This design allows for an extreme RV stability down to a target precision of 10\,cm/s. We refer to \cite{2021A&A...645A..96P} for further details.

\subsection{Data acquisition, calibration, and pipeline reduction}

Observations were acquired in high-resolution (HR) mode, with light fed by the 1\arcsec-diameter fibre, and delivering a resolution of $\sim$140\,000; the detectors were read without applying spatial or spectral direction binning (sampling \texttt{1x1}). Given the brightness of the stars and thus a comparatively low contribution of readout noise to the error budget, we selected a \texttt{FAST} readout speed. 

ESPRESSO's wavelength calibration uses a Thorium-Argon lamp and a Fabry-Perot (FP) illuminated by a flat spectral source \citep{2010SPIE.7735E..4XW}. Combining the absolute reference from the atomic lines with the regular grid of FP lines enables the derivation of an homogeneous and precise wavelength solution across ESPRESSO's wavelength range.

On top of precise wavelength solution methodology, ESPRESSO uses a simultaneous reference technique, described in detail for its precursor ELODIE in \cite{1996A&AS..119..373B}. For precise RVs the second fibre is illuminated by the FP to measure the instrument's drift relative to the last wavelength solution; this provides for a way to quantify and correct for any instrument's internal stability variation. This drift value is calculated independently for the two detectors. 

The observations were acquired with a cadence of up to 10\% of the star's HZ inner edge period, as a trade-off between time investment and ability to detect planets up to the HZ. All observations were reduced with the ESPRESSO pipeline version \texttt{3.3.1}\footnote{Publicly available on the ESO site \url{https://www.eso.org/sci/software/pipelines/espresso/espresso-pipe-recipes.html}.}. The pipeline provides a complete and automatic reduction chain, from the raw files to the post-processing analysis such as the radial velocities and associated quantities. The main improvements of the major version \texttt{3} are: the calculation of time-dependent chromatic variation of the etalon spacing and potential lamp flux variations \citep{2022A&A...664A.191S}, along with improved masks weights that take into account the line's full-width at half maximum (FWHM) to calculate the Doppler information content (instead of the contrast only). In version \texttt{3.3.1,} the pipeline also accounts for the influence of the variable Sun-Earth distance and observer’s velocity in the barycentric correction computation (an effect with an amplitude smaller than 20\,cm/s).

As an additional step, we calculated the parallax-dependent term of the barycentric correction to apply to RVs\citep[see. e.g.][]{2010PASP..122..935E}. We did so by subtracting from the \texttt{barycorrpy}\footnote{\url{https://github.com/shbhuk/barycorrpy}} \citep{Kanodia_2018} computation (using the Gaia DR3 parallax) the correction value calculated by the ESPRESSO pipeline. For reference, the value of this term is below 1\,cm/s for HD\,10700, peak-to-peak. Nonetheless, we added this differential barycentric correction term to the RVs, \textit{a posteriori}, for the sake of completeness in correction\footnote{We are fully aware that adding the RV is formally incorrect, but the error incurred in this approximation will be below the cm/s and and as such negligible for our target precision.}.

Finally, for each acquired spectra the ESPRESSO pipeline also produces an extracted spectrum corrected for telluric H$_{\mathrm{2}}$O absorption, computed as described in \cite{2022A&A...666A.196A}. This telluric-corrected spectrum is also provided with associated RVs and ancillary quantities.

\subsection{Instrument interventions and technical issues}

In June 2019 a new fibre feed was installed in ESPRESSO. While the instrument focus did not need an adjustment, we cannot exclude the possibility that the intervention introduced an offset in the RVs calculation. As such, we consider throughout the paper two data sets \texttt{ESPR18} and \texttt{ESPR19} corresponding to before and after the intervention, respectively The two optomechanical configurations are treated then as two different instruments, which introduces an offset parameter on our RV analysis. For more on the topic we refer to the equivalent situation on HARPS, described in \cite{2015Msngr.162....9L}. 

Previous works, such as that of \cite{2022A&A...658A.115F}, considered the  \textit{\texttt{ESPR21}} dataset, corresponding to the data acquired after the COVID 19 observations interruption. This dataset was differentiated by the use of a different FP light source with a markedly different flux distribution than its predecessor. However, since pipeline version \texttt{3}, calibration lamp flux differences are taken into account and absorbed by the improved wavelength calibration; a dataset separation is thus no longer necessary.

From January to April 2021, ESPRESSO atmospheric dispersion corrector (ADCs) were affected by a critical communication issue. The ADC prisms were not positioned at the correct configuration, which led to the atmospheric dispersion not being corrected. A comparison of RVs taken with the ADC operating correctly showed that the compromised dataset was affected by an additional scatter $\gtrsim$1\,m/s. Programs aiming at high spectral fidelity are also compromised by systematic errors on the flux and line-spread function determinations. The severity of these effects will vary from spectrum to spectrum, and without having access to the ADC's true position a correction is not possible. We used a zero value of the header ADC2 keywords on right ascension, pressure and temperature to identify the observations affected, that were then discarded\footnote{For more information, we refer to the instrument user manual and news at \url{https://www.eso.org/sci/facilities/paranal/instruments/espresso.html}.}.

Since we are aiming at the highest precision in RV, we restrict our analysis to spectra in which the pipeline general quality control, flux correction of atmospheric absorption, and drift correction were all correctly executed\footnote{These correspond to the header quality control (QC) keywords \tiny{\texttt{QC\,SCIRED\,CHECK}}, \texttt{QC\,SCIRED\,FLUX\,CORR\,CHECK}, and \texttt{QC\,SCIRED\,DRIFT\,CHECK}.}.

We identified three nights corrupted by localised issues or events. On the 30$^{\mathrm{th}}$ October 2018 at 2UT a pressure gauge stopped working and there was a sudden increase of pressure within ESPRESSO vacuum vessel. This affected a series of HD\,10700 observations taken around 3UT (MJD\,=\,58421.65\,d), that was discarded from our analysis. The pressure gauge was replaced on the following day and did not affect any other measures. On the 28$^{\mathrm{th}}$ July 2022, around 06 UT a block of HD\,10700 observations was interrupted halfway, with the astronomer on duty signalling an interruption due to weather. A check on seismic activity logs of that day showed an earthquake with magnitude 6.2 and epicentre on Calama, only $\sim$280\,km from the observatory, triggering multiple aftershocks. We discard also this incomplete series of observations (MJD\,=\,59788.7\,d). 

ESPRESSO was designed to centre the image of the star on the entrance of the fibre after the spectrally dispersed image is spatially corrected by the ADC. The operational limit of this device is at a maximum airmass of 2.2; observations at higher airmass will contain an uncorrected component, with the extreme blue and red wavelengths being slightly off-centred. Up to an airmass value of 2.3, this uncorrected term remains very small, with the maximum difference between the bluest and reddest star image centre of 0.15\arcsec. In this case, and for the specific case of ESPRESSO design, the loss in flux is of 1.5\% in the blue and 0.5\% in the red, and the RV effect is expected to be much smaller than 1 cm/s \citep{2020MNRAS.491.3515W}. As such, we choose to keep the spectra taken up to an airmass of 2.3 in the analysis, affecting only 12 individual integrations spread over 3 nights. However, we choose to discard a series of observation taken in the airmass range of 2.5-2.9, on the 7$^{\mathrm{th}}$ December 2020 at 6UT (MJD\,=\,59190.25\,d, fully outside the operation range of the ADC and for which the effect in RV can be larger than 10\,cm/s, due to a considerable reduction in the flux of the bluest orders.

\subsection{RV calculation}

The RV used in this paper were calculated from the ESPRESSO pipeline extracted spectra, and followed two parallel methodologies:

\begin{itemize}
    \item \emph{Cross-correlation function method} (CCF), in which a mask representative of the spectral lines for a spectral type is cross-correlated with the measured spectra. For the original reference on the method and upgrade to weighted masks as a function of the spectral type, we refer to \cite{1996A&AS..119..373B} and \cite{2002A&A...388..632P}, respectively.
    \item \emph{Correlation with a template mask}, where a template is first constructed by stacking a large number of spectra form the star, and the correlation is calculated between the individual spectra and this empirical template. We use the methodology described and demonstrated for ESPRESSO in \cite{2022A&A...663A.143S}, named \texttt{S-BART}.
\end{itemize}

The specificities of the application of \texttt{S-BART} to ESPRESSO high-precision RV are given in Appendix\,\ref{App:SBART}. For the following analysis, it is important to retain that by default \texttt{S-BART} discards wavelength regions contaminated by telluric features deeper than 1\%. 

On top of the standard \texttt{S-BART} run, we calculate the RVs using the spectra taken from either the blue or the red detector only. These chromatic RVs enable additional diagnosis for activity and the study of detector-based systematic effects. In the remainder of the paper the RV time series corresponding to the cross-correlation, \texttt{S-BART} with two detectors, \texttt{S-BART} blue detector, and \texttt{S-BART} red are labelled \texttt{CCF}, \texttt{TM}, \texttt{TMb}, and \texttt{TMr}, respectively. 

The ESPRESSO spectra, \texttt{CCF} RVs and associated quantities were retrieved via the \texttt{DACE} interface\footnote{\url{https://dace.unige.ch/radialVelocities/?}}. To execute the {\tt DACE} queries we used the \texttt{arvi}\footnote{\url{https://github.com/j-faria/arvi}} package, which allows for command line manipulation and data download, and performs convenient tasks such as secular acceleration correction using \texttt{Simbad} stellar parameters and to calculate the photon-noise weighed average RV.
The complete time series for the four stars, including reduced and ancillary data is made publicly available to all via the \texttt{DACE} interface.

\subsection{Photospheric line profile and chromospheric activity indicators}

On the CCF one can measure directly the contrast, FWHM, and bisector inverse slope (BIS). The BIS was early on identified as a useful line profile indicator for tracking photospheric stellar activity \citep{2001A&A...379..279Q}, and has seen extensive use. For stabilised spectrographs in which the instrumental profile can be assumed as fixed, the FWHM can be used as an indicator of stellar photospheric activity as well. This has been demonstrated for ESPRESSO on the works of \cite{2020A&A...639A..77S} and \cite{2022A&A...658A.115F} for the M dwarf Proxima.

The most extensively used chromospheric indicator is the log($R'_{HK}$) calculated from the Mount Wilson $S$ index data and originally developed with the objective of studying long-term stellar activity \citep{1984ApJ...279..763N}. We used the \texttt{ACTIN} package \citep{2018JOSS....3..667G} to implement both log($R'_{HK}$) and H$_{\alpha}$. For a detailed discussion of log($R'_{HK}$) as used in \texttt{ACTIN,}   we refer to \cite{2021A&A...646A..77G}; for the H$_{\alpha}$ implementation and in particular the choice of wavelength bandwidth around the line, we\ refer to \cite{2022A&A...668A.174G}.

For the stars that have more than one spectrum per night, the activity indexes value is calculated independently for each spectra and the uncertainty-weighted average value taken as representative for the night, in parallel with what is done on RV.

%--------------------------------------------------------------------

\section{Intra-night radial velocity variations}\label{Sec:RVintra}

When observing very bright stars at a sub-m/s RV precision, it is common to acquire multiple consecutive short integrations. This allows us to cover the typical timescales of stellar pulsations and average out their RV signal, while keeping the photoelectrons on individual exposures below CCD saturation level. We employed this observational strategy for the G - K stars HD\,10700, HD\,20794, and HD\,102365 studied in this work. We obtained between 7 and 15 spectra at very high S/N (ranging between 200 and 660), leading to photon noise RV uncertainties below 30\,cm/s and well adapted to the search for low-mass planets at high RV precision. Observations of the fainter M dwarf HD\,304636 employ a single integration per night, and the star is not discussed in this section.

\begin{figure*}[t]
\center
\includegraphics[width=0.9\textwidth]{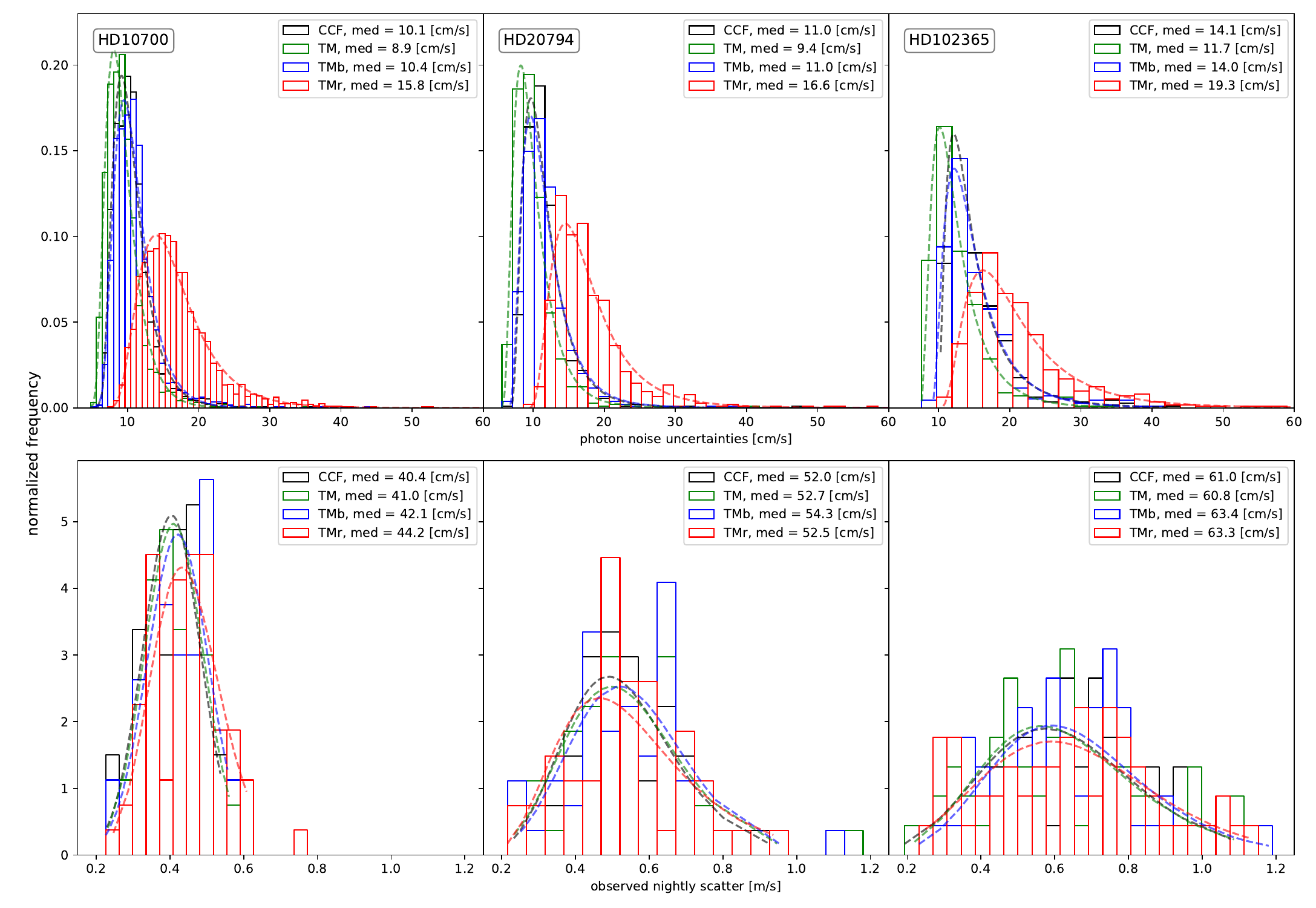}
\caption{Distribution of the photon noise uncertainty values (\textit{top}) and observed nightly scatter (\textit{bottom}) for the stars HD\,10700 (\textit{left}), HD\,20794 (\textit{centre}), and HD\,102365 (\textit{right}). The solid-line histograms represent the data and the dashed lines the cumulative log-normal distribution fits.}\label{Fig:uncsca}
\centering
\end{figure*}

The acquisition of several consecutive RV points each night allows for an evaluation of the RV precision on the short timescales they cover, ranging from minutes to around half an hour. In Fig\,\ref{Fig:uncsca} we plot the distribution of photon-noise uncertainties and the observed intra-night scatter (as root-mean-square or rms) for the four different RV time series \texttt{CCF}, \texttt{TM}, \texttt{TMb}, and \texttt{TMr} for the three stars. We considered only nights when the complete number of exposures was acquired, when the sequence was not interrupted by weather or operational issues. The cumulative distribution functions are fitted by a cumulative log-normal distribution. 

The time series \texttt{CCF}, \texttt{TM}, \texttt{TMb}, and \texttt{TMr} RVs display a very low photon noise, between 10 and 20 cm/s. The intra-night scatter is significantly larger and dependent on the star, being of $\sim$40\,cm/s for HD\,10700, $\sim$50\,cm/s for HD\,20794, and $\sim$60\,cm/s for HD\,102365.

\subsection{P-mode pulsations modelling}\label{Sec:Puls}

The main source of RV scatter for main sequence G-K stars on timescales of minutes is arguably the p-mode waves. These stellar pulsations are characterised by the quantities maximum pulsation frequency, $\nu_{max}$, and large frequency separation, $\Delta \nu$, known to scale with stellar fundamental parameters \citep[e.g.][]{2019LRSP...16....4G} as

\begin{equation}
     \hspace{0.5cm} \frac{\nu_{max}'}{\nu_{max}} = \frac{M'}{M} \left( \frac{R'}{R} \right)^{-2} \left( \frac{T_{\mathrm{eff}}'}{T_{\mathrm{eff}}} \right)^{-1/2} 
,\end{equation}\label{Eq:numax}

\begin{equation}
     \hspace{0.5cm} \frac{\Delta \nu'}{\Delta\nu} = \left( \frac{M'}{M} \right)^{1/2} \left( \frac{R'}{R} \right)^{-3/2}
,\end{equation}

\noindent
as a function of stellar mass $M$, radius $R$, and effective temperature, $T_{\mathrm{eff}}$. The maximum pulsation frequency is simply the frequency of the strongest pulsation mode and the large frequency separation is, to a first approximation, twice the frequency separation between consecutive modes.

For RV pulsation signals, the mode amplitude envelope is well approximated by a Gaussian function and its FWHM scales linearly with large frequency separation \citep{2017ApJ...835..172L}. For our cases we consider FWHM$_{env}$ = 10\,$\Delta\nu$, in line with theoretical works such as \cite{2019AJ....157..163C} (see their Fig.\,2). Pulsation studies using photometry showed independently that FWHM$_{env}$ = $\nu_{max}$/2 \citep{2016ApJ...830..138C}, again in line with our assumptions.

The RV modes' envelope has an amplitude, $A_{env}$, that depends on stellar luminosity, $L$, and mass, $M,$ \citep[see ][]{2007A&A...463..297S} as 

\begin{equation}
     \hspace{0.5cm} \frac{A_{env}'}{A_{env}} = \left( \frac{L'\,M}{L\,M'} \right)^{0.7}\, .
\end{equation}

Dedicated asteroseismology campaigns have allowed for the accurate measurement of these asteroseismic quantities for main sequence stars, which we have used as our benchmark; we chose $\alpha$\,Cen\,A and $\mu$\,Ara, studied in detail in \cite{2002A&A...390..205B} and \cite{2005A&A...440..609B}, respectively. The stellar parameters and average reference values are presented in Table\,\ref{Tab:PmodeSimul}.

We simulated the p-mode signals contribution to the RV measured on each star; we use the simulation to calculate both the intra-night scatter generated and represented in Fig.\,\ref{Fig:uncsca} and the additional variance introduced on the nightly averaged RVs as it will be of interest to long-term RV analysis. We proceeded as follows.

For the star's maximum pulsation frequency, $\nu_{max}$ -- the strongest p-mode -- we created a sinusoidal RV signal of amplitude $A_{env}$ with a random phase. Then, for the neighbouring modes of frequencies $\pm\,\Delta\nu$/2, we calculated the mode amplitude from the envelope value at the corresponding frequency and simulate two sinusoidal signals, each with a random phase. We repeated this procedure for increasingly higher-order modes $m$ of frequency $\pm\,m\,\Delta\nu$/2 stopping when the mode amplitude as drawn from the envelope is smaller than 1\,cm/s.

The RV signal of the calculated modes was co-added and its value projected on a temporal grid representing the detector integration sequence for each star in one night, with the integrations separated by the detector readout time and the individual photon noise drawn from the observed distribution median $e_{ph. noise}$. The simulations were then repeated 50\,000 times to obtain the same number of individual sequences, or nights. For each sequence, we calculated the RV offset introduced by the pulsation signal on this particular observation as the RV weighted mean and the intra-night dispersion from the intra-night rms. The dispersion of the RV offset, $e_{puls}$, is representative of the uncertainty introduced by the pulsation on the average value, while the median of the intra-night dispersion distribution, $\sigma_{puls}$, can be directly compared to the observed one. For visual representations of key steps of the simulation, we refer to Appendix\,\ref{App:Puls}; namely, we depict how the frequencies are drawn, an example of pulsations during one night and a distribution of scatter and distance to the mean when repeating the process a large number of times.

For the stellar parameters we use the values presented in Table\,\ref{Tab:FinalParams}; the measured and simulated quantities are presented in Table\,\ref{Tab:IntraRVs}, namely the distribution median photon noise, $e_{ph.\,noise}$, the measured scatter, $\sigma_{meas}$, the final uncertainty introduced by pulsations on the nightly average, $e_{puls}$, and the dispersion created by pulsation on the intra-night set $\sigma_{puls}$. The results are found to be repeatable and thus consistent down to 1-2\,cm/s. 

\begin{table}[h]

\centering
\caption{Intra night-statistics and pulsation simulations for each of the three stars.}
\label{Tab:IntraRVs}

\begin{tabular}{ c c |  c c | c c }
 \hline \hline

\multirow{2}{*}{Star} &  & \multicolumn{2}{c|}{\textit{\small{(measured)}}} & \multicolumn{2}{c}{\textit{\small{(simulated)}}} \\
 \multirow{2}{*}{\tiny{(sequence)}} & & $e_{ph.\,noise}$ & $\sigma_{meas}$ & $e_{puls}$ & $\sigma_{puls}$ \\
 & & \multicolumn{2}{c|}{\tiny{[cm/s]}} & \multicolumn{2}{c}{\tiny{[cm/s]}} \\
\hline 

\multirow{3}{*}{HD\,10700}  & \texttt{CCF} & 10.1 & 40.4 & 4 & 42 \\
&  \texttt{TM} & 8.9 & 41.0 & 4 & 42 \\
 \multirow{2}{*}{\small{(15$\times$40\,s)}} & \texttt{TMb} & 10.4 & 42.1 & 4 & 42 \\
&  \texttt{TMr} &  15.8 & 44.2 & 5 & 43 \\

\hline

\multirow{3}{*}{HD\,20794}  & \texttt{CCF} & 11.1 & 52.0 & 6 & 48 \\
& \texttt{TM} & 9.3 & 52.8 & 6 & 47 \\
\multirow{2}{*}{\small{(10$\times$60\,s)}} & \texttt{TMb} & 11.2 & 54.9 & 6 & 48 \\
& \texttt{TMr} & 15.3 & 52.4 & 7 & 49 \\
\hline 

\multirow{3}{*}{HD\,102365} & \texttt{CCF} & 14.1 & 61.0 & 9 & 49 \\            

& \texttt{TM} & 11.8 & 61.0 & 8 & 49 \\
\multirow{2}{*}{\small{(7$\times$100\,s)}} & \texttt{TMb} & 14.0 & 63.4 & 8 & 49  \\
& \texttt{TMr} & 19.2 & 63.4 & 10 & 51 \\

\hline
\end{tabular}
\end{table}

The first takeaway is that $e_{puls}$ is of 5-10\,cm/s. This is expected, in line with the common assumption that averaging the p-modes over a timescale of a few pulsation periods is enough to correct for their contribution to the nightly averaged RV. We show here it is also achieved at a precision of better than 10\,cm/s.

The simulations predict a scatter $\sigma_{puls}$ that follows the observed $\sigma_{meas}$. The pulsation-induced scatter for HD\,10700 matches the observations to 1-2\,cm/s, and for HD\,20794 to 4-5\,cm/s. For the earlier spectral type (G5) HD\,102365 there is a larger and clearly non-negligible difference of 10\,cm/s between predicted and observed values. ESPRESSO is expected to have a RV precision of better than 10\,cm/s over one hour, and if instrument stability was the source of scatter it would be present in all stars, having been easily detected for at least HD\,10700. The discrepancy can be attributed to poor stellar parameter determination or an additional source of scatter.

\subsection{Granulation signal modelling}

The convective phenomena on stellar photospheres are commonly referred to as granulation. Contrary to pulsations, the physical mechanisms governing their RV signals remains poorly understood. 
Solar observations enabled the measurement of the RV signal of convection cells, or granules, and of the coordinated motion of groups of granules, often know as super-granules \citep{1964ApJ...140.1120S}, which create an additional RV component know as super-granulation. While the granules lifetime  and renovation time has  long been estimated at $\sim$10\,min \citep[e.g.][]{1961ApJ...134..312B, 2013ApJ...770L..36G}, super-granulation timescales are still the object of active research, with \cite{2023AstBu..78..606S} showing the values range from $\sim$24\,h for the quiet to $\sim$54\,h for the active regions. For a review on super-granulation, including studies on timescales, we refer to \cite{2018LRSP...15....6R}.

There has been significant focus on the photometric observation of granulation on giant stars, as their extended convective region and higher brightness allow for a more accurate signal characterisation than on main sequence stars. A star's power spectrum density (PSD) is fitted with Harvey-type functions PSD($\nu$)\,=\,$A$/(1 + $(\tau\,\nu)^\alpha)$), in which $A$ is the amplitude, $\tau$ is the characteristic timescale, and $\alpha$ is the power-law slope.

In the literature one can find different implementations. The photometric granulation signal is represented as a sum of up to three Harvey functions, and different functional forms are used to represent the pulsation signal (Gaussian vs. Lorentzian), with some authors choosing not to fit the pulsation-dominated frequencies. 
The work of \cite{2011ApJ...741..119M} reviewed the different operational choices applied to Kepler data and concluded that the results are qualitatively similar, but the quantitative differences still leave room for different interpretations on the underlying physics. The authors showed then that from basic principles one has $\tau\propto1\,/\nu_{max}$, and found evidence for the relationship between the power of granulation and $\nu_{max}$, $P_{gran}\propto1\,/\nu_{max}^2$, as proposed by \cite{2011A&A...529L...8K}. The Kepler mission data is used to fit the slopes of these dependencies as $\tau\propto\nu_{max}^{-0.89}$ and $P_{gran}\propto\nu_{max}^{-1.90}$, but the residuals around the fitted models exhibit non-negligible scatter.  

Comparatively, there have been much fewer works attempting to characterise granulation on RV; we summarise these in Appendix\,\ref{App:Granul}. For our simulations we follow the PSD characterisation of \cite{2023A&A...669A..39A} on Sun RV data. The authors used two components and fix the slope $\alpha$ at 2. Since the importance of scaling the parameters is still a matter of debate, we performed a first set of simulations with the quantities measured for the Sun and for a second we applied the scaling relationships on Eq.\,\ref{Eq:numax} to derive the amplitude and granulation timescales.
For the sake of simplicity and in accordance with the literature, we refer to the two components as granulation and super-granulation.

The RVs were calculated as in \cite{2023A&A...669A..39A}, decomposing the power spectra into frequencies, $\nu,$ with width, $\Delta\nu,$ that generate signals with amplitude $\sqrt{PSD(\nu)\,\Delta\nu,}$ and randomised phase. The integration limits are set by the limits of our knowledge: since we cannot characterise granulation on timescales longer than 100\,d, our lowest frequency was set at 0.1\,$\mu$Hz; given we cannot sample signals with higher cadence than 1 minute (our Nyquist frequency), we set the upper limit at 6\,mHz. The integration is done in steps of 1\,$\mu$Hz. 
For the sake of simplicity, we perform the simulations for \texttt{CCF} and \texttt{TM} only. The results for the effect of the granulation on the long-term uncertainties, $e_g$, and on the nightly scatter, $\sigma_g$, are presented in Table\,\ref{Tab:granulsimul}.

\begin{table}[h]
\centering
\caption{Effect of granulation simulated with \cite{2023A&A...669A..39A} prescriptions.}

\begin{tabular}{ l c | c c c c | c c c c }
 \hline \hline

\multirow{4}{*}{Star} & & \multicolumn{4}{c|}{Solar values} & \multicolumn{4}{c}{Scaled values} \\
 & & \multicolumn{2}{c}{$\tau_{gran}$} & \multicolumn{2}{c|}{$\tau_{s-gran}$} & \multicolumn{2}{c}{$\tau_{gran}$} & \multicolumn{2}{c}{$\tau_{s-gran}$} \\
 & & $e_g$ & $\sigma_g$ & $e_g$ & $\sigma_g$ & $e_g$ & $\sigma_g$ & $e_g$ & $\sigma_g$ \\
 & & \multicolumn{4}{c|}{\tiny{[cm/s]}} & \multicolumn{4}{c}{\tiny{[cm/s]}} \\
\hline 

\multirow{2}{*}{\small{HD\,10700}} & \small{\texttt{CCF}} & 23 & 21 & 65 & 17 & 19 & 21 & 59 & 17 \\
& \small{\texttt{TM}} & 22 & 21 & 64 & 16 & 20 & 20 & 58 & 16 \\
\hline

\multirow{2}{*}{\small{HD\,20794}}  & \small{\texttt{CCF}} & 25 & 21 & 65 & 16 & 26 & 21 & 68 & 16 \\
& \small{\texttt{TM}} & 24 & 20 & 68 & 15 & 26 & 20 & 69 & 15 \\
\hline 

\multirow{2}{*}{\small{HD\,102365}}  & \small{\texttt{CCF}} & 25 & 22 & 66 & 18 & 28 & 22 & 71 & 18 \\     
& \small{\texttt{TM}} & 25 & 20 & 66 & 16 & 27 & 21 & 71 & 16 \\

\hline
\end{tabular}

\label{Tab:granulsimul}
\end{table}

The results with the solar and scaled values are similar, with the scaled values showing a larger variation across stars, up to 5\,cm/s. 
The granulation component introduces an uncertainty of 19-28\,cm/s on the nightly averaged RVs; on the other hand, the super-granulation component introduces an uncertainty of 58-71\,cm/s. Either component introduces an intra-night scatter of 15-20\,cm/s.

With this model, the granulation or super-granulation components impact on the intra-night scatter remains of around 25\,cm/s across stars, and scaled / unscaled modelling choices. When one discounts the effects of pulsations, this estimation is in excess of the values measured for HD\,10700 but remains compatible with the scatter measurements on the other two stars. However, the effect of super-granulation, of 65-70\,cm/s, is not averaged out by our observation strategy and should be easily detectable as a parasitic effect on the night-to-night RV variation.

%--------------------------------------------------------------------

\section{Nightly averaged time series}\label{Sec:RVnight-to-night}

We perform an analysis of the nightly averaged RV time series of the targets HD\,10700 and HD\,102365, and on those of HD\,304646, for which a single observation per night was taken. The analysis of nightly averaged time series of HD\,20794 was presented in \cite{2025A&A...693A.297N}.

\subsection{Dataset properties and simple statistics}\label{Sec:RVstats}

\begin{figure*}[t]
\center
\includegraphics[width=0.95\textwidth]{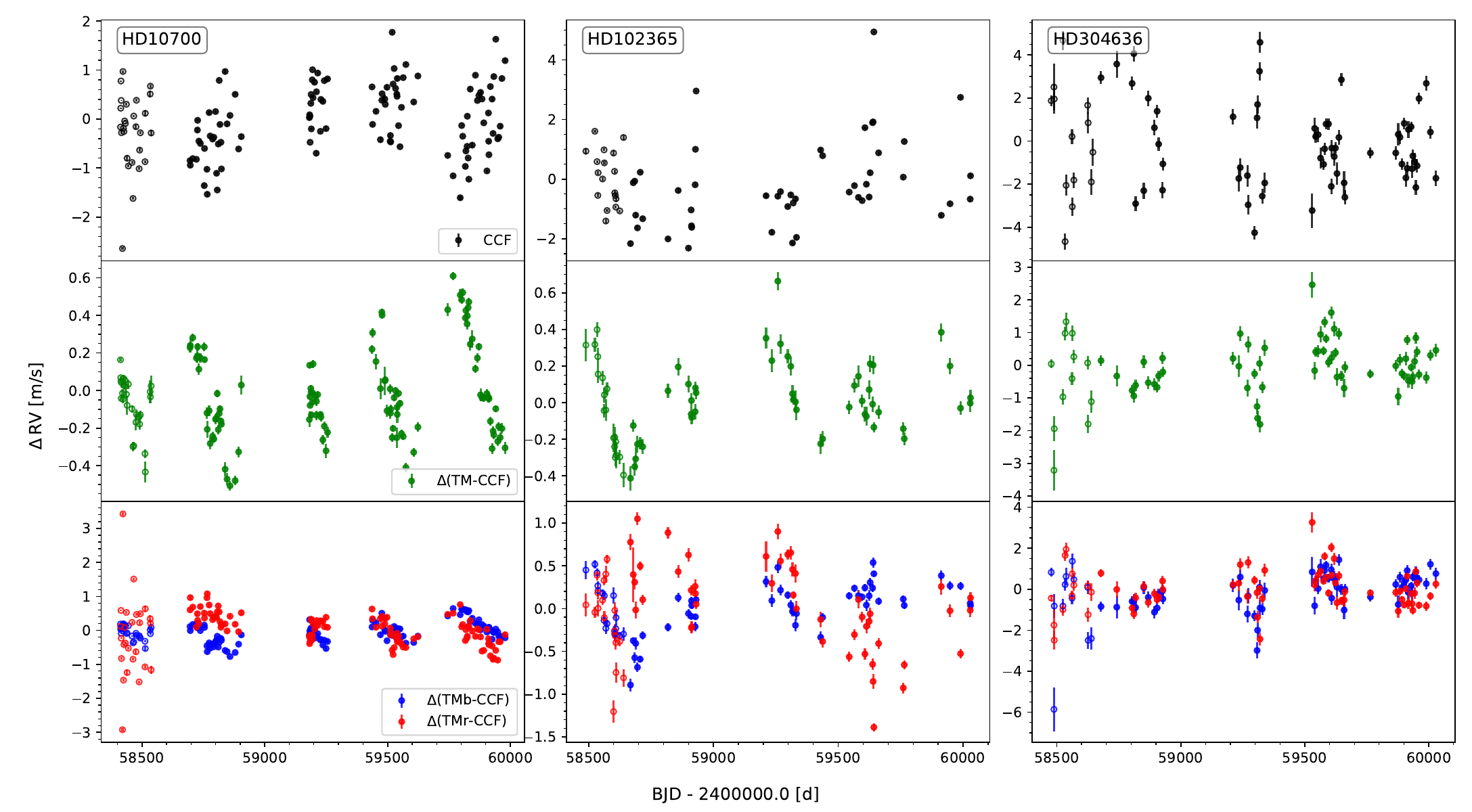}
\caption{RV time series using the different RV calculation methods \texttt{CCF} (\textit{top}) and difference relative to it of \texttt{TM}(\textit{center}), \texttt{TMb}, and \texttt{TMr} (\textit{bottom}) for the stars HD\,10700 (\textit{left}), HD\,102365 (\textit{center}), and HD\,304636 (\textit{right}). \texttt{ESPR18} data are represented by open circles while \texttt{ESPR19} data are represented by filled circles.}\label{Fig:timeseries}
\centering
\end{figure*}

The RV time series are shown in Fig.\,\ref{Fig:timeseries} and simple statistics listed in Table\,\ref{Tab:RVstimeseries}. 
For each star, dataset, and RV timeseries we list the number of nights and observing time span, plus the average RV uncertainty (calculated only from the photon noise) and weighted rms of the nightly averaged RVs. 

\begin{table*} 
\centering

\caption{Simple statistics for night-averaged RVs using the different RV calculation methods.}
\label{Tab:RVstimeseries}

\begin{tabular}{c c c c|c c|c c c |c c c }
\hline  \hline

 \multirow{3}{*}{Star} & \multirow{3}{*}{dataset} & \multirow{3}{*}{\#\,pts} & \multirow{3}{*}{$\Delta t$} & \multicolumn{2}{c|}{\texttt{CCF} RVs} & \multicolumn{6}{c}{\texttt{S-BART} RVs}  \\
    &   &   &   & avg. err  & w. rms & \texttt{TM}                                   & \texttt{TMb} & \texttt{TMr} & \texttt{TM}  & \texttt{TMb} & \texttt{TMr} \\
    &   &   &  \tiny{[d]} &   \multicolumn{2}{c |}{\tiny{[cm/s]}}   & \multicolumn{3}{c |}{avg. err \tiny{[cm/s]}}  & \multicolumn{3}{ c}{w. rms \tiny{[cm/s]}}  \\ 
    \hline 

 \multirow{3}{*}{HD\,10700}                & full              & 136                      & 1565.9                      & 3                                           & 68                      & 3                                             & 3            & 5            & 67                         & 70           & 69           \\
                & \texttt{ESPR18}        & 27                       & 124.9                       & 3                                           & 66                      & 3                                             & 3            & 5            & 70                         & 68           & 76           \\
                & \texttt{ESPR19}        & 109                      & 1282.7                      & 3                                           & 69                      & 3                                             & 3            & 5            & 66                         & 70           & 66           \\ 
 \multirow{3}{*}{HD\,102365}              & full              & 65                       & 1541.0                      & 6                                           & 151                     & 5                                             & 6            & 8            & 152                        & 164          & 127          \\
               & \texttt{ESPR18}        & 18                       & 150.8                       & 6                                           & 89                      & 5                                             & 6            & 8            & 100                        & 108          & 95           \\
               & \texttt{ESPR19}        & 47                       & 1362.3                      & 5                                           & 166                     & 4                                             & 5            & 8            & 165                        & 177          & 136          \\ 
 \multirow{3}{*}{HD\,304636}              & full              & 74                       & 1550.8                      & 40                                          & 187                     & 21                                            & 34           & 27           & 173                        & 197          & 163          \\
               & \texttt{ESPR18}        & 13                       & 167.7                       & 51                                          & 251                     & 28                                            & 44           & 36           & 211                        & 255          & 187          \\
               & \texttt{ESPR19}        & 61                       & 1350.2                      & 38                                          & 176                     & 20                                            & 32           & 25           & 167                        & 187          & 159          \\

 \hline

\end{tabular}
\end{table*}

On HD\,10700, \texttt{CCF} \texttt{TM} and \texttt{TMb} show an average uncertainty of 3\,cm/s, with \texttt{TMr} being larger at 5\,cm/s. The weighted rms is of 67-70\,cm/s for the different RV calculations; the similarity between \texttt{TM}, \texttt{TMb} and \texttt{TMr} suggests that the dominant source of scatter is achromatic.

For HD\,102365 there is a clear difference between the \texttt{ESPR18} and \texttt{ESPR19} datasets. While the average photon noise is similar, the rms of the nightly averaged RVs increases significantly from \texttt{ESPR18} to \texttt{ESPR19}. In contrast, HD\,304636 \texttt{ESPR18} shows a significantly higher rms than \texttt{ESPR19}. For both stars \texttt{TMr} shows a significantly lower scatter than \texttt{TMb}.  
This colour-dependent rms points towards an activity-induced flux-dominated RV variation, that is known to be mitigated at redder wavelengths, where the photometric contrast (or flux) effect between active regions is smaller. A variation of the instrument stability is excluded by the contemporaneous RVs of HD\,10700, that show a scatter of only 60-70\,cm/s over approximately the same period.

It is worth noting that for the two G-K stars HD\,10700 and HD\,102365 the effect of pulsations on the nightly average was estimated to be slightly larger than when considering photon noise alone: for HD\,10700 we estimate the uncertainties to increase up to 5\,cm/s, and for HD\,102365 up to 8\,cm/s.

\begin{figure*}
\center
\includegraphics[width=0.95\textwidth]{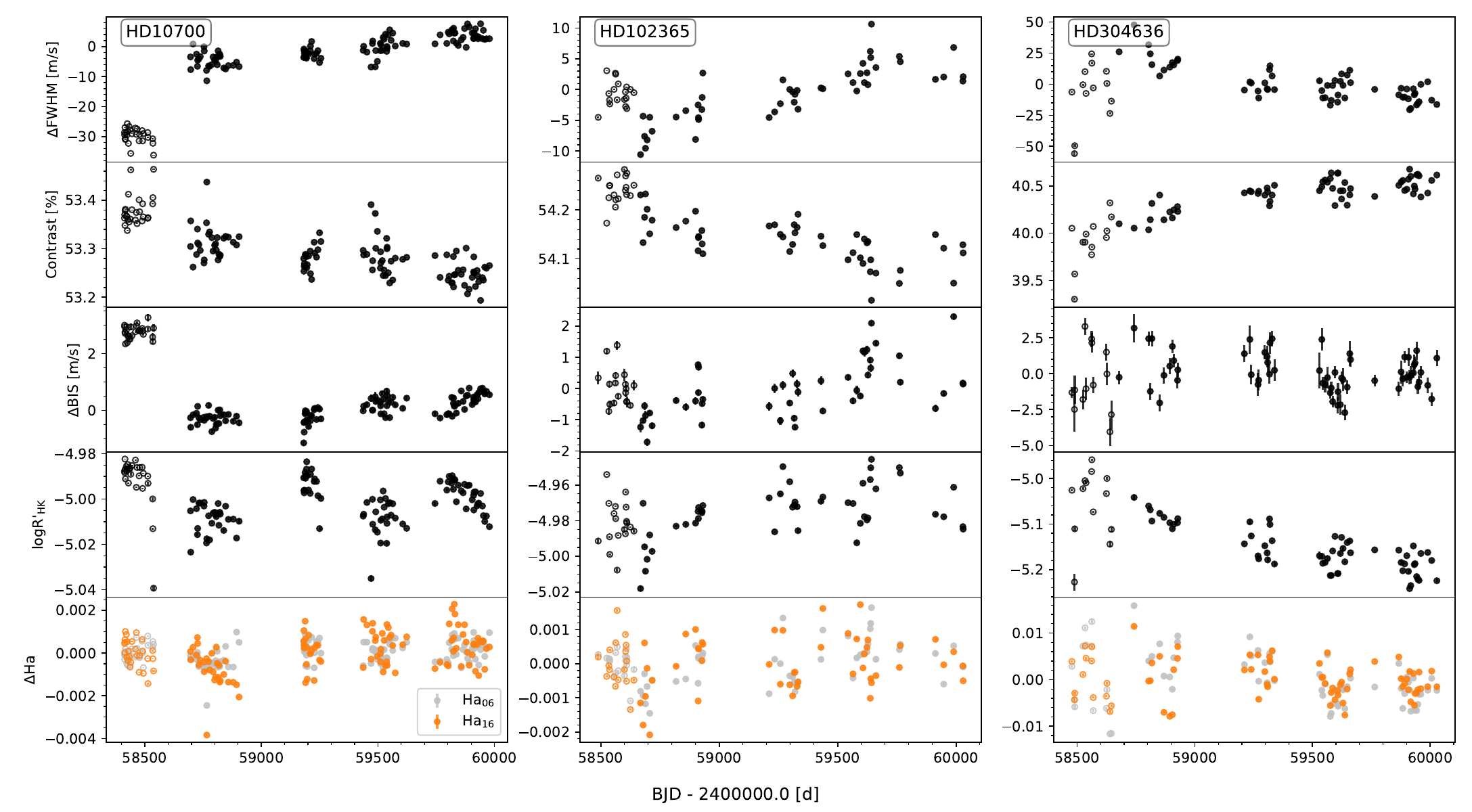}
\caption{Time series for the line profile and chromospheric indicators for the stars HD\,10700 (\textit{left}), HD\,102365 (\textit{centre}), and HD\,304636 (\textit{right}). \texttt{ESPR18} data are represented by open circles, while \texttt{ESPR19} data are represented by filled circles.}\label{Fig:Indtimeseries}
\centering
\end{figure*}

The temporal evolution of the stars in RV does not define a clear shape. In Fig.\,\ref{Fig:timeseries} for HD\,10700 we have what seems to be a periodic signal in the relative variation $\Delta$(\texttt{TM}\,-\,\texttt{CFF}), which we discuss in Sect.\,\ref{Sec:telcor}. The temporal evolution of the photospheric and chromospheric indicators is presented in Fig.\,\ref{Fig:Indtimeseries}. Their scatter is much larger than their formal uncertainties. The line asymmetry and width variations measured by the BIS and FWHM are not expected to vary for observations of very quiet stars with RV-dedicated spectrograph, that were built to have stable instrumental profiles. As such, the variability shown could actually represent an activity effect impacting the RV. 

\cite{2011arXiv1107.5325L} had already noted that long-term RV trends caused by the stellar magnetic cycles would be detected as opposite sign trends on the FWHM and the contrast indicators; this might be at play here.
However, a very slow monotonic variation in resolution over the years\footnote{As can be seen using the data from ESPRESSO \href{https://www.eso.org/observing/dfo/quality/ESPRESSO/reports/HEALTH/trend_report_WAVERESO_HR_HC.html}{ESO Health Check} webpage.} can explain small-amplitude indicator variations as seen on HD\,10700. 

The log($R'_{HK}$) chromospheric indicator is known to show variability even for the lowest activity stars. The Sun shows a variation in log($R'_{HK}$) of 0.05 due to rotational modulation during its low-activity phase \citep[e.g.,][]{2019ApJ...874..107M, 2019A&A...627A.118M}, and low-activity solar-like stars, like the ones targeted commonly by RV searches, show similar amplitudes \citep{1985ApJ...294..310B}. \cite{2021A&A...646A..77G}, showed that even HD\,10700, one of the lowest variability stars within their HARPS sample, has a variability of 1.6\%, that corresponds to 0.017. This matches well the ESPRESSO observations presented here.

Overall, the long-term ($\gtrsim$100\,d) evolution of FWHM, Contrast, BIS, and log($R'_{HK}$) on different stars and with different slopes is suggestive of a long-term activity evolution of the stars over the 4 years of our observations. 

\subsection{Presence of periodic signals}\label{Sec:RVPeriods}

\begin{figure*}
\center
\includegraphics[width=0.95\textwidth]{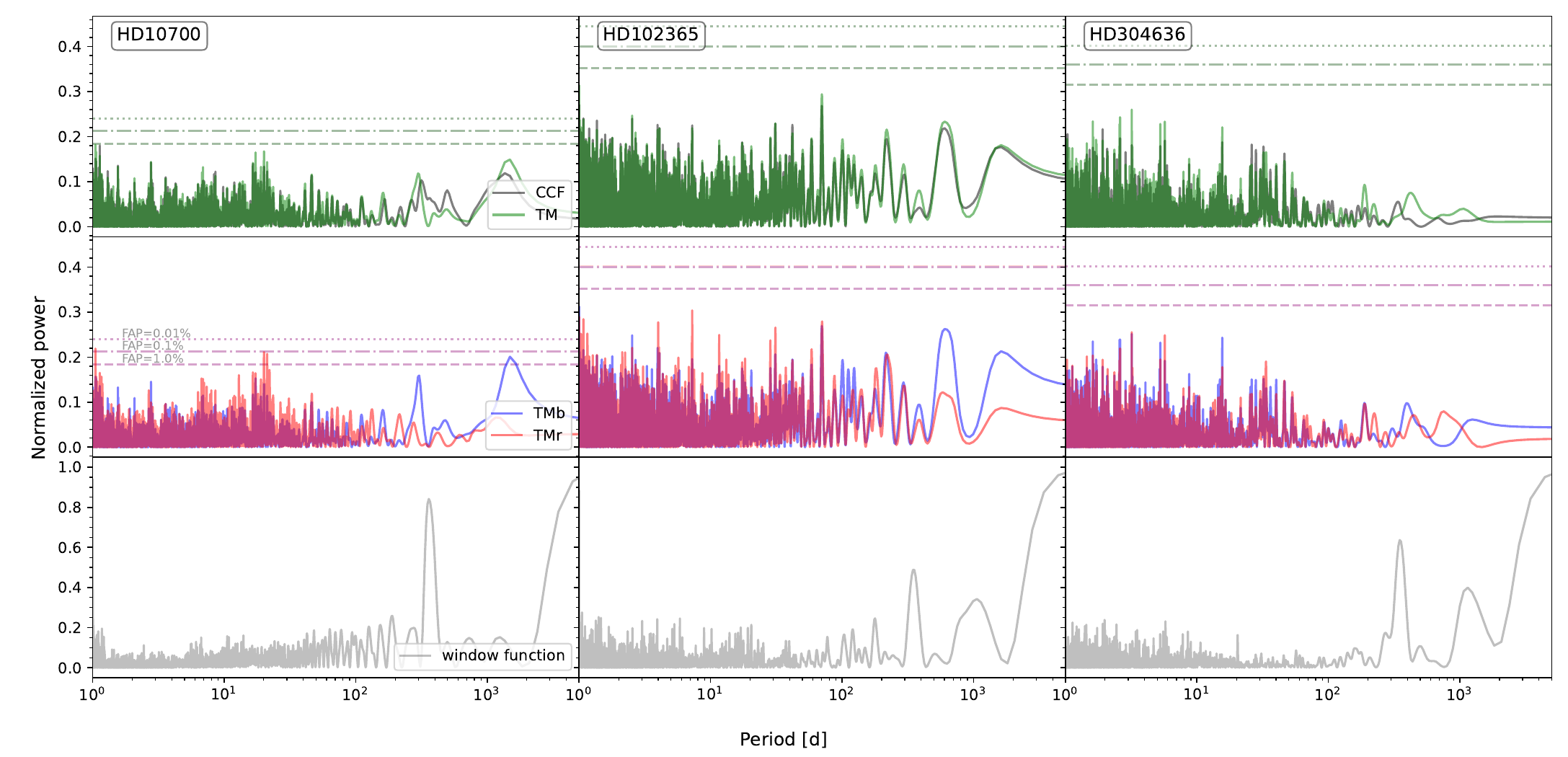}
\caption{Lomb-Scargle periodogram for the different methods of RV calculation and associated window function for the stars HD\,10700 (\textit{left}), HD\,102365 (\textit{center}), and HD\,304636 (\textit{right}). The FAP levels of 1, 0.1, and 0.01\% are shown as horizontal grey lines.}\label{Fig:glsRVs}
\centering
\end{figure*}

We computed the Lomb-Scargle periodogram \citep[LS, ][]{1976Ap&SS..39..447L, 1982ApJ...263..835S} in its floating-mean form (known as generalised Lomb-Scargle, GLS) as introduced by \cite{2009A&A...496..577Z} and implemented in \texttt{astropy}. The false-alarm probability (FAP) is computed as defined on \cite{2008MNRAS.385.1279B}. We chose three FAP threshold probabilities of 1\%, 0.1\% and 0.01\%, and a maximum frequency of 1\,d$^{-1}$ for the GLS and FAP calculations. The LS periodogram for the different RV sets is shown in Fig\,\ref{Fig:glsRVs}, along with the different FAP. Given the observations time span is $\sim$1500\,d we consider the longest period that can be confidently detected is of the order of 1000\,d; variations longer than this value are better modelled as linear or quadratic slopes and a frequency-oriented discussion is not particularly insightful.

For HD\,10700 the different RV time series exhibit different significant peaks: \texttt{CCF} at 1.13\,d, \texttt{TM} at 1.05\,d and 1.13\,d (the three at (1\% FAP) and \texttt{TMr} a doublet at 20.07 and 21.23\,d (reaching 0.1\% FAP). \texttt{TMb} shows a significant peak at 1500\,d (between 1\% 0.01\% FAP). No significant signals are detected on HD\,102365 or HD\,304636.

The peaks of the LS periodogram do not represent directly the best-fit matches to a sinusoidal signal. The amplitude of the periodogram at a given frequency depends not only on the signals present at this frequency but but also on signal spectral leakage, that depends directly on the temporal sampling. This effect is represented by the Window function, estimated by setting the measured RV signal at 1.0 \citep[see][]{2018ApJS..236...16V}; the window functions for the different stars is shown on the bottom row of Fig\,\ref{Fig:glsRVs}. 
The strongest individualised peak is at approximately one year (360\,d for HD\,10700, 350\,d for the other two stars), only topped by the broad signals with periods longer than 2000\,d. The window functions of HD\,102365 and HD\,304636 also show strong power at 1000\,d. It is worth noting the doublet seen in HD\,10700 \texttt{TMr} are aliases if one considers a sampling frequency of approximately year.

The FAP is known to be a reliable estimator, as long as the uncertainties and the frequency grid are correctly defined \citep[see e.g.][for a discussion]{2020A&A...635A..83D}. Unfortunately, so far we have only considered photon-noise uncertainties for the error budget. If we repeat the GLS and FAP calculations adding quadratically 10\,cm/s to the individual photon noise error bars the signals of HD\,10700 become non-significant, and the signals at 310-330\,d in \texttt{CCF}, \texttt{TM}, and \texttt{TMb} approach 1\% FAP. This serves as a word of caution on the interpretation of the frequency analysis.

\vspace{0.3cm}

The LS periodogram evaluates the least-square fit of a single sinusoidal function to the data and assumes that the formal uncertainties describe the variance of the dataset. These two assumptions can be lifted by considering the $\ell_1$ periodogram \citep{2017MNRAS.464.1220H}, that uses sparse recovery tools to represent a RV times series as linear combinations of a small number of sinusoidal functions. The tool also allows us to define the noise model via a covariance matrix $V$, and we adopt the representation with element $V_{kl}$ of the form

\begin{equation}
    \hspace{0.5cm} V_{kl}\,=\,\delta_{kl}(\sigma_k^2 + \sigma_W^2) + \sigma_R.e^{-\frac{(t_k-t_l)^2}{2\tau^2}}
,\end{equation}

\noindent 
where (on top of the individual nominal uncertainties, $\sigma_k$), we consider a white noise, $\sigma_W$, common to all data and a red noise with standard deviation, $\sigma_R$, and a timescale, $\tau$.

To solve the basis pursuit model we used the default choice of the least-angle regression (LARS)\footnote{\url{https://en.wikipedia.org/wiki/Least-angle_regression}} algorithm. Significance is estimated with the FAP, defined for the sequence of peaks, namely, the first value evaluates the presence of the highest amplitude signal, the second value evaluates the presence of the two highest amplitude signals, and so on, all calculated relative to the null hypothesis of no signals present in the data. We used $\ell_1$ on our RV datasets, restricting our analysis to periodicities longer than 1\,d.

\begin{figure*}
\center
\includegraphics[width=0.8\textwidth]{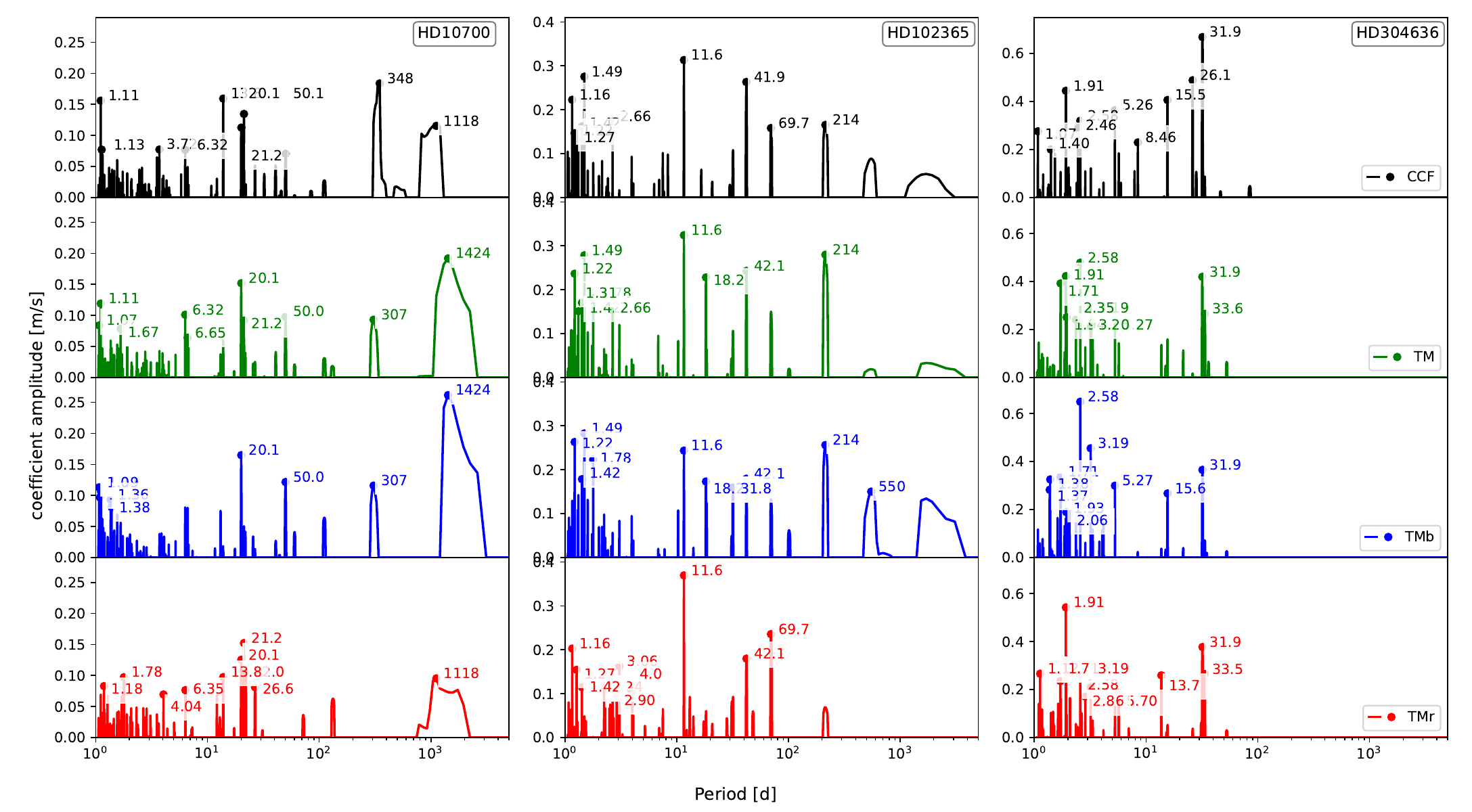}
\caption{$\ell_1$ periodogram for the stars HD\,10700 (\textit{left}), HD\,102365 (\textit{centre}), and HD\,304636 (\textit{right}). The ten strongest peaks periods are identified in the image; the associated FAP is listed in Table\,\ref{App:l1FAP}.}\label{Fig:l1}
\centering
\end{figure*}

In a first analysis we considered a white noise component of 10\,cm/s. The resulting periodograms are shown in Fig\,\ref{Fig:l1}, with the ten strongest peaks represented. The calculated FAP values for the three stars and different RV calculation methods are listed in App.\,\ref{App:l1FAP}\,.

The detected signals have either a FAP larger than 1\% or a period very similar to our temporal baseline. For HD\,10700 \texttt{CFF} show a 3\% FAP signal close to one year, while \texttt{TM} and \texttt{TMb} show a significant signal at 1423.5\,d. A signal with 20\,d in \texttt{TM} and \texttt{TMb} and a very similar signal in \texttt{TMr} show a FAP of a few percent. For HD\,102365 there is no formally significant signal and for HD\,304636, we would need a combination of 10 (\texttt{CCF}), 6 (\texttt{TMb}), or the 10 peaks (\texttt{TMr}) to reach an FAP of 1\%. 

The forest of peaks on the $\ell_1$ periodogram is characteristic of noisy datasets that do not have a sparse representation in frequency, as discussed in \cite{2020A&A...636L...6H}. In such a case, we  should evaluate the impact of the noise model and to test models including correlated noise. We followed the cross-validation methodology described in Appendix B.1 of \cite{2020A&A...636L...6H} and compared models with white noise $\sigma_W$ of 10, 20, 40, and 60\,cm/s combined with $\sigma_R$ with a value of 0 (no red noise component), 20, and 60\,cm/s and timescales of 0.02 and 0.54\,d. These amplitudes and timescales correspond roughly to the two timescales and amplitudes of granulation measured on the Sun. An average of the 20\% of the models with the highest cross-validation ranking still led to FAP that were, at best, much larger than 1\%; the only exception were the already detected signals at 1423.5\,d for HD\,10700.

\vspace{0.3cm}

In summary, there is only one periodic signal detected in RV with a period below 1000\,d. It is present in the different template matching RV time series of HD\,10700 and has an approximate period of 20\,d. The signal is statistically significant when measured with an LS analysis but only if photon-noise error bars are considered, and with varying significance between 1\% and 0.05\%, depending on the RV reduction. On $\ell_1$ the signal is present but with FAP of a few percent, and cannot be considered significant. The other signals detected are a consequence of the window function effect. 

\subsection{Activity indicators study: periodicity and correlations}\label{sec:actinds}

\begin{figure*}
\center
\includegraphics[width=0.95\textwidth]{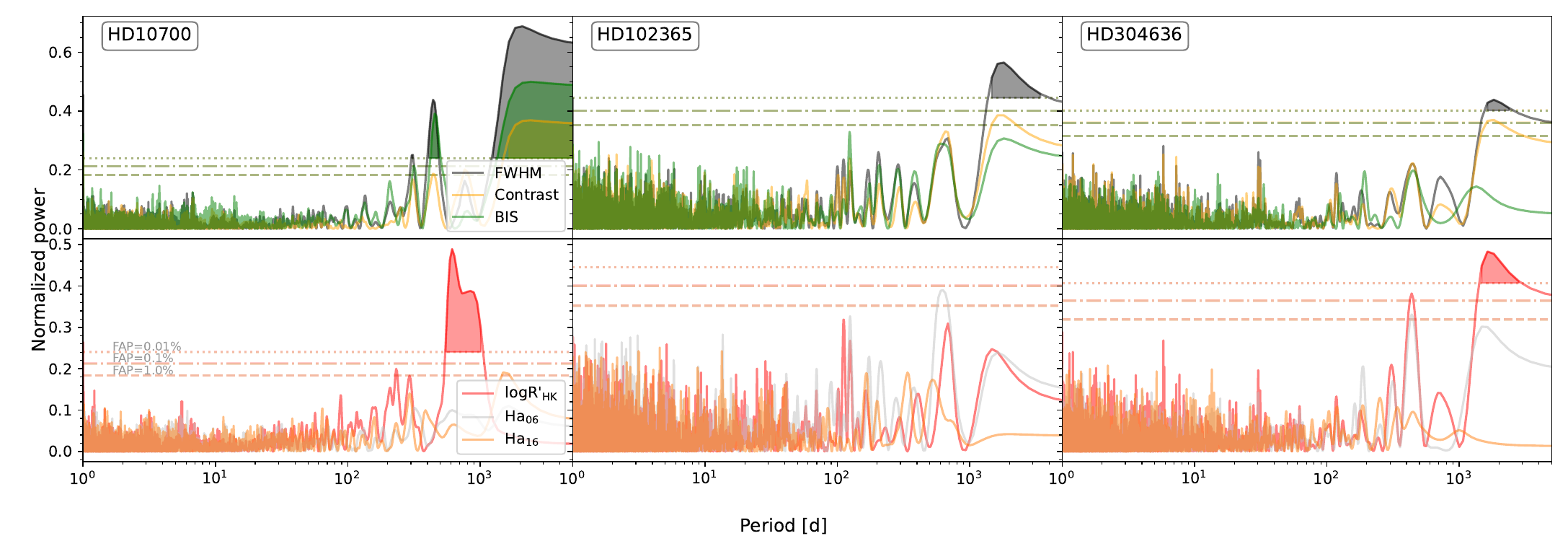}
\caption{Lomb-Scargle periodogram for the activity indicators for the stars HD 10700 (left), HD 102365 (centre),
and HD 304636 (right). On the upper panels we have the photospheric indicators FWHM, line contrast, and BIS, and on the bottom on we have log($R'_{HK}$) and the two Ha parametrisations. The FAP levels of 1, 0.1, and 0.01\% are shown as horizontal grey lines. }\label{Fig:glsInd}
\centering
\end{figure*}

We perform an LS periodogram analysis for the indicators as previously done for RVs. The results are presented in Fig.\,\ref{Fig:glsInd}.
By construction, the window function for the indicators is almost identical to that of the RV data and was therefore omitted.

For HD\,10700, the different photospheric indicators show peaks at common periods: 307\,d, and 440\,d, and a wide peak from 1250\,d approximately ($<$0.01\% FAP). The log($R'_{HK}$) indicator has a peak at 230\,d (1\% FAP) and a much wider and stronger peak from 550 to 1050\,d ($<$0.01\% FAP).

HD\,102365 shows highly significant power for periods longer than 1380\,d in FWHM ($<$0.01\% FAP) and between 570 and 690\,d in Ha$_{06}$ (between 0.1\% and 1\% FAP). While never significant at 1\%, it is interesting to note all indicators show a peak at 125\,d; log($R'_{HK}$) even shows a doublet at 119\,d, which are the aliases pair for a sampling frequency of 1000\,d. 

For HD\,304636, we have a peak in log($R'_{HK}$) and Ha$_{06}$ at 437\,d (0.1\% and 1\% FAP, respectively). The FWHM and log($R'_{HK}$) show significant power above 1500\,d.

\begin{figure*}
\center
\includegraphics[width=\textwidth]{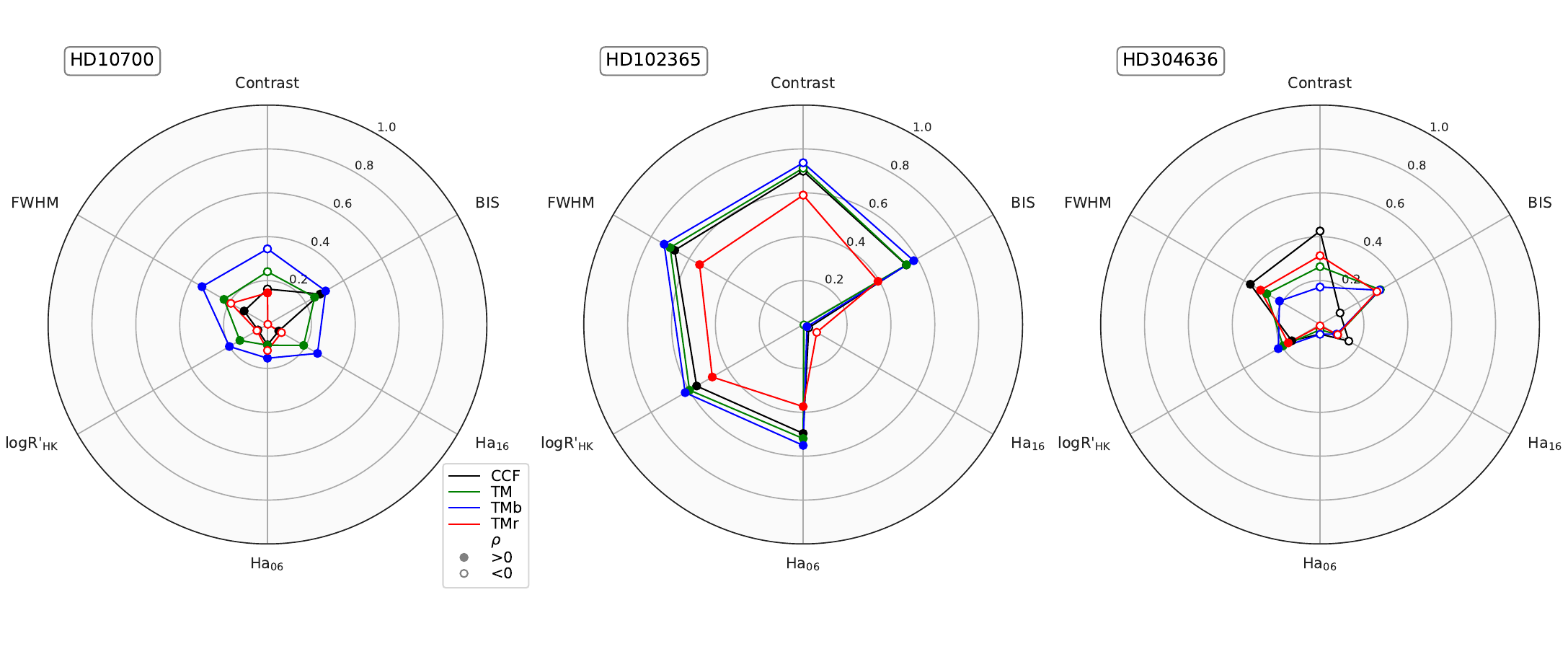}
\caption{Radar chart of the correlation between nightly averaged RV and each of the indicators, for the four RV calculation methods and three stars. The different colours represent the different RV datasets, with filled or open dots representing positive or negative signed Pearson correlation coefficients.}\label{Fig:radarcorr}
\centering
\end{figure*}

The presence of significant signals on a timescale similar to the observing span motivates a linear fitting on the indicators and analysis of the residuals. For HD\,102365 on log($R'_{HK}$) and Ha$_{\mathrm{06}}$, we detect a significant signal at $\sim$123\,d, very similar to the announced planet period at 122\,d. On HD\,3046336, the period of 30.3\,d is detected on log($R'_{HK}$), close to the expected rotation period.
We checked for correlations between nightly averaged RVs and indicators using
the Pearson correlation coefficient and display these values in the radar
chart in Fig\,\ref{Fig:radarcorr}.

\vspace{0.3cm}

The correlation coefficients absolute values for HD\,10700 and HD\,304636 are below 0.4, showing weak correlations, if present at all. On the other hand, for HD\,102365 the coefficients values range from 0.4 to 0.8. The only exception is Ha$_{16}$, that is arguably poorly adapted to this spectral type. Interestingly, for \texttt{CCF}, \texttt{TM}, and \texttt{TMb} the photospheric indicators FWHM and contrast show a correlation of 0.7. The explained variance is given by R$^2$, that in the simplistic case of a linear model is given by the square of the coefficient of correlation. In this approximation, the correlation coefficient value of 0.7 sets an important threshold: half of the variance is explained by the correlation. We have then clear evidence that HD\,102365 RVs are contemporaneous with activity.

The study of periodicities of RVs and their correlations with activity indicators is insightful but incomplete. Activity-induced signals may not be periodic, and remain undetected in a periodicity analysis; they can also show a time (or phase) lag relative to the associated RV signal and thus lead to a null correlation coefficient while being physically connected \citep[see e.g.][for an example on the Sun]{2019MNRAS.487.1082C}. For a complete analysis one must model simultaneously activity and potential planetary signals.

\section{Radial velocity analysis with \texttt{kima}}\label{Sec:RVkima}

We used the {\tt kima} package\footnote{\url{https://github.com/kima-org/kima}} \citep{2018JOSS....3..487F} to evaluate the number of planetary signals present in the data while considering a realistic modelling of activity. \texttt{kima} uses the diffusive nested sampling (DNS) algorithm from \cite{10.1007/s11222-010-9198-8} to sample from the posterior distribution of the model parameters; this method delivers an estimate for the marginal likelihood for each model and enables direct model comparisons \citep[see e.g.][]{2014arXiv1411.3921B, 2011MNRAS.415.3462F}. In this context, the number of detected planets $N_p$ is a parameter that can be left free or fixed for the analysis. For each model we obtain at least 50 000 effective samples from the posterior, which allows for a robust sampling of the different parameters.

\subsection{Modelling with a quasi-periodic GP}\label{sec:kimabestmatch}

\texttt{kima} is able to make use of a Gaussian process (GP) with a quasi-periodic kernel as a noise model to accurately represent activity-induced signals; the hyper-parameters $\eta_1$-$\eta_4$ of the GP are inferred together with the planetary orbital parameters, star background velocity, and instrument-specific noise and offset values. $\eta_1$ is the amplitude of the GP quasi-periodic kernel, $\eta_2$ is the evolution timescale, $\eta_3$ the rotational period it tries to represent and $\eta_4$ the signal's harmonic complexity. This activity signal representation has had its origin in photometric activity modelling and has seen a large number of applications and success on RV modelling. For an early implementation see \cite{2014MNRAS.443.2517H}; for a recent revision work see \cite{2022MNRAS.515.5251N}. The choice of priors for each of the parameters is discussed in detail in Appendix\,\ref{App:priors} and listed in Table\,\ref{Tab:kimapriors}.

The best-fit parameters for the different stars and RV calculation methods is presented in Table\,\ref{Tab:paramsGPs}. Using the quasi-periodic kernel and the previously mentioned priors we could reproduce a rotation-modulated activity signal with a period close to the expected $P_{rot}$ value, as listed in Table\,\ref{Tab:FinalParams}. 

\begin{figure*}[h]
\centering
\includegraphics[width=0.8\textwidth]{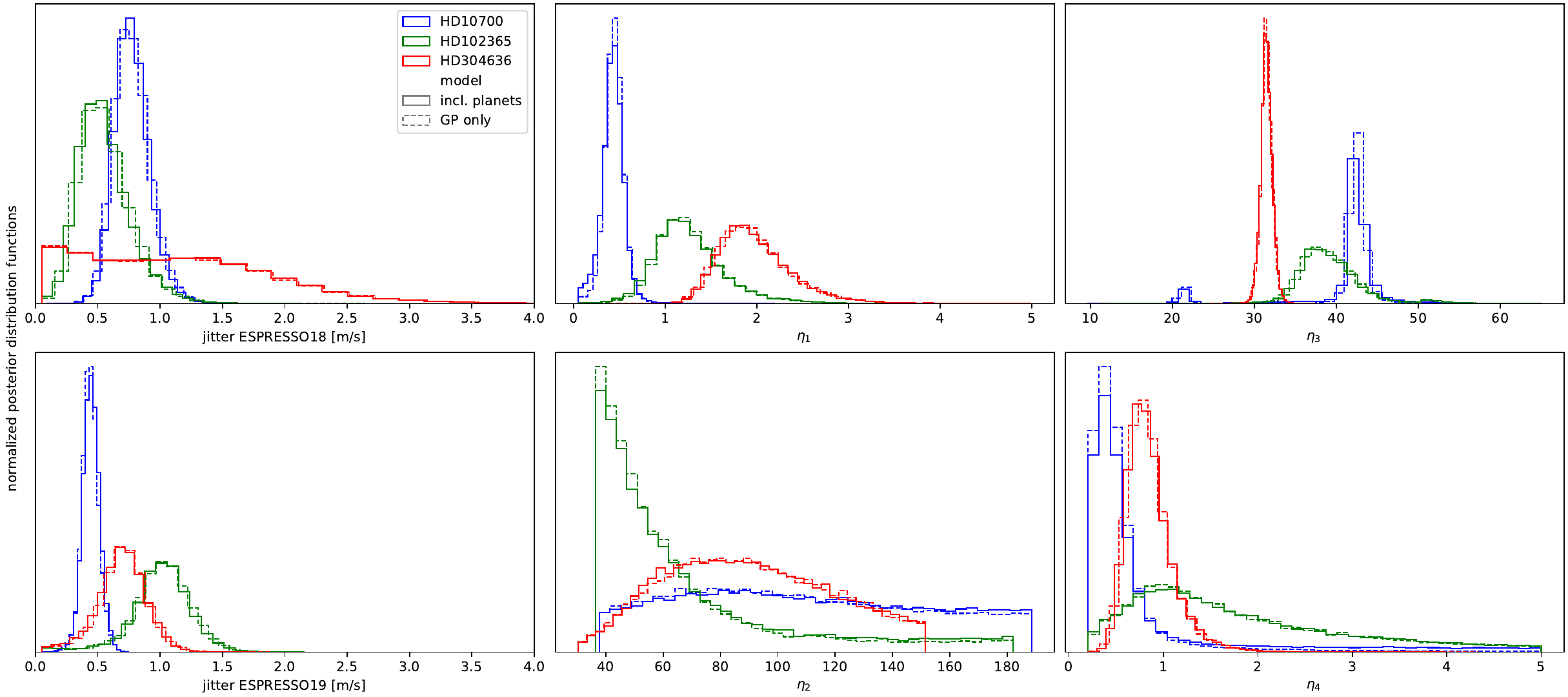}
\caption{Normalised posterior distribution of the two instrument jitter and four GP parameters as estimated by \texttt{kima}, for the three stars studied in the section and the \texttt{TM} RV calculation method. The distributions of the posteriors for GP-only (\textit{dashed line}) and up to one planet runs (\textit{solid line}) are both shown.}\label{Fig:postGPs_TM}
\centering
\end{figure*}

Letting the number of planets free between zero and one, no significant planetary signals were detected on any of the stars using any of the RV calculation methods. We repeated the analysis fixing the number of planets at zero to obtain the best characterisation of stellar activity; the posteriors for the case with variable planet number and fixing planet number at zero are almost identical. For a given star, the results across RV calculation methods are also very similar; the only exception are for $\eta_3$ and $\eta_4$ that for the \texttt{CCF} RV of HD\,10700 are very poorly constrained. The distribution of the key parameters for the \texttt{TM} RV calculation methods, with and without planets, are shown in Figure\,\ref{Fig:postGPs_TM}.

The modelling on HD\,10700 can be interpreted as successful for several reasons. The best-fit rotational period $\eta_3$ is of approximately 42\,d, halfway between the 38\,d $P_{rot}$ value estimate from spectroscopic calibrations and the 46\,d value from interferometric studies (see Sect\,\ref{Sec:Host}). Not only a rotation period but a first harmonic are detected, and an evolution timescale $\eta_2$ maximum is found at 2.5 times the rotation period, compatible with what is observed on the Sun \citep{2019MNRAS.487.1082C}. Moreover, the jitter on the long timespan dataset \texttt{ESPR19} and $\eta_1$ parameter distributions are approximately normal and have values of 45-50\,cm/s. However, $\eta_4$ is of 2.2 for the \texttt{CCF} dataset, which can lead to a degeneracy in the activity signal characterisation (as discussed on Appendix\,\ref{App:priors}). The RV offsets for the three stars are of the order of m/s, compatible with zero, and mutually compatible across RV calculation methods.

The corner plots are presented on Appendix\,\ref{App:corners}, and show correlation between different parameters. The linear slope and quadratic term of the stellar systemic trend are mutually correlated and correlated with the instrumental offset; the quadratic term is correlated with the systemic velocity. The jitter parameters for the three stars are slightly correlated with their GP parameters $\eta_2$ and $\eta_4$, with the correlations being less pronounced for HD\,10700. The correlation between jitter and $\eta_2$ / $\eta_4$ occurs when the jitter is very small and the timescales very short (with $\eta_2$ similar to the rotation period), and $\eta_4$ is at its lowest limit, creating a signal with very low complexity.

\vspace{0.3cm}

The classification of a detection as significant depends on the (Keplerian) planet model being favoured, which in turn depends on a chosen metric crossing an a-priori threshold. For exoplanet detection within the Bayesian Framework, the metric is typically the Bayes factor and the threshold 150 \citep{KassRaftery95}.
While no star shows a significant detection, the posterior on the number of planets of HD\,10700 has a similar number of samples for $N_p=$\,1 than for $N_p=$\,0. The posterior of the orbital period and the joint posterior of semi-amplitude period and eccentricity period for \texttt{TM} are shown in Fig.\,\ref{Fig:postorbsHD10700TM}. There is a pile-up of samples at periods of $\sim$1\,d and $\sim$20\,d; when looking at the semi-amplitude, we see that these extend from 15 to 60\,cm/s, with the eccentricity extending up to around 0.6. This cluster of posterior parameters does not correspond then to a well-characterised and well-defined signal, but are very variable parametrisations of tentative detections. The posteriors for the \texttt{CCF} RVs are similar, with a third clustering at a period of one year.

\subsubsection{Changing the priors: Highly constrained planet searches}

The comparison between planetary model and non-planetary model depends on the assumed priors. In particular, an orbital period constrained to regions of high likelihood (e.g., periodogram peaks) can boost the posterior evidence for the planetary model when compared to a wider parameter search. Whether the narrow priors have a physically sound justification remains a separate question.

To test the presence of planets in this skewed scenario we consider three cases of narrow searches around 20, 122, and 7.57\,d for HD\,10700, HD\,102365, and HD\,304636, respectively. We consider Gaussian priors around these periods, with a standard deviation of 20\% of the chosen period value. For HD\,10700 we choose the \texttt{TM} and for HD\,102365 and HD\,304636 we choose the \texttt{TMr} datasets as optimistic cases. The results remain largely unchanged; there is no significant planet detection even in this scenario. 

\subsubsection{Changing the model: Quasi-periodic joint kernel with FWHM}

We tested a joint modelling of RV and FWHM, as used in \cite{2022A&A...658A.115F}. The FWHM is known to track activity changes in a wide range of scenarios, but in our study it only shows a significant correlation with RVs for HD\,102365. None of the three stars has an RV or indicator time series with a significant periodicity at the expected stellar rotational period. This model is tried thus for purely exploratory reasons, without having any \textit{a priori} support from data inspection.

In our joint model the parameters $\eta_2$ and $\eta_3$ are shared between the RVs and FWHM variation, as a way of anchoring the rotational period and the evolution timescale. The parameter $\eta_{FWHM,1}$ is drawn from a log-Uniform between 0.1 and the maximum FWHM variation and the parameter $\eta_{FWHM,4}$ is drawn from the same prior as $\eta_4$. The rest of the analysis remains unchanged.

The joint model results are presented in Table\,\ref{Tab:paramsGPs}, and the posterior distributions are very similar to those seen in the previous section. For HD\,10700 the only noticeable difference occurs for $\eta_3$, that now has a very skewed distribution towards small values. For HD\,102365 $\eta_1$ and $\eta_4$ show larger values than seen in the previous section; $\eta_4$ being larger than 2.0 is associated with a higher model degeneracy. For HD\,304636 the posteriors are compatible, but the parameters $\eta_2$, $\eta_3$ and $\eta_4$ show narrower posteriors, indicating this model provides increased constraints.

\subsection{Telluric correction and its impact on RVs}\label{Sec:telcor}

Telluric water absorption features are present across the optical domain and as such cannot be masked out without a considerable loss in RV precision. With a small relative depth ($<$2\%) these are often referred to as micro-tellurics, and are expected to have an impact on RVs that is larger than 10\,cm/s over a wide range of conditions (target spectral types, airmass, precipitable water vapour content on the atmosphere), reaching up to 80\,cm/s on extreme cases. For more on the topic, we  refer to \cite{2014A&A...568A..35C}. 

\cite{2022A&A...666A.196A} presented a telluric absorption correction model of ESPRESSO and tested it on a smaller dataset on HD\,10700 than present here, identifying an effect on RVs that was of the order of 10\,cm/s rms, could reach a maximum value of 58\,cm/s, and had a significant impact on the detection of signals with a period close to one year. 

Since \texttt{CCF} and \texttt{TM} are subject to different telluric masking criteria, water micro-tellurics are a prime candidate for the modulation seen in $\Delta$(\texttt{CCF}\,-\,\texttt{TM}) (see Sect.\,\ref{Sec:RVstats}).
To test for this hypothesis and attempt a gain in RV precision we apply the CCF and \texttt{S-BART} RV computation to telluric-corrected spectra of HD\,10700.

We tested different telluric masking conditions on \texttt{S-BART}; on top of the baseline choice of masking regions affected by tellurics deeper than 1\%, we alternatively masked water features to deeper than 25\%, 40\%, and 50\%. The masking of non-water features remained fixed at 1\%. Since there is a larger density of water features in the redder optical wavelengths, the scatter of \texttt{TMr} is the most efficient metric for an improved correction. We notice that for the 25\% threshold there is a reduction in 6-8\,cm/s, and a negligible reduction for deeper thresholds. For 25\% one also has a reduction of 2-3\,cm/s rms in the \texttt{TM}. This is in line with the correction being at its most effective for shallow water lines, as intended. We keep then this threshold as our nominal \texttt{S-BART} correction threshold.

The night scatter for the telluric corrected spectra is presented in Table\,\ref{Tab:rmstelcorr}, and the rms can be compared directly with that of Table\,\ref{Tab:RVstimeseries}. The scatter on \texttt{CCF}$_{\mathrm{tc}}$ remains the same when compared with the uncorrected one very probably because the correlation masks were not extended in wavelength to make use of the now corrected wavelength range. 
An LS of the uncorrected-corrected \texttt{CCF} RVs and indicators show clear peaks at one year and fractions of it (Fig\,\ref{Fig:HD10700glstellcorr}), illustrating nonetheless the positive impact of the correction.

The \texttt{kima} analysis using the same model and priors as in Sect.\,\ref{Sec:RVkima} on the telluric corrected RVs is presented in Table\,\ref{Tab:paramsGPstc}. 
The telluric correction allows us to reach lower \texttt{ESPR19} jitter values for the different \texttt{TM} runs, between 41 and 45\,cm/s.

\begin{table}[H]
\centering
\caption{Scatter of night-averaged RVs HD\,10700 when telluric correction is applied.}\label{Tab:rmstelcorr}
\begin{tabular}{l|c c c c}
\hline \hline
  \multirow{2}{*}{dataset}  & \texttt{CCF}$_{\mathrm{tc}}$ & \texttt{TM}$_{\mathrm{tc}}$ & \texttt{TMb}$_{\mathrm{tc}}$ & \texttt{TMr}$_{\mathrm{tc}}$ \\ 
   & \multicolumn{4}{c}{(w. rms) \tiny{[cm/s]}} \\ \hline 
  full   & 67   & 64  & 68  & 62  \\
  \texttt{ESPR18}  & 67 &  68  &  67 & 68 \\
  \texttt{ESPR19}  & 68 & 63   & 69  & 59 \\
  \hline
\end{tabular}
\end{table}

\subsection{Compatibility limits}\label{Sec:compatlim}

\begin{figure}[h]
\includegraphics[width=8.2cm]{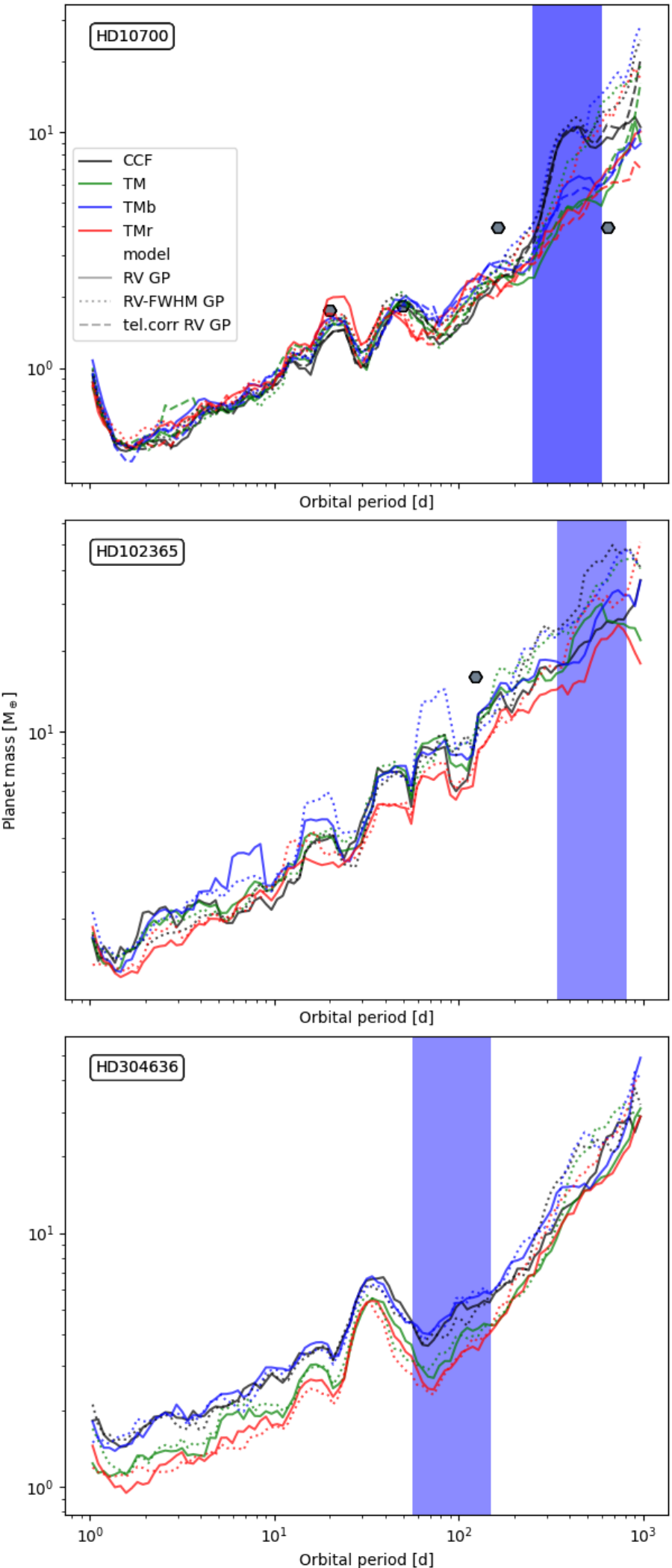}
\caption{Compatibility limits for the three stars studied, HD\,10700 (\textit{top}), HD\,102365 (\textit{middle}), and HD\,304636 (\textit{bottom}). The maximum semi-major RV amplitude of the samples is plotted for each of the RV calculation methods model, and telluric correction set; HZ limits as presented in Tab.\,\ref{Tab:FinalParams} and orbital properties of announced planed described in Sect.\,\ref{Sec:Targets} are shown.}\label{Fig:CompatLimitsRuns}
\centering
\end{figure}

When running \texttt{kima} one can fix $N_p$ to the number of detected planets plus one to calculate the parameters of undetected planets that are compatible with the data \citep[see][]{2022MNRAS.511.3571S}. The samples with the largest planet semi-major axis amplitude, $K_0$, in a given period bin will yield the most massive planet compatible with the data and with the priors. The compatibility limits produced in this way have important advantages when compared with the traditional detection limits from the injection-recovery method: the different orbital parameters are marginalised by construction and the results do not depend on signal injection properties. We calculate and compare the compatibility limits for the different RV time series in Fig.\,\ref{Fig:CompatLimitsRuns}; for HD\,10700 and HD\,102365 we mark the parameters of the planets announced in the literature.

For a given star the results for the different RV calculation methods are similar but not identical. 
For HD\,10700, as we increase in period we have a first bump (increase in equivalent planetary mass) at around 20\,d, the value of the posterior pile-up we saw in Sect.\,\ref{sec:kimabestmatch}, and the period of an announced planet. For periods longer than 300\,d the \texttt{CCF} RV calculation method shows a lower detection capability than the others, unfortunately for periods covering the star's HZ. However, using the telluric-corrected version of the RVs corrects for this effect, creating an almost linear limit between 100 and 1000\,d. While the gain is the most pronounced for \texttt{CCF}, it is also present for the other reductions.
For HD\,102365 and HD\,304636 \texttt{TMr} provides the lowest (i.e., most sensitive) limits. For HD\,102365 this is interpreted as the red wavelengths being less sensitive to star's activity. The different RV time series and different models yield slightly different compatibility limits, with the bumps showing a different amplitude; this points to these being created by the spurious fitting of activity signals. For HD\,304636 we obtain very similar detection limits when comparing pairwise \texttt{CCF} -- \texttt{TMb} and \texttt{TM} -- \texttt{TMr}. We can speculate that this is probably due to the template matching being more efficient at extracting RVs from M-dwarfs, that have most of the spectral information on the red detector, with the \texttt{CCF} being more similar to the information on the blue detector. This could be the effect of activity at play, leading to slightly lower limits measured with \texttt{TMr} than \texttt{TM}. For this star we have a clear bump at 30-40\,d, the expected rotational period of the star.

For each star we need to choose an RV calculation method to define the most representative compatibility limit. For HD\,10700 we choose \texttt{TM} as is the one that best constraints the GP parameters, while allowing for a lower detection limit to be reached up to the HZ. 
For HD\,102365 and HD\,304636 the lowest detection limits are obtained with \texttt{TMr}, that we thus select. A summary of the compatibility limits mass for key periods is shown in Table\,\ref{Tab:CompatLimits}. For HD\,10700 we are below 1\,M$_{\oplus}$ for periods shorter than 10\,d and at 1.7\,M$_{\oplus}$ for periods shorter than 100\,d; within the HZ we are sensitive to planets with a minimum mass above 2.5 to 5\,M$_{\oplus}$, approximately.

\begin{table}[H]
\centering

\caption{Compatibility limits in units of Earth mass for reference periods and the HZ period range.}
\label{Tab:CompatLimits}

\begin{tabular}{ l c c c c c c}
 \hline \hline

\multirow{2}{*}{Star} & 5\,d & 10\,d & 20\,d & 50\,d & 100\,d & HZ\\
 & \multicolumn{6}{c}{\tiny{[M$_{\oplus}$]}} \\
\hline 

HD\,10700 & 0.7 & 0.8 & 1.5 & 1.7 & 1.6 & 2.4-5.2 \\
HD\,102365 & 1.9 & 2.4 & 3.2 & 5.6 & 5.9 & 14-25 \\
HD\,304636 & 1.4 & 1.7 & 2.4 & 3.4 & 3.6 & 2.6-4.0 \\

\hline
\end{tabular}

\end{table}

\subsection{RV stability and precision floor}

Any RV variability not parametrised by $\eta_1$\,-\,$\eta_4$ will be captured by the jitter term of each dataset. While the dataset \texttt{ESPR18} spans only 150 days, \texttt{ESPR19} spans almost 4 years, with the data clustered into the months when a star is observable from Paranal observatory.

\begin{figure*}
\center
\includegraphics[width=0.9\textwidth]{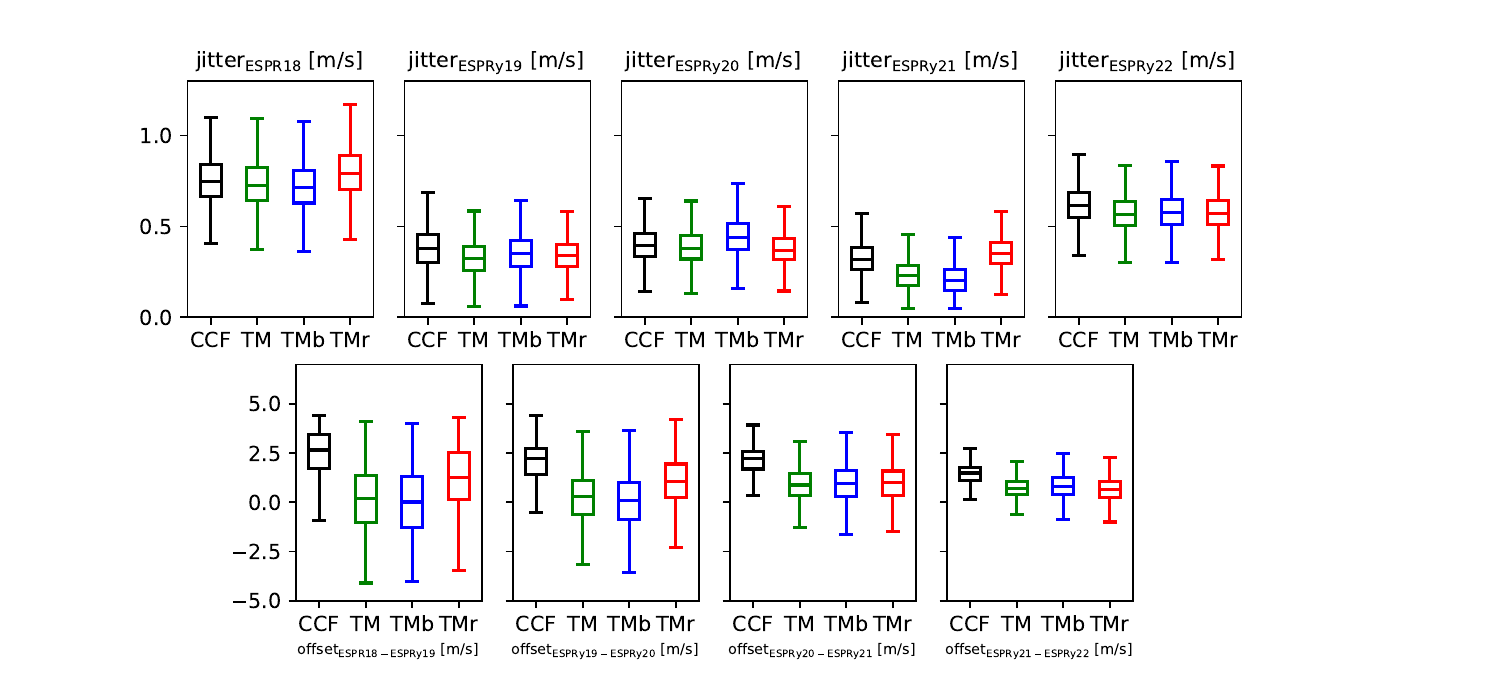}
\caption{Boxplots for the posteriors of the jitter (\textit{top}) and offsets between datasets (\textit{bottom}) for the different RV calculation methods with five datasets of telluric-corrected RVs of HD\,10700.}\label{Fig:5datasets}
\centering
\end{figure*}

For a further inspection of ESPRESSO stability and the RV precision limit of our model we take HD\,10700 telluric corrected data and subdivide \texttt{ESPR19} into four yearly datasets \texttt{ESPRy19} - \texttt{ESPRy22}, with 20 to 33 points each, timespans in the range 70-230\,d, and temporal gaps between subsets from 120 to 270 days, approximately. We use the previously described quasi-periodic kernel and priors within the GP to reproduce the RVs derived with the different methods. The \textit{ad-hoc} separation leads to the derivation of five jitter values and four offset ones; their posteriors are presented as box plots in Fig\,\ref{Fig:5datasets}. The dataset properties and measured values are presented in Table\,\ref{Tab:RVprec}.

\begin{table}
\centering

\caption{Subdivided datasets for the telluric-corrected spectra of HD\,10700 and associated jitter estimated via GP.}
\label{Tab:RVprec}

\begin{tabular}{ l c c | c c c c c}
 \hline \hline

\multirow{2}{*}{dataset} & \#\,pts & $\Delta t$ & \texttt{DRS} & \texttt{TM} & \texttt{TMb} & \texttt{TMr}\\
 & & \tiny{[d]} & \multicolumn{4}{c}{jitter \tiny{[cm/s]}}\\[3pt]
\hline 

\texttt{ESPR18} & 27 & 124.9 &74.6 & 72.7 & 71.2 & 79.1 & \\
\texttt{ESPRy19} & 30 & 208.6 &37.8 & 32.2 & 35.1 & 33.9 & \\
\texttt{ESPRy20} & 20 & 73.9 &39.4 & 37.9 & 44.0 & 37.0 & \\
\texttt{ESPRy21} & 26 & 186.7 &32.0 & 22.8 & 20.1 & 35.0 & \\
\texttt{ESPRy22} & 33 & 233.7 &61.2 & 56.7 & 57.6 & 57.3 & \\

\hline
\end{tabular}

\end{table}

The result from the different RVs is, once again, compatible; we focus our analysis on the results for \texttt{TM}. While the jitter for \texttt{ESPR18} remains at 73\,cm/s, for \texttt{ESPRy19} - \texttt{ESPRy22} this is of 32, 38, 23, and 57\,cm/s, respectively. The datasets have different number of points and span, with \texttt{ESPRy20} being particularly small and short; the results on it can be affected by small-number statistics. Nonetheless, the different sub-datasets show variable jitter, up to a factor of 2 to 3.5, depending on the reduction. The reason for the difference is unclear, and can come from either a variation in instrument stability or induced stellar activity. The different reductions have a minimum yearly timescale jitter between 20\,cm/s and 30\,cm/s.

%--------------------------------------------------------------------

\section{Discussion}\label{Sec:Disc}

\subsection{A forward model of pulsations}

The asteroseismology of main sequence stars experienced significant developments over recent decades. We refer to \cite{2019LRSP...16....4G} for a comprehensive review of the topic.
Our work uses simple principles and scaling equations to estimate the contribution of p-mode pulsations to averaged intra-night RVs and intra-night RV scatter. We take into consideration the time intervals covered by individual exposures and the detector readout-time gaps in the acquisition of consecutive integrations, as used in RV searches, and use individual p-mode frequencies instead of an envelope to represent the ensemble of oscillations. These two aspects are certainly behind the very good match with observations, and key differences between our others such as \cite{2019AJ....157..163C}.  

Some important shortcomings apply. There have been reports of significant correlations between surface activity indicators and solar-like oscillations properties for the Sun, with the oscillation frequencies increasing during higher activity periods. The commonly accepted explanation is that magnetic fields induce perturbations in the outer layers of the star, influencing the oscillation cavity of the p-modes. The exact mechanism behind this frequency shift remains unknown; see \cite{2008AdSpR..41..846H} and references therein. In addition, several other p-mode characteristics such as amplitude and widths change within an activity cycle \citep[e.g.][]{2011JPhCS.271a2030S}. A similar mechanism may be at play and detectable at very high RV precision. 

Asteroseismology enables the measurement of stellar parameters with exquisite accuracy. We could compare our own parameter estimation against those provided by asteroseismology runs for HD\,10700 and reached an excellent agreement. It was precisely on this very well-characterised star that we were able to reproduce pulsations scatter to better than 10\,cm/s. We cannot know if the mismatch between simulations and observed values for HD\,102365 and HD\,20794 comes from a higher activity impact on pulsations or from poorly determined parameters; either can be responsible for the mismatch, without one having to invoke additional sources of intra-night scatter.

In the last section we saw that for the GP model on telluric-corrected HD\,10700 RVs using \texttt{TM}\,/\,\texttt{TMb}\,/\,\texttt{TMr} the best estimate for the jitter of \texttt{ESPR19} (measured over 3.5\,yr) was of approximately 40\,cm/s. This value is very similar the intra-night scatter generated by pulsations. The similarity might be coincidental, but in the worst-case scenario can be the consequence of the pulsations intra-night scatter not being averaged, and thus being present in the nightly averaged dataset. To test that the pulsations were effectively averaged out, we binned the data of HD\,10700 considering only 3 points per night for the time series \texttt{TM}\,/\,\texttt{TMb}\,/\,\texttt{TMr} (with the actual points selected being randomly drawn from the available within each night). While the photon noise of the averaged points increases, it remains below 15\,cm/s. The jitter on \texttt{ESPR18} and \texttt{ESPR19} increases to 85 and 45\,cm/s, approx, pointing to an additional source of noise of 30-40\,cm/s, with the other parameters remaining very similar. These tests reinforce the idea that the stellar pulsations signal were significantly averaged out.

\subsection{Granulation's signature and impact}

Our understanding of the detailed physical mechanisms creating granulation RV signatures remains poor and has historically been influenced by the phenomenological characterisation of granulation coming from photometry. In the direct comparison in Fig.\,4 of \cite{2019LRSP...16....4G} the photometry shows a very clear shape and knees that allow for the fitting of Harvey functions for granulation, while for RV the location, amplitude and presence of the different components is still not firmly established.

A Harvey function with $\alpha$\,=\,2, as commonly used for parametrising granulation, is a Cauchy distribution. This is a well-known pathological distribution, with both expected value and variance undefined; if we draw samples from this distribution, the sample average does not converge. This provides a simple mathematical explanation as to why the many works that simulate granulation from such prescription do not measure an increase in planetary detection capability when stacking up a very large number of observations \citep[e.g.][]{2023A&A...676A..82M}.

The expected granulation signature is clearly not detected in the intra-night scatter of HD\,10700, being still compatible with those of HD\,102365 and HD\,304636. On the other hand, the predicted impact of super-granulation on average RVs, of 65-75\,cm/s rms, is above the long-term stability of 40\,cm/s measured on HD\,10700. Our jitter measurement imposes then stringent constraints on granulation signals modelling and parametrisations.

\subsection{A different signature: low-amplitude white-light flares}

Main sequence stars with spectral types A to M exhibit flares, associated with events of magnetic reconnection in the solar corona. It has long been know that flares are detected predominantly on M dwarfs \citep[e.g.][]{2019ApJS..241...29Y}. However, a comparison of flare size distributions on different hosts show these are mutually compatible across spectral types and with the distribution measured on the Sun \citep{2021ApJ...910...41A}. This points towards detectability as the main source for the difference. The study of \cite{2022ApJ...935..143P} on TESS data of 300 000 stars and using a photometric cadence of 2\,min detected that 7.7\% of stars show flares. The distribution of the duration of the flares peaks at 50\,min and has a long queue up to several hours. The amplitude of the vast majority of the detected flares is of lower than 10\% of the continuum flux, with the number of flares as a function of energy following a power-law. 

In their seminal work on the physics of flares, \cite{1991ApJ...380L..89L} argue that the observed power-law dependence stems from the solar coronal magnetic field being in a self-organised critical state. In this scenario, solar flares are avalanches of many small reconnection events, where the size of a given flare is determined by the number of elementary reconnection events, and explaining why the phenomenon extends over more than five orders of magnitude in peak flux.

White-light flares are expected to extend to much lower amplitude than currently detected, and at a much higher frequency. These low-amplitude flares are expected to have an RV signature, that unfortunately has not been characterised yet. The recent work of \cite{2024A&A...682A..46P} compares flaring activity with contemporaneous HARPS-N and concludes that flares, if present, should have an RV signature below 50\,cm/s. The timescale of 50\,min measured on TESS photometry matches the knee on the power spectrum of solar RVs as measured by \cite{2023A&A...669A..39A}, and the power spectrum of these phenomena have been characterised by a Harvey function with $\alpha$\,=\,2 \citep[see][]{1991SoPh..133..357H}. One can speculate that small-amplitude white light flares contribute to the RV variability measured by ESPRESSO at 40\,cm/s. However, without a detailed study or benchmark on the impact of small-amplitude white-light flares it remains impossible to quantify the contribution.

\subsection{Stellar signals or planets}

The RV analysis of periodic signals only showed a significant signal at around 20\,d for HD\,10700, and under specific noise assumptions (without other sources of variance than photon noise). Being located close to an harmonic of the expected rotational period, at 42\,d, the physical origin of this signal remains hard to assess. The best-match planet parameters posteriors show an over density of samples at this period. However, the RV amplitude of the signal varies from 15 to 60\,cm/s, and is at the edge of our compatibility limits. The posterior of the GP $\eta_3$ parameter shows also a local maximum at 20\,d, which could be a spurious fit to a planetary signal or simply the harmonic of the star rotational period at 42\,d. The detection in RV of the first harmonics of the rotation period is well documented in the literature \citep[e.g.,][]{2017A&A...598A.133D, 2017MNRAS.468.4772S, 2023A&A...674A.117G} and is even expected under some stellar configurations \citep{2011A&A...528A...4B}. The recent work of \cite{2023AJ....166..123K} shows that HD\,10700 is observed practically pole-on, and there are examples on the literature of pole-on stars like Vega that show a first-harmonic RV signature of the rotational period \citep{2021AJ....161..157H}. However, without a detailed characterisation of the star's active regions, the modelling of the activity signal and confirmation of this hypothesis remains out of reach. The GP delivers a very low jitter value of $\sim$40 cm/s to \texttt{ESPR19} over 3.5\,yr, to which contribute uncharacterised RV effects and undetected signals.

It is possible to make a comparison of these results with those obtained by \cite{2017AJ....154..135F} using the compatibility limits shown in Figure\,\ref{Fig:CompatLimitsRuns}. The planets announced at 20.0\,d (1.75\,M$_{\oplus}$) and 49.4\,d (1.83\,M$_{\oplus}$) are very close to the detection limit. The 3.93\,M$_{\oplus}$ planet at 162\,d should have been detected with our analysis, while the announced planet at 636\,d is below our detection limits. In a nutshell, our data and analysis remains compatible with \cite{2017AJ....154..135F} on the potential presence of a signal at 20\,d. However, we do not find significant evidence to call it a detection, and cast some doubts on its planetary origin (even if confirmed). Our results do not support the presence of a 3.93\,M$_{\oplus}$ planet at 162\,d as previously announced. When interpreting the discrepancy in detected planets and detectability estimates between HARPS and ESPRESSO one should consider that HARPS can reach an intrinsic precision of down to 60 cm/s \citep[see][for one of the first results]{2006Natur.441..305L}. As such the detection of Earth-mass planets up to hundred days remains extremely difficult when done with HARPS, even with extensive datasets and the most advanced analysis.

The detailed study of HD\,102365 by \cite{2011ApJ...727..103T} was motivated by the large measured RV scatter ($\sim$3\,m/s) when compared to the expected RV scatter on the star. The work of \cite{2017MNRAS.468.4772S} yields an an activity-induced RV semi-amplitude of only 41\,cm/s for the mean value of log($R'_{HK}$), giving the star a privileged status for RV studies. 

However, our ESPRESSO data shows clear evidence for activity-induced RV. First of all, the RV scatter doubled from \texttt{ESPR18} to \texttt{ESPR19}. The LS of BIS, log($R'_{HK}$) and Ha$_{06}$ show signals almost reaching an FAP of 1\% ate 123\,d. Very importantly, the correlation between nightly averaged indicators and RV shows that half of the RV variance can be attributed to the correlation between an indicator and RV.

Finally, our \texttt{kima} analysis shows no evidence for the presence of a planet of 16\,M$_{\oplus}$ on a 122\,days period orbit. The compatibility limits, calculated under the assumption of rotation-period informed GP model, show no candidate signals at this period and amplitude compatible with the ESPRESSO RV data. Therefore, we consider the 122-123\,d signal as having a stellar origin.

The M star HD\,304636 does not show a clear rotational period modulation in photometry, or on any spectral activity indicator. The \texttt{kima} analysis lead to a non-Gaussian jitter distribution for \texttt{ESPR18} and poor detection limits around the expected rotational period of 30-40\,d. 
A natural question is if the ESPRESSO data can provide further information on the TESS candidate found at 7.57\,d. We used \texttt{Forecaster} \citep{2017ApJ...834...17C} that for the radius of $\sim$0.87\,R$_{\oplus}$ delivers a planetary mass of $\sim$0.58\,M$_{\oplus}$. This value is well below the 1.5\,M$_{\oplus}$ (for 5 days) and 1.7\,M$_{\oplus}$ (for 10 days) compatibility limits found in this work, showing that the detection of such a planet is beyond our ability. 

The \texttt{kima} model of the three stars delivers a compatible RV slope; this common behavior suggests a long-term variation of instrumental RV. The best characterisation is provided by HD\,10700, delivering an RV slope of around 30\,cm/s per year with an uncertainty of 10 cm/s.

It should be stated that the different RV calculations \texttt{CCF}, \texttt{TM}, \texttt{TMb} and \texttt{TMr} have different sensitivities to telluric contamination. This comes first as a consequence of the different method's selection or exclusion of contaminated wavelength regions, but also due to the template matching calculating a unique template for a dataset. We showed that using telluric correction provides appreciable gains for HD\,10700, but a finer analysis on a wide range of targets can be done.

As a concluding remark, we note that while HD\,10700 and HD\,102365 have very similar activity values as measured by the most common indicators, the stars show a very different RV scatter and a different correlation level between indicators and RV. Moreover, they are modelled with \texttt{kima} with very different success, both in terms of the posteriors reached and the final jitter value reached. This exemplifies how the  a priori measurement of activity indicators is still insufficient when selecting the most suitable stars for the detection of Earth-mass planets. 
This conclusion expands the results of the systematic study of \cite{2021A&A...646A..77G}, where it was shown that low-activity stars are not necessarily low-variability stars.

%--------------------------------------------------------------------

\section{Conclusions}\label{Sec:Conc}

We present a study of main sequence stars using ESPRESSO GTO data. We started with a stellar characterisation using complementary methods to draw a complete picture of the stars. The RVs were derived from the ESPRESSO spectra with the CCF and TM methods and ancillary photospheric and chromospheric activity indicators.

We were able to study the intra-night RV behaviour on the three G-K stars with multiple consecutive exposures per night. We compared the measured RV signals with the expected scatter from pulsations and granulation. We reached a very good match between the predicted RV variability from oscillations and measured RV down to a level of better or equal to 10\,cm/s, with an additional scatter that might be attributed to poor stellar parameter determination or the impact of activity on oscillations.
   
A \texttt{kima} analysis was used to evaluate the number of planets supported by the data, under the assumption that a quasi-periodic GP can model the activity signal. A rotationally modulated signal close to the expected period was modelled for the three stars, with the analysis for HD\,10700 being particularly successful in reproducing the observed data. A planetary signal at a period of 20\,d could be modelled along with it, but the evidence for its presence is not significant. Moreover, its physical origin remain uncertain due to the similarity between this period and the first harmonic of rotation. ESPRESSO data on their own do not provide conclusive evidence for planets around HD\,10700, HD\,102365, or HD\,304636. This is at odds with the detection by \cite{2011ApJ...727..103T} of a 123\,d period planet with 16M$_{\oplus}$ around HD\,102365, which would clearly have been revealed by our data. The latest analysis of \cite{2017AJ....154..135F} on HD\,10700 HARPS archive data also predicts a planet with 3.93\,M$_{\oplus}$ planet at 162\,d that while within our detection capabilities is not detected, with our analysis casting a different interpretation on the 20\,d signal. Our RV analysis allowed us to detect planets around HD\,10700 down to 1.7\,M$_{\oplus}$ for periods up to 100\,d and of 2-5M$_{\oplus}$ within the star's habitable zone. As data analysis progresses, these values are expected to decrease.  

In this work, ESPRESSO demonstrates an on-sky RV precision of better 10 cm/s on short time-scales ($<$1\,h) and of around 40\,cm/s on a timescale of 3.5\,yr on HD\,10700 telluric-corrected spectra. By further further splitting the datasets, we estimated that the RV stability over a timescale of one year can be as low as 25-30\,cm/s, variable from year to year. 

The convective phenomena of granulation were modelled with literature prescriptions that unfortunately do not reproduce the level of scatter we observed in this work. In particular, the parametrisations of super-granulation overestimate the scatter measured in our RVs for HD\,10700, which provides strong constraints for detailed modelling in future studies. The analysis presented here illustrates the significant differences among  stellar RV signatures when analysed at sub-m/s level. It underlines the fact that no current activity indicator  allows us to select  RV-stable stars in advance for searches at better than 1\,m/s.

\section*{Data availability}

The RV data for the targets discussed in this work are available in electronic form at the CDS via anonymous \texttt{ftp} to \url{cdsarc.u-strasbg.fr} (130.79.128.5) or via \url{http://cdsweb.u-strasbg.fr/cgi-bin/qcat?J/A+A/}. The ESPRESSO data, reduced with the latest version of the pipeline, can also be accessed through the \texttt{DACE} platform via \url{https://dace.unige.ch/openData/?record=10.82180/dace-3js7g1l9}. 

In parallel, the ESPRESSO Legacy data release can be found at \url{https://www.doi.org/10.82180/dace-3g5jf9h8}.

\begin{acknowledgements}

    In memory of Denis M\'{e}gevand, project manager of ESPRESSO and dear friend, who passed away in 2022.

    We thank the anonymous referee for a review and comments that very much improved the readability of the manuscript.

    PF and JPF thank Patrick Eggenberger for very fruitful discussions on the impact on RV of p-mode oscillations, \^{A}ngela Santos for her expertise on solar physics and stellar flaring, Khaled Al-Moulla for discussing the temporal analysis of HARPS and HARPS-N data, and Alex Pietrow for a discussion on the state-of-the art on flaring observations.  

    This work has been carried out within the framework of the National Centre of Competence in Research PlanetS supported by the Swiss National Science Foundation under grant 51NF40\_205606. The authors acknowledge the financial support of the SNSF.

    FPE and CLO would like to acknowledge the Swiss National Science Foundation (SNSF) for supporting research with ESPRESSO through the SNSF grants nr. 140649, 152721, 166227, 184618 and 215190. The ESPRESSO Instrument Project was partially funded through SNSF’s FLARE Programme for large infrastructures.

    Co-funded by the European Union (ERC, FIERCE, 101052347). Views and opinions expressed are however those of the author(s) only and do not necessarily reflect those of the European Union or the European Research Council. Neither the European Union nor the granting authority can be held responsible for them. This work was supported by FCT - Fundação para a Ciência e a Tecnologia through national funds and by FEDER through COMPETE2020 - Programa Operacional Competitividade e Internacionalização by these grants: UIDB/04434/2020; UIDP/04434/2020.

    A.C.-G., O.B.-R. and J.L.-B. are funded by the Spanish Ministry of Science through MCIN/AEI/10.13039/501100011033 grants PID2019-107061GB-C61. HMT, PID2023-150468NB-I00 and CNS2023-144309.

    HMT acknowledges support from the "Tecnolog\'{i}as avanzadas para la exploraci\'{o}n de universo y sus componentes" (PR47/21 TAU) project funded by Comunidad de Madrid, by the Recovery, Transformation and Resilience Plan from the Spanish State, and by NextGenerationEU from the European Union through the Recovery and Resilience Facility. HMT also acknowledges financial support from the Agencia Estatal de Investigaci\'{o}n (AEI/10.13039/501100011033) of the Ministerio de Ciencia e Innovaci\'{o}n through projects PID2022-137241NBC41 y PID2022-137241NB-C44.

    JIGH, ASM, and RR acknowledge financial support from the Spanish Ministry of Science, Innovation and Universities (MICIU) projects PID2020-117493GB-I00 and PID2023-149982NB-I00, and from the Government of the Canary Islands project ProID2020010129.

    X.D acknowledges the support from the European Research Council (ERC) under the European Union’s Horizon 2020 research and innovation programme (grant agreement SCORE No 851555) and from the Swiss National Science Foundation under the grant SPECTRE (No 200021\_215200).

    This work has made use of data from the European Space Agency (ESA) mission Gaia (\url{https://www.cosmos.esa.int/gaia}), processed by the Gaia Data Processing and Analysis Consortium (DPAC, \url{https://www.cosmos.esa.int/web/gaia/dpac/consortium}). Funding for the DPAC has been provided by national institutions, in particular the institutions participating in the Gaia Multilateral Agreement.
        
    This work made use of {\tt astropy}, a community-developed core Python package and an ecosystem of tools and resources for astronomy, described in \href{https://ui.adsabs.harvard.edu/abs/2013A%26A...558A..33A/abstract}{astropy collaboration (2013)}, \href{https://ui.adsabs.harvard.edu/abs/2018AJ....156..123A/abstract}{astropy collaboration (2018)}, and \href{https://ui.adsabs.harvard.edu/abs/2022ApJ...935..167A/abstract}{astropy collaboration (2022)}. Corner plots were created using the package \texttt{corner.py}, described in \href{https://doi.org/10.21105/joss.00024}{Foreman-Mackey (2016)}.
            
\end{acknowledgements}

\bibliographystyle{aa} % style aa.bst
\bibliography{bibliography.bib} % references .bib

%--------------------------------------------------------------------

\begin{appendix} 

\section{\texttt{ARIADNE} priors and input photometry}\label{App:ARIADNE}

For the effective temperature, log(g), and [Fe/H] \texttt{ARIADNE} uses as priors the observed parameter distributions from the fifth RAVE survey data release \citep{2017AJ....153...75K}. A Normal distribution centred around the \cite{2018AJ....156...58B} value and with a standard deviation of 5 times the associated uncertainty is used as prior for the distance. As we wanted to retain uninformative priors we kept these default choices; for a full description we refer to \cite{2022MNRAS.513.2719V}. Since the distance to these stars is at maximum 10\,pc, we fixed the extinction coefficient $A_v$ at 0, in what is our only informed choice.  

The photometry and associated uncertainties retrieved for the four stars is presented on Table\,\ref{Tab:ARIADNEinput}. Since 2MASS J, H, and K$_s$ photometry is saturated for magnitudes below 4.5, 4, and 3.5, respectively, we exclude from the analysis of HD\,10700, HD\,20794, and HD\,102365 the 2MASS data \footnote{See \url{https://www.ipac.caltech.edu/2mass/releases/allsky/doc/sec1_6b.html} for detailed information.}. 

Gaia DR3 photometric measurements are also affected for saturation for bright targets. Nonetheless, the consortium provides a correction formula readily applicable to our case, described in \cite{2021A&A...649A...3R}, Appendix\,C1. Following this correction all Gaia magnitudes could be used.

\begin{table*}
\centering

\caption{\texttt{ARIADNE} input photometry, ordered by central wavelength.}
\label{Tab:ARIADNEinput}
\begin{tabular}{ l | c c c c }
 \hline \hline
 
 Filter &  HD\,10700 & HD\,20794 & HD\,102365 & HD304636 \\ 
  \hline

STROMGREN u & 5.6490$\pm$0.1308& 6.3280$\pm$0.0310 & 6.8190$\pm$0.0160 &        --- \\   
GROUND JOHNSON U  & 4.4380$\pm$0.0171 & 5.1880$\pm$0.0095&               5.6550$\pm$0.0121 & 12.1300$\pm$0.0121 \\  
STROMGREN v    & 4.6660$\pm$0.0732& 5.3640$\pm$0.0168 &          5.9220$\pm$0.0099 & --- \\  
TYCHO B MvB     & --- & --- &    5.6970$\pm$0.0140 & 11.3690$\pm$0.0570 \\  
GROUND JOHNSON B  & 4.2220$\pm$0.0122 & 4.9710$\pm$0.0081 &      5.5480$\pm$0.0100  & 10.9200$\pm$0.0560\\  
STROMGREN b    & 3.9570$\pm$0.0361  & 4.6970$\pm$0.0099 &                5.3030$\pm$0.0050  &--- \\  
GaiaDR3 BP    & 3.7995$\pm$0.0113$^c$ & 4.4409$\pm$0.0034&               5.0517$\pm$0.0028  & 9.6998$\pm$0.0029  \\  
TYCHO V MvB     & --- & --- &    4.9620$\pm$0.0090  & 9.6620$\pm$0.0220 \\ 
STROMGREN y     & 3.4990$\pm$0.0030 &  4.2600$\pm$0.0070 &       4.8920$\pm$0.0030  &      --- \\
GROUND JOHNSON V  & 3.4950$\pm$0.0100 & 4.2630$\pm$0.0070 &      4.8810$\pm$0.0080  &      9.4800$\pm$0.0130\\  
GaiaDR3 G     & 3.3004$\pm$0.0030$^c$ & 4.0639$\pm$0.0029$^c$ &          4.7017$\pm$0.0028$^c$  &        8.6806$\pm$0.0028 \\  
GaiaDR3 RP    & 2.7483$\pm$0.0053$^{c}$ & 3.5083$\pm$0.0050&     4.1954$\pm$0.0045  &      7.6769$\pm$0.0038 \\  
TESS      & --- &       --- & 4.2702$\pm$0.0067  &      7.6687$\pm$0.0073 \\  
2MASS J       & 2.1490$\pm$0.3100$^{\dagger}$ & 3.0320$\pm$0.2620$^{\dagger}$ &        3.9310$\pm$0.2760$^{\dagger}$  & 6.4420$\pm$0.0230 \\  
2MASS H       & 1.8000$\pm$0.2340$^{\dagger}$ & 2.7090$\pm$0.2340$^{\dagger}$ &        3.4900$\pm$0.2380$^{\dagger}$  & 5.7930$\pm$0.0330     \\  
2MASS Ks      & 1.7940$\pm$0.2740$^{\dagger}$ & 2.6360$\pm$0.2780$^{\dagger}$ &        3.4890$\pm$0.2780$^{\dagger}$  &       5.5870$\pm$0.0210 \\  
WISE RSR W1      & --- & --- & --- &    5.4160$\pm$0.1820 \\  
WISE RSR W2      & --- & --- & --- &    5.2300$\pm$0.0710 \\  

\hline
\multicolumn{5}{l}{\textit{c}: \footnotesize{Gaia Data affected by saturation: corrected as described in the text.}}\\
\multicolumn{5}{l}{$^{\dagger}$ : \footnotesize{2MASS data below saturation threshold: not used.}}
\end{tabular}

\end{table*}

%--------------------------------------------------------------------

\section{Rotational period derived assuming different spectral types}\label{App:Prot}

For the sake of completeness and comparison, we computed the rotational period using the calibrations of \cite{2016A&A...595A..12S} for each star's estimated spectral type, closest class, plus GKM's as a whole. For log($R'_{HK}$) we use the value derived by \cite{2019A&A...629A..80H}. The results are shown in Table\,\ref{Tab:Prot}.

\begin{table}[H] 
\centering

\caption{Stellar rotational periods when assuming different spectral classes.}
\label{Tab:Prot}

\begin{tabular}{ l c c c | c}
 \hline \hline
 Star & P$_{rot,\,G}$ [d] &  P$_{rot,\,K}$ [d]  &  P$_{rot\,M}$ [d] & P$_{rot,\,GKM}$ [d] \\ 
  \hline
  
HD\,10700 & 26.1$\pm$5.2 & 37.7$\pm$7.5 & --  & 28.2$\pm$5.6 \\
HD\,20794 & 27.7$\pm$5.5 & 44.5$\pm$8.9 & --  & 33.9$\pm$6.8 \\ 
HD\,102365 & 25.8$\pm$5.2 & 36.4$\pm$7.3 & -- & 27.2$\pm$5.4 \\ 
HD\,304636 & -- & 40.2$\pm$8.0 & 30.3$\pm$6.1 & 30.4$\pm$6.1 \\ 

\hline
\end{tabular}
\end{table}

%--------------------------------------------------------------------

\section{\texttt{S-BART} application to ESPRESSO data for high-precision radial velocities}\label{App:SBART}

The specifics of our application of S-BART to ESPRESSO data can be summed up as follows:

\begin{itemize}
    \item the method assumes a common RV value for the individual orders and slices of the instrument. In this context, the two ESPRESSO slices of each interference order are treated as two independent measurements, ignoring the correlations between the two orders flux values and their  uncertainties;
    \item To decrease the impact of micro-telluric features on the stellar template, only observations taken at an airmass smaller than 1.6 are considered for the template construction. 
    \item \texttt{S-BART} masks wavelength regions to prevent spectral contamination from Earth's atmosphere on the spectra, that leads to parasite effects on RV. A reference synthetic transmittance profile is constructed with \texttt{TELFIT} \citep{2014AJ....148...53G} at a resolution of 140000 for the wavelength range of ESPRESSO; the airmass, stellar RV and precipitable water vapour of the observation with the highest relative humidity are assumed as worst-case scenario for each star. Wavelength bins with a transmission lower than 99\% relative to the continuum and within the maximum yearly BERV variation ($\sim$ 30 km/s) are masked out. This mask is applied to both the construction of the stellar template and the calculation of the individual RVs. This criteria leads to discarding 36 out of the 170 slices of ESPRESSO.
    \item two independent templates are generated for the two independent datasets \texttt{ESPR18} and \texttt{ESPR19}; 
    \item When discarding wavelength domains, and to ensure that the RV measurements are informed by the same spectral regions, we consider only orders accepted for both \texttt{ESPR18} and \texttt{ESPR19}.

\end{itemize}

%--------------------------------------------------------------------

\section{P-mode simulations}

\subsection{Reference values used in the scaling relationship}

The reference values for asteroseismic and stellar parameters are presented in Table\,\ref{Tab:PmodeSimul}, with the corresponding bibliography.

\begin{table*} 
\centering

\caption{Asteroseismic and stellar parameters for our reference stars $\alpha$\,Cen\,A and $\mu$\,Ara, and values used in this study.}
\label{Tab:PmodeSimul}
\begin{tabular}{ l | c c c c | c c c | l }
 \hline \hline
 \multirow{2}{*}{Star} & T$_\mathrm{eff}$ & L$_{\mathrm{star}}$ &  M$_{\mathrm{star}}$ & R$_{\mathrm{star}}$  & $\nu_{max}$\, & $\Delta\nu$ & $A_{env,\,RV}$ & \multirow{2}{*}{Sources}  \\ 
 & \tiny{[K]} & \tiny{[L$_{\odot}$]} & \tiny{[M$_{\odot}$]} & \tiny{[R$_{\odot}$]}& \multicolumn{2}{c}{\tiny{[mHz]}} & \tiny{[cm/s]} & \\
  \hline
  
\multirow{3}{*}{$\alpha$\,Cen\,A} & \multirow{3}{*}{5795$^1$} & \multirow{3}{*}{1.521$^1$} & \multirow{3}{*}{1.1055$^1$} & \multirow{3}{*}{1.2234$^2$} & \multirow{3}{*}{2.3$^3$} & \multirow{3}{*}{0.106$^3$} & \multirow{3}{*}{30$^3$} & $^1$:  \tiny{\cite{2016A&A...594A.107K}} \\
 & & & & & & & & $^2$: \tiny{\cite{2017A&A...597A.137K}} \\
  & & & & & & & & $^3$ \tiny{\cite{2002A&A...390..205B}} \\ 

\multirow{2}{*}{$\mu$\,Ara} & \multirow{2}{*}{5820$^4$} & \multirow{2}{*}{1.90$^4$} & \multirow{2}{*}{1.10$^4$} & \multirow{2}{*}{1.36$^4$} & \multirow{2}{*}{1.9$^5$} & \multirow{2}{*}{0.090$^5$} & \multirow{2}{*}{30$^5$} & $^4$: \tiny{\cite{2010A&A...513A..49S}} \\ 
  & & & & & & & & $^5$: \tiny{\cite{2005A&A...440..609B}}  \\ 

ref. values & 5808 & 1.71 & 1.10 & 1.29 & 2.1 & 0.100 & 30 &  \\[2pt]

\hline
\end{tabular}
\end{table*}

\subsection{Simulation diagrams and examples}\label{App:Puls}

\begin{figure}[H]
\center
\includegraphics[width=6.3cm]{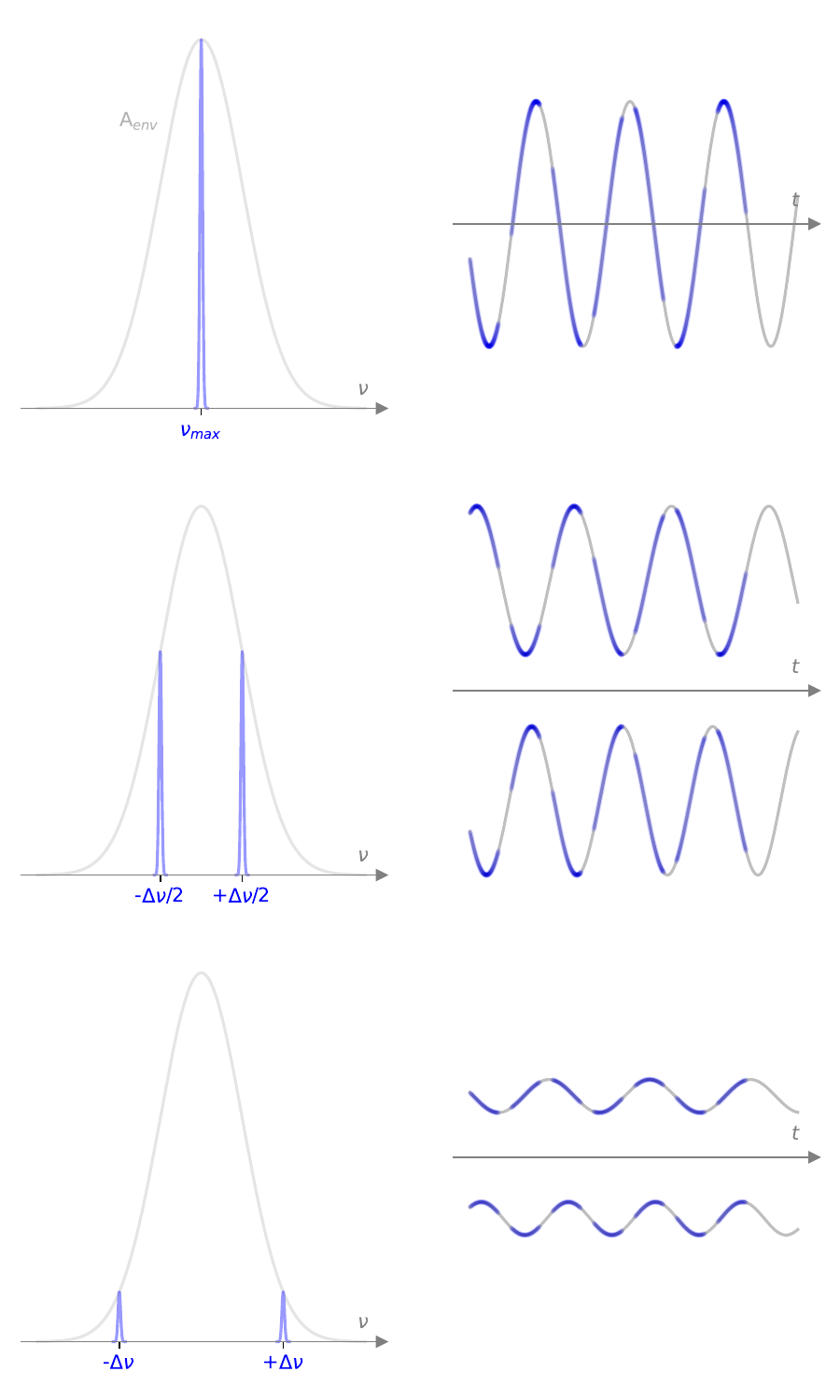}
\caption{Representation on the frequencies and amplitudes drawn from the modes envelope (\textit{left}), and how these plus random phases create the signals in the temporal domain (\textit{right}). From top to bottom we have the main frequency $\nu_{\mathrm{max}}$, plus first and second order terms. In the temporal domain, open shutter is represented in blue.}\label{Fig:pulsdiagram}
\centering
\end{figure}

A representation of how the different frequencies and amplitudes generate different pulsation components is shown in Fig.\,\ref{Fig:pulsdiagram}.

\begin{figure}[H]
\center
\includegraphics[width=7.5cm]{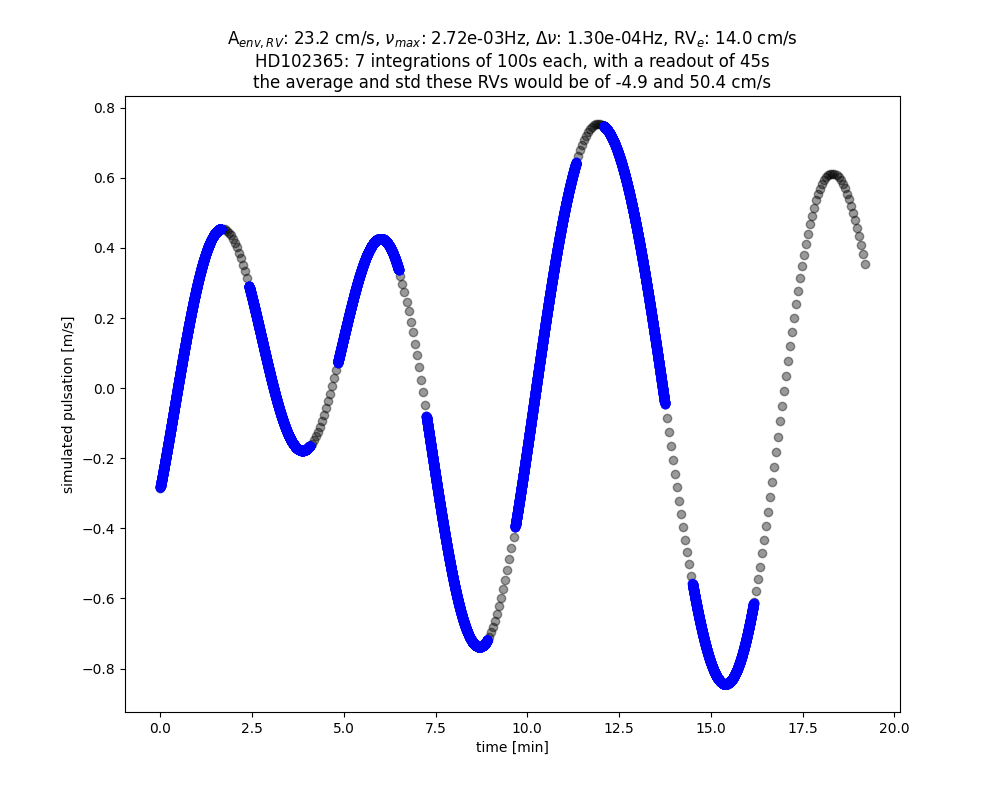}
\caption{Example of a simulated pulsation signal for HD\,102365. The open shutter time intervals are represented in blue.}\label{Fig:pulssimul}
\centering
\end{figure}

\begin{figure}
\center
\includegraphics[width=6.5cm]{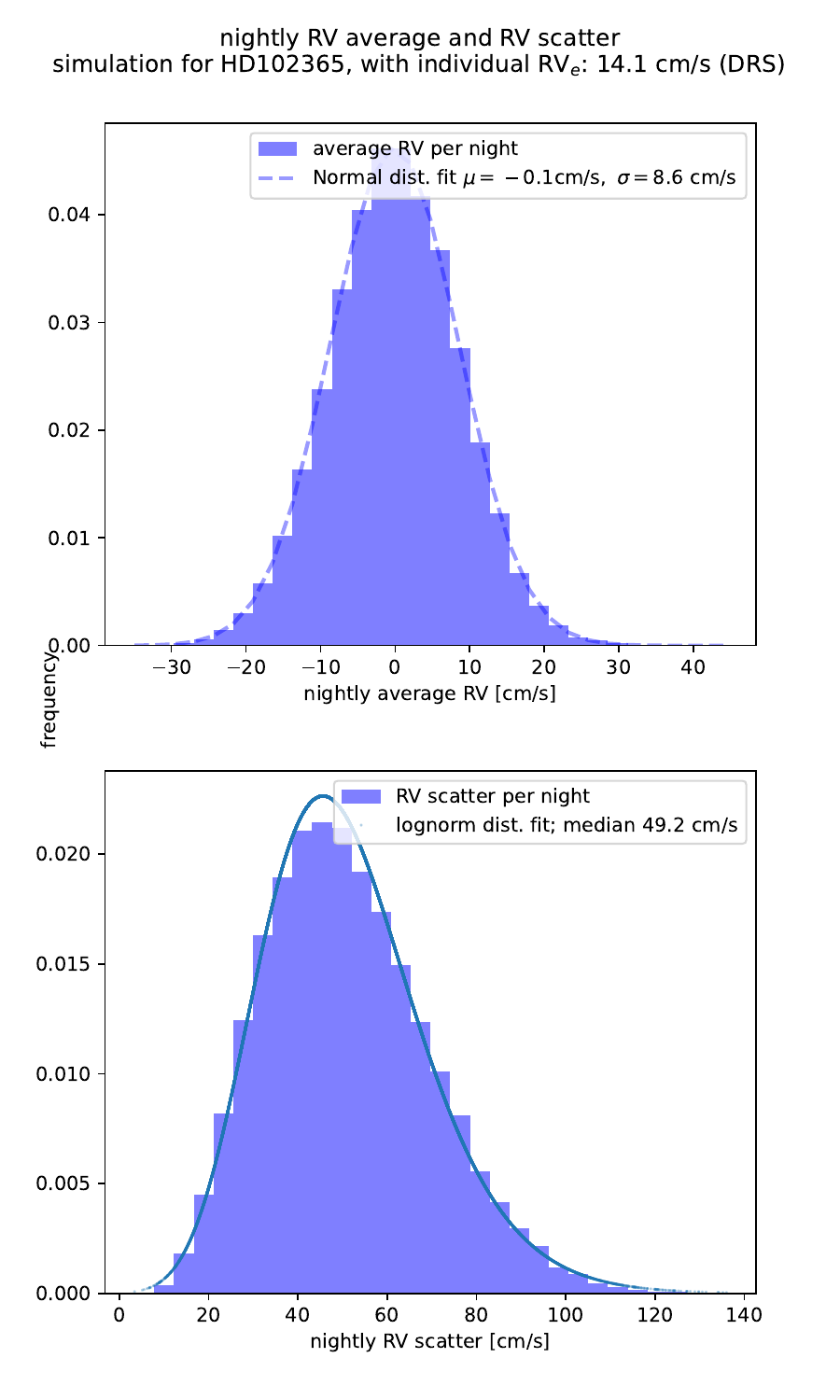}
\caption{Distributions of the nightly averaged RV (\textit{top}) and intra-night scatter (\textit{bottom}) introduced by pulsations, calculated for 50\,000 runs for the star HD\,102365.}\label{Fig:pulsdist}
\centering
\end{figure}

An example of a simulated night sequence for HD\,102365 is represented in Fig\,\ref{Fig:pulssimul}, depicting how the open shutter time covers the final pulsation signal. Once a large number of realisations of this simulation is done, one can evaluate the distribution of nightly averaged RVs and RV scatter (rms), as seen in Fig.\,\ref{Fig:pulsdist}.

%--------------------------------------------------------------------

\section{Granulation measurements and simulations}\label{App:Granul}

\cite{2011A&A...525A.140D} used HARPS asteroseismology data on the stars $\beta$\,Hyi, $\alpha$\,Cen\,A, $\mu$\,Ara, and $\tau$\,Cet and fitted the PSD with Harvey and Lorentzian function, applying to RV the same principles studied in detail in photometry. In a more recent work \cite{2023A&A...669A..39A} used HARPS and HARPS-N solar telescopes RV data and fitted two Harvey functions to represent granulation on the RV's power density spectrum. While \cite{2011A&A...525A.140D} fitted three granulation signals, \cite{2023A&A...669A..39A} fitted two, and the power-law slope $\alpha$ was fixed at 2.0 to avoid degeneracy and because this value reproduced well the observational data. The results are summarised in Table\,\ref{Tab:granulfits}.

From the proposed scaling relationships one gets that the amplitude and granulation timescales are proportional to $\Delta\nu$; yet, for $\alpha$\,Cen\,A and $mu$\,Ara, two stars with very similar large frequency separations, the parametrisations of granulation obtained by \cite{2011A&A...525A.140D} is very different. This can be a consequence of poor time sampling of the phenomenon or the poor granulation modelling, but one should also consider that the scaling model could be inaccurate, as suggested already in \cite{2011ApJ...741..119M} .

\begin{table*} 
\centering

\caption{Harvey function parametrisation of Granulation taken from \cite{2011A&A...525A.140D} and \cite{2023A&A...669A..39A}.}
\label{Tab:granulfits}
\resizebox{1.8\columnwidth}{!}{%
\begin{tabular}{ l | c c c | c c c | c c c }
 \hline \hline

 \multirow{2}{*}{Star} & A$_1$ & $\tau_1$ & \multirow{2}{*}{$\alpha_1$} & A$_2$ & $\tau_2$ & \multirow{2}{*}{$\alpha_2$} & A$_3$ & $\tau_3$ & \multirow{2}{*}{$\alpha_3$} \\ 
  & \tiny{[(m/s)$^2$\,Hz$^{-1}$]} & \tiny{[h]} & & \tiny{[(m/s)$^2$\,Hz$^{-1}$]} & \tiny{[h]} & & \tiny{[(m/s)$^2$\,Hz$^{-1}$]} & \tiny{[h]} & \\[3pt]

  \hline
$\alpha$\,Cen\,A & 3$\times10^2$ & 0.30 & 8.9 & 3$\times10^3$ & 1.2 & 3.9 & 2.7$\times10^4$ & 7.4 & 3.1 \\[3pt]
$\mu$\,Ara & 1.1$\times10^3$ & 0.73 & 4.5 & 2.7$\times10^4$ & 3.4 & 5.0 & 2.9$\times10^4$ & 13.0 & 6.0 \\[3pt]
HD\,10700 & 3$\times10^2$ & 0.31 & 19.8 & 2$\times10^3$ & 1.2 & 8.9 &  2.7$\times10^4$ & 6.7 & 2.6 \\[3pt]
Sun & (4.11$\pm$1.35)$\times10^2$ & 0.911$\pm$0.200 & 2.0 & -- & -- & -- & (2.79$\pm$0.28)$\times10^4$ & 13.3$\pm$1.1 & 2.0 \\[3pt]

\hline
\end{tabular}
}
\end{table*}

%--------------------------------------------------------------------

\section{FAP from $\ell_1$ analysis}\label{App:l1FAP}

The FAP as estimated from $\ell_1$ analysis for the three stars is presented in Tables\,\ref{Tab:l1FAP_HD10700}, \ref{Tab:l1FAP_HD102365}, and \ref{Tab:l1FAP_HD304636}. 

\begin{table*} 
\centering

\caption{$\ell_1$ FAP for the first ten sinusoidal functions detected on HD\,10700 and shown in Fig.\,\ref{Fig:l1}.}
\label{Tab:l1FAP_HD10700}

\resizebox{1.8\columnwidth}{!}{%

\begin{tabular}{ l | c c c | c c c | c c c | c c c}
\hline \hline

 dataset & \multicolumn{3}{c}{\texttt{CCF}} & \multicolumn{3}{c}{\texttt{TM}} & \multicolumn{3}{c}{\texttt{TMb}} & \multicolumn{3}{c}{\texttt{TMr}}  \\

 \multirow{2}{*}{peak \#} & \small{Period}                   & \small{Amp.}  & \small{log$_{\mathrm{10}}$FAP} & \small{Period}                  & \small{Amp.}  & \small{log$_{\mathrm{10}}$FAP} & \small{Period}                   & \small{Amp.}  & \small{log$_{\mathrm{10}}$FAP} & \small{Period}                   & \small{Amp.}  & \small{log$_{\mathrm{10}}$FAP} \\
         & \tiny{[d]}                       & \tiny{[cm/s]} & \tiny{[ ]}                     & \tiny{[d]}                      & \tiny{[cm/s]} & \tiny{[ ]}                     & \tiny{[d]}                       & \tiny{[cm/s]} & \tiny{[ ]}                     & \tiny{[d]}                       & \tiny{[cm/s]} & \tiny{[ ]}                     \\
\hline

 1       & 348.0                            & 18.4          & -1.5e+00                       & 1423.5                          & 19.2          & -2.2e+00                       & 1423.5                           & 26.2          & -4.3e+00                       & 21.2                             & 15.2          & -1.7e+00                       \\
 2       & 13.9                             & 15.9          & -6.1e-03                       & 20.1                            & 15.2          & -1.6e+00                       & 20.1                             & 16.5          & -1.1e+00                       & 20.1                             & 12.5          & 0.0e+00                        \\
 3       & 1.1                              & 15.6          & -7.4e-01                       & 1.1                             & 11.9          & -4.6e-03                       & 50.0                             & 12.2          & -2.7e-01                       & 22.1                             & 10.9          & -2.4e-11                       \\
 4       & 21.2                             & 13.5          & -8.8e-01                       & 6.3                             & 10.1          & -1.3e-02                       & 307.0                            & 11.6          & -8.8e-08                       & 13.9                             & 9.7           & -1.1e-01                       \\
 5       & 1118.5                           & 11.5          & -1.9e-03                       & 50.0                            & 9.8           & -2.5e-03                       & 1.1                              & 11.3          & -3.0e-05                       & 1.8                              & 9.7           & -8.9e-12                       \\
 6       & 20.1                             & 11.3          & -1.6e-05                       & 307.0                           & 9.3           & -1.1e-03                       & 1.1                              & 11.2          & -3.8e-09                       & 1118.5                           & 9.5           & -4.5e-01                       \\
 7       & 3.7                              & 7.7           & -5.6e-08                       & 21.2                            & 9.2           & 0.0e+00                        & 1.1                              & 10.6          & -1.4e-01                       & 1.2                              & 8.3           & -3.2e-02                       \\
 8       & 1.1                              & 7.7           & 0.0e+00                        & 1.1                             & 8.4           & -8.0e-01                       & 1.1                              & 9.7           & 0.0e+00                        & 26.7                             & 8.1           & -4.0e-01                       \\
 9       & 6.3                              & 7.7           & -1.1e+00                       & 1.7                             & 7.9           & -1.5e-09                       & 1.4                              & 9.3           & 0.0e+00                        & 6.4                              & 7.6           & 0.0e+00                        \\
 10      & 50.2                             & 7.0           & 0.0e+00                        & 6.7                             & 6.4           & -1.5e-03                       & 1.4                              & 8.1           & -2.1e-01                       & 4.0                              & 6.9           & 0.0e+00                        \\

\hline

\end{tabular}
}
\end{table*}

\begin{table*} 
\centering

\caption{$\ell_1$ FAP for the first ten sinusoidal functions detected on HD\,102365 and shown in Fig.\,\ref{Fig:l1}.}\label{Tab:l1FAP_HD102365}

\resizebox{1.8\columnwidth}{!}{%

\begin{tabular}{ l | c c c | c c c | c c c | c c c}
\hline \hline

 dataset & \multicolumn{3}{c}{\texttt{CCF}} & \multicolumn{3}{c}{\texttt{TM}} & \multicolumn{3}{c}{\texttt{TMb}} & \multicolumn{3}{c}{\texttt{TMr}}  \\

  \multirow{2}{*}{peak \#} & \small{Period}                   & \small{Amp.}  & \small{log$_{\mathrm{10}}$FAP} & \small{Period}                  & \small{Amp.}  & \small{log$_{\mathrm{10}}$FAP} & \small{Period}                   & \small{Amp.}  & \small{log$_{\mathrm{10}}$FAP} & \small{Period}                   & \small{Amp.}  & \small{log$_{\mathrm{10}}$FAP} \\
         & \tiny{[d]}                       & \tiny{[cm/s]} & \tiny{[ ]}                     & \tiny{[d]}                      & \tiny{[cm/s]} & \tiny{[ ]}                     & \tiny{[d]}                       & \tiny{[cm/s]} & \tiny{[ ]}                     & \tiny{[d]}                       & \tiny{[cm/s]} & \tiny{[ ]}                     \\
\hline 

 1       & 11.6                             & 31.3          & -2.4e-02                       & 11.6                            & 32.4          & -5.5e-02                       & 1.5                              & 28.3          & -1.3e-02                       & 11.6                             & 36.9          & -7.1e-01                       \\
 2       & 1.5                              & 27.5          & -3.6e-09                       & 214.0                           & 28.0          & 0.0e+00                        & 1.2                              & 26.3          & -5.9e-01                       & 69.7                             & 23.6          & 0.0e+00                        \\
 3       & 42.0                             & 26.3          & -2.2e-04                       & 1.5                             & 27.8          & -4.4e-04                       & 214.0                            & 25.6          & -1.8e-01                       & 1.2                              & 20.3          & -9.3e-06                       \\
 4       & 1.2                              & 22.3          & -7.9e-05                       & 42.1                            & 24.3          & -1.3e-09                       & 11.6                             & 24.4          & 0.0e+00                        & 42.1                             & 18.0          & -2.6e-01                       \\
 5       & 2.7                              & 17.4          & -2.0e-07                       & 1.2                             & 23.6          & -8.6e-04                       & 1.8                              & 22.3          & -9.7e-05                       & 3.1                              & 16.3          & 0.0e+00                        \\
 6       & 214.0                            & 16.5          & -9.2e-02                       & 18.3                            & 22.8          & -9.8e-01                       & 42.1                             & 18.0          & 0.0e+00                        & 1.3                              & 15.4          & -2.6e+00                       \\
 7       & 1.4                              & 16.2          & -1.6e-01                       & 1.8                             & 18.1          & 0.0e+00                        & 1.4                              & 17.8          & -4.2e-05                       & 2.2                              & 11.9          & -1.1e-05                       \\
 8       & 69.7                             & 15.8          & 0.0e+00                        & 1.4                             & 17.0          & -1.1e-12                       & 18.3                             & 17.3          & -4.3e-01                       & 1.4                              & 11.5          & 0.0e+00                        \\
 9       & 1.2                              & 14.7          & -8.3e-04                       & 2.7                             & 15.3          & -5.0e-01                       & 31.8                             & 16.0          & -2.2e-06                       & 4.0                              & 10.5          & 0.0e+00                        \\
 10      & 1.3                              & 13.3          & -2.8e-02                       & 1.3                             & 15.1          & -5.8e-16                       & 550.3                            & 15.0          & 0.0e+00                        & 2.9                              & 9.8           & -6.7e-02                       \\

\hline

\end{tabular}
}
\end{table*}

\begin{table*} 
\centering

\caption{$\ell_1$ FAP for the first ten sinusoidal functions detected on the star HD\,304636 and shown in Fig.\,\ref{Fig:l1}.}\label{Tab:l1FAP_HD304636}

\resizebox{1.8\columnwidth}{!}{%
\begin{tabular}{ l | c c c | c c c | c c c | c c c}
\hline \hline

 dataset & \multicolumn{3}{c}{\texttt{CCF}} & \multicolumn{3}{c}{\texttt{TM}} & \multicolumn{3}{c}{\texttt{TMb}} & \multicolumn{3}{c}{\texttt{TMr}}  \\

  \multirow{2}{*}{peak \#} & \small{Period}                   & \small{Amp.}  & \small{log$_{\mathrm{10}}$FAP} & \small{Period}                  & \small{Amp.}  & \small{log$_{\mathrm{10}}$FAP} & \small{Period}                   & \small{Amp.}  & \small{log$_{\mathrm{10}}$FAP} & \small{Period}                   & \small{Amp.}  & \small{log$_{\mathrm{10}}$FAP} \\
         & \tiny{[d]}                       & \tiny{[cm/s]} & \tiny{[ ]}                     & \tiny{[d]}                      & \tiny{[cm/s]} & \tiny{[ ]}                     & \tiny{[d]}                       & \tiny{[cm/s]} & \tiny{[ ]}                     & \tiny{[d]}                       & \tiny{[cm/s]} & \tiny{[ ]}                     \\
\hline

 1       & 32.0                             & 66.8          & -2.0e-02                       & 2.6                             & 47.9          & -2.3e-01                       & 2.6                              & 64.9          & -4.5e-01                       & 1.9                              & 54.2          & -8.8e-02                       \\
 2       & 26.2                             & 48.9          & -5.6e-01                       & 1.9                             & 42.3          & -2.9e-01                       & 3.2                              & 45.5          & -3.0e-02                       & 32.0                             & 37.7          & -2.5e-06                       \\
 3       & 1.9                              & 44.5          & -1.2e-03                       & 31.9                            & 42.0          & -5.5e-02                       & 32.0                             & 36.5          & -1.3e-02                       & 1.1                              & 26.6          & -2.3e+00                       \\
 4       & 15.6                             & 40.6          & -1.3e+00                       & 1.7                             & 39.2          & -5.8e-16                       & 1.7                              & 33.6          & -2.2e-02                       & 33.6                             & 26.5          & -5.8e-01                       \\
 5       & 5.3                              & 36.4          & -5.9e-05                       & 3.2                             & 30.7          & -3.6e-01                       & 1.4                              & 32.5          & -4.1e-03                       & 2.6                              & 26.1          & -2.6e-05                       \\
 6       & 2.6                              & 32.0          & -5.7e-03                       & 33.6                            & 26.5          & -9.0e-01                       & 5.3                              & 29.9          & -2.1e+00                       & 13.7                             & 25.9          & 0.0e+00                        \\
 7       & 2.5                              & 28.9          & 0.0e+00                        & 5.3                             & 25.3          & 0.0e+00                        & 1.4                              & 28.2          & -2.9e-16                       & 1.7                              & 23.6          & -3.5e-01                       \\
 8       & 1.1                              & 27.6          & -1.0e-01                       & 1.9                             & 25.1          & -3.6e-01                       & 15.6                             & 26.7          & -1.8e-02                       & 3.2                              & 21.0          & -1.8e-05                       \\
 9       & 8.5                              & 22.9          & -1.1e-02                       & 2.4                             & 24.1          & -2.6e-11                       & 1.9                              & 18.9          & -2.7e-01                       & 5.7                              & 18.7          & 0.0e+00                        \\
 10      & 1.4                              & 20.0          & -2.8e+00                       & 3.2                             & 21.1          & -7.7e-02                       & 2.1                              & 18.2          & 0.0e+00                        & 2.9                              & 17.1          & -3.6e-02                       \\
\hline

\end{tabular}
}
\end{table*}

%--------------------------------------------------------------------

\section{\texttt{kima} analysis} 

\subsection{Description of prior distributions}\label{App:priors}

We use non-informative priors as much as possible, including only information available from simple summary statistics and the previously determined rotation periods listed in Table\,\ref{Tab:FinalParams}. 

The modified log-uniform distribution is a log-uniform distribution with parameters knee $a$ and upper limit $b$, introduced by \cite{2005ApJ...631.1198G}. Using the modified version allows us to extend the prior to zero, assigning non-zero probability to values below the knee value. 
For the planet semi-major amplitude we use a modified log-uniform prior with a knee at 10 cm/s and an upper limit of 10\,m/s.

For the period we use a log-uniform prior from 1 to 1000\,d. The eccentricity is represented with a \cite{1980JHyd...46...79K} distribution which is an approximation to the $\beta$ distribution proposed in \cite{2013MNRAS.434L..51K}. For the mean anomaly and argument of periastron we use uniform distributions between 0 and 2$\pi$.

For the stars systemic RV and the instrument-related parameters we derive the priors from the datasets: the systemic RV is taken from a uniform distribution between the RV extreme values, and the jitter from a uniform distribution between 5\,cm/s to 5\,m/s. The offset is drawn from a uniform distribution between minus and plus the RV span value. The linear and quadratic slopes are zero-centred normal distributions with variance given by the data dispersion. 

As the GP we use \texttt{kima} parametrisation of the quasi-periodic kernel

\begin{equation}\label{Eq:kQP}
     \hspace{0.5cm} k_{QP}(t, t') = \eta_1^2 \, \mathrm{exp} \left[ - \frac{2}{\eta_4^2} \,\mathrm{sin}^2 \left( \frac{\pi(t-t')}{\eta_3} \right) - \frac{(t-t')^2}{2 \eta_2^2} \right]  
,\end{equation}

where $\eta_1$ is the amplitude of the kernel, $\eta_2$ is the evolution timescale, $\eta_3$ is the rotational period and $\eta_4$ is the harmonic complexity. $\eta_2$ represents the timescale of coherence of the active regions. The adimensional parameter $\eta_4$ governs the short-scale variation of the GP relative to the rotation timescale $\eta_3$; a smaller value of $\eta_4$ indicates a higher number of short-scale structures within a single stellar rotation period. For further information of the quasi-periodic kernel GP and on GPs in general, we refer to \cite{2006gpml.book.....R} and \cite{2012RSPTA.37110550R}.

It is worth mentioning that this GP model has several potentially degenerate variables. Equation\,\ref{Eq:kQP} is basically the product of two exponential functions: the term with $\eta_3$ and $\eta_4$, designed to represent the periodic part with higher harmonic complexity, and the term with $\eta_2$, intended to represent the temporal evolution. However, for a small temporal difference relative to the putative period $(t-t') / \eta_3$, when 2$\pi\,\eta_2$\,=\,$\eta_3\eta_4$ there is a potential issue\footnote{This is obtained by making the change of variable x=$\pi\,(t-t')/\eta_3$, and approximating sin(x)\,$\sim$\,x.}. Since $\eta_2 \geq \eta_3$, for $\eta_4<1$ there is no confusion; however, when $\eta_4\,>\,1$ there is a degeneracy between the terms that intend to represent temporal evolution and periodic behaviour. Moreover, for very low or high values of harmonic complexity $\eta_4$, the final function is not quasi-periodic. For $\eta_4\ll1$ the function will contain a very large number of peak-to-valley variations per timescale $\eta_3$, while for $\eta_4\gg1$ the function will be constant over timespans much longer than $\eta_3$, and including the GP corresponds to simply adding a second component of white noise.
The term of harmonic complexity $\eta_4$ has its origin from the temporal evolution of spot complexes as seen in photometry, but its usage in RV is far from trivial. The exact value is hard to recover even when injected on simulated RV datasets, and its interpretation at the physical level is not trivial; for a discussion on the subject, we refer to \cite{2022MNRAS.515.5251N}.

For $\eta_1$ we consider a uniform distribution between 5\,cm/s and 5\,m/s. For the rotation timescale $\eta_3$ we use a Normal distribution centred at P$_{rot}$ and with standard deviation $\sigma$ as the period's estimated uncertainties, with the two variables as presented in Table\,\ref{Tab:Prot}. For the evolution timescale parameter $\eta_2$ we consider a Uniform distribution between 1 and 5$\times$ the rotational period. There is little information about this parameter on stars other than the Sun, but on our star the values of $\eta_2$ are assumed to be a few times the rotation period. For $\eta_4$ we choose a uniform between 0.2 and 5.

All priors are listed in Table\,\ref{Tab:kimapriors}. 

\begin{table} 
\centering

\caption{Priors used in the \texttt{kima} analysis.} 
\label{Tab:kimapriors}

\begin{tabular}{ c c c c }

\hline \hline

\multirow{2}{*}{Group} & \multirow{2}{*}{Parameter} & \multirow{2}{*}{Prior} & \multirow{2}{*}{\small{units}} \\
 & & &  \\

\hline 

\multirow{4}{*}{Planet} & $K_p$ & $\mathcal{MLU}$(0.1,\,10) & m/s \\[2pt]
 & $P$ & $\mathcal{LU}$(1.0, 1e5) & d \\[2pt]
 & $e$ & $\mathcal{K}(0.867,\,3.03)$ & -- \\[2pt]
 & $\phi$, $\omega$ & $\mathcal{U}(0,\,2\pi)$ & -- \\[2pt]

 \hline

\multirow{2}{*}{Instruments}  & jitters & $\mathcal{U}(0.05,\,5)$ & \multirow{2}{*}{m/s} \\[2pt]
& offset & $\mathcal{U}(-\Delta RV, \Delta RV)$ & \\[2pt]

\hline

\multirow{4}{*}{Stellar GP} & $\eta_1$ & $\mathcal{U}(0.05,\,5)$ & m/s \\[2pt]
 & $\eta_2$ & $\mathcal{U}$(P$_{rot}$,\,5P$_{rot}$) & d \\[2pt]
 & $\eta_3$ & $\mathcal{N}$(P$_{rot}$,\,$\sigma$) & d \\[2pt]
 & $\eta_4$ & $\mathcal{U}$(0.2,\,5) & -- \\[2pt]

 \hline
\multirow{3}{*}{Star} & linear slope & $\mathcal{N}$(0,\,10$^{\frac{\Delta RV}{\Delta t}}$) & \multirow{3}{*}{m/s} \\[2pt]
 & quadratic slope & $\mathcal{N}$(0,\,10$^{\frac{\Delta RV}{\Delta t^2}}$) & \\[2pt]
 & systemic RV & $\mathcal{U}(RV_{min}, RV_{max})$ & \\[2pt]
 
\hline
\end{tabular}
\end{table}

\subsection{Posterior samples for HD\,10700 \texttt{TM} run}

\begin{figure}[h]
\center
\includegraphics[width=7cm]{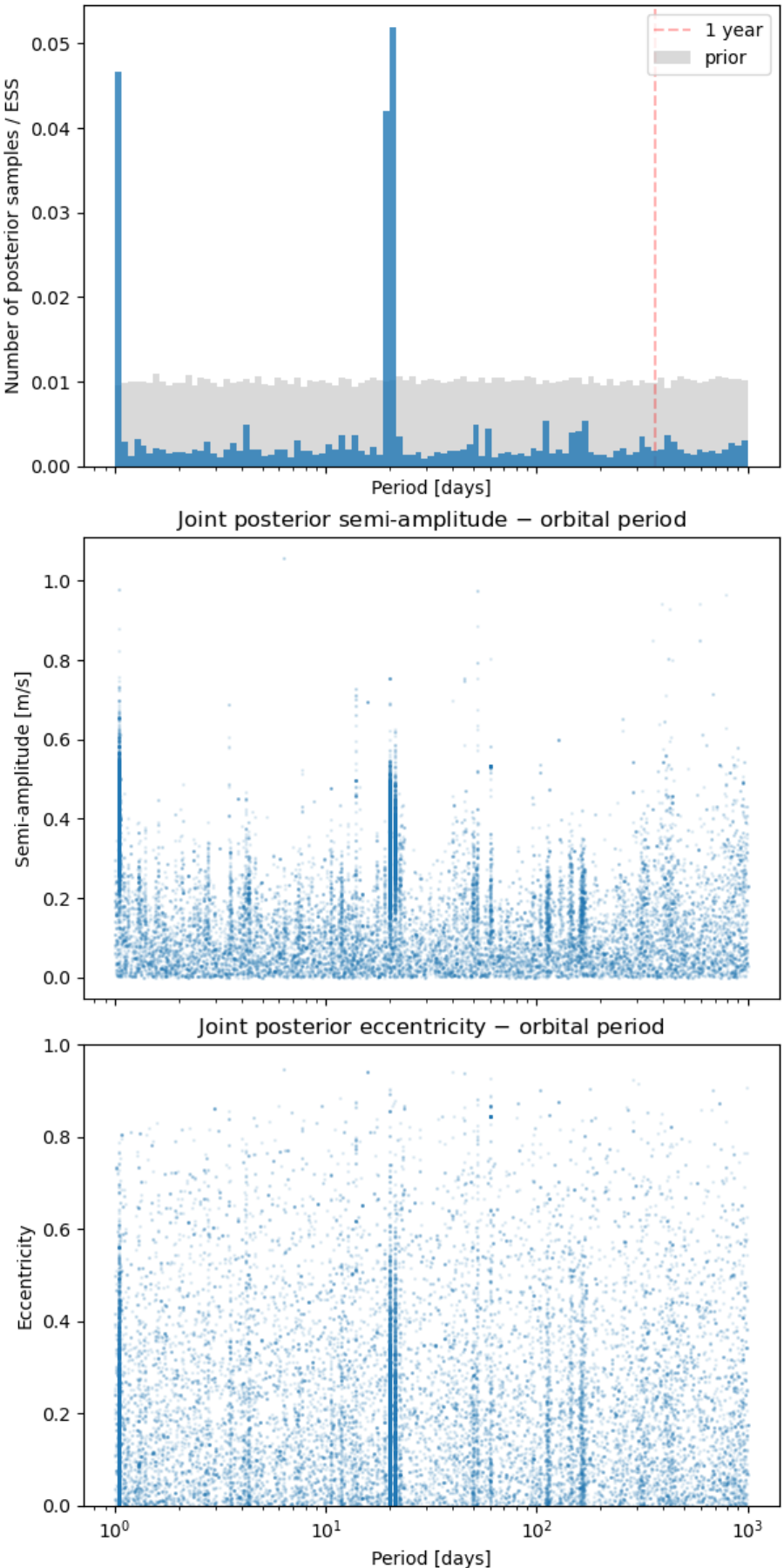}
\caption{Posterior distribution for the orbital period (\textit{top}) and joint posteriors semi-amplitude-period (\textit{middle}) and eccentricity period (\textit{bottom}) for the \texttt{TM} RV calculation method of HD\,10700.}\label{Fig:postorbsHD10700TM}
\centering
\end{figure}

\subsection{Telluric corrected HD\,10700 time series}

\begin{figure}[H]
\center
\includegraphics[width=8cm]{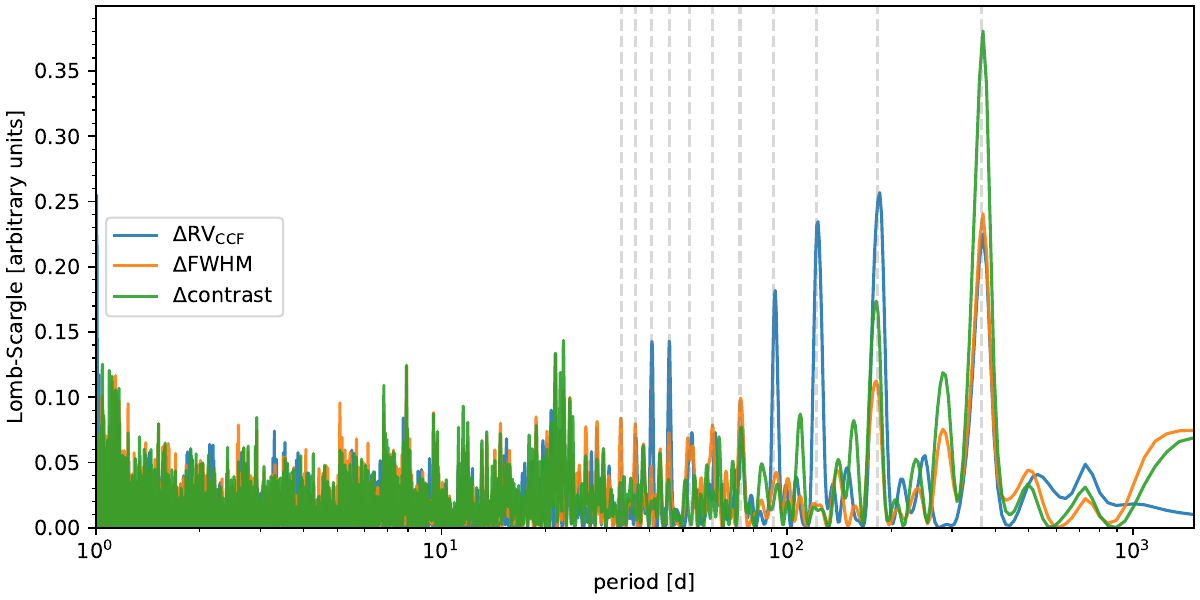}
\caption{Lomb-Scargle periodogram for the quantities $\Delta$RV, $\Delta$FWHM and $\Delta$contrast calculated by subtracting telluric corrected time series from uncorrected ones for HD\,10700. The vertical dashed lines represent one year and fractions of it, from 1/2 to 1/10.}\label{Fig:HD10700glstellcorr}
\centering
\end{figure}

\subsection{Best-fit parameters}

\subsection{Corner plots}\label{App:corners}

\begin{sidewaystable*}
\setlength{\extrarowheight}{6pt}

\centering
\caption{Median and confidence interval posterior values for the \texttt{kima} analysis and no-planet models}.\label{Tab:paramsGPs}
\resizebox{\columnwidth}{!}{%
\begin{tabular}{c| c c c c| c c c c | c c c c | c }
\hline \hline

    \multirow{2}{*}{Parameter}                             & \multicolumn{4}{c}{HD\,10700} & \multicolumn{4}{c}{HD\,102365}  & \multicolumn{4}{c|}{HD\,304636}  & \multirow{2}{*}{Units}\\ 
                                 & \texttt{CCF}                               & \texttt{TM}                                         & \texttt{TMb}                                        & \texttt{TMr}                                              & \texttt{CCF}                                        & \texttt{TM}                          & \texttt{TMb}                                        & \texttt{TMr}                               & \texttt{CCF}                               & \texttt{TM}                                               & \texttt{TMb}                                        & \texttt{TMr}                         &     \\

\hline

\multicolumn{14}{c}{\large{\textit{GP on RVs with quasi-periodic kernerl}}} \\[8pt]

$j^{RV}_{\rm \texttt{ESPR18}}$       & $0.77 ^{+0.14} _{-0.11}$                   & $0.76 ^{+0.15} _{-0.13}$                         & $0.75 ^{+0.15} _{-0.13}$                   & $0.86 ^{+0.16} _{-0.13}$                         & $0.54 ^{+0.22} _{-0.16}$                   & $0.53 ^{+0.23} _{-0.17}$                   & $0.61 ^{+0.26} _{-0.20}$                   & $0.58 ^{+0.21} _{-0.17}$                   & $1.7 ^{+0.8} _{-0.6}$                            & $1.1 ^{+0.8} _{-0.7}$                & $2.2 ^{+0.8} _{-0.7}$                      & $1.4 ^{+0.6} _{-0.5}$                & m/s     \\
 $j^{RV}_{\rm \texttt{ESPR19}}$       & $0.51 ^{+0.06} _{-0.07}$                   & $0.44 ^{+0.07} _{-0.06}$                         & $0.46 ^{+0.08} _{-0.07}$                   & $0.48 ^{+0.07} _{-0.05}$                         & $1.05 ^{+0.21} _{-0.20}$                   & $1.02 ^{+0.21} _{-0.21}$                   & $1.06 ^{+0.22} _{-0.24}$                   & $1.00 ^{+0.18} _{-0.19}$                   & $1.02 ^{+0.23} _{-0.25}$                         & $0.70 ^{+0.17} _{-0.18}$             & $0.9 ^{+0.3} _{-0.4}$                      & $0.60 ^{+0.15} _{-0.16}$             & m/s     \\
 RV offset                       & $0.9 ^{+0.8} _{-0.8}$                      & $0.7 ^{+0.5} _{-0.5}$                            & $0.9 ^{+0.5} _{-0.6}$                      & $0.1 ^{+0.6} _{-0.5}$                            & $1.5 ^{+1.1} _{-1.0}$                      & $1.5 ^{+1.1} _{-1.0}$                      & $1.9 ^{+1.2} _{-1.1}$                      & $0.1 ^{+1.0} _{-0.9}$                      & $-2.6 ^{+1.8} _{-1.8}$                           & $-2.2 ^{+1.5} _{-1.5}$               & $-3.0 ^{+1.8} _{-1.9}$                     & $-2.1 ^{+1.5} _{-1.5}$               & m/s     \\
 $\eta_1$                        & $0.57 ^{+0.19} _{-0.13}$                   & $0.44 ^{+0.10} _{-0.10}$                         & $0.46 ^{+0.11} _{-0.11}$                   & $0.42 ^{+0.11} _{-0.11}$                         & $1.2 ^{+0.4} _{-0.3}$                      & $1.2 ^{+0.4} _{-0.3}$                      & $1.3 ^{+0.4} _{-0.3}$                      & $0.98 ^{+0.35} _{-0.28}$                   & $2.0 ^{+0.5} _{-0.4}$                            & $1.9 ^{+0.4} _{-0.3}$                & $1.9 ^{+0.4} _{-0.3}$                      & $1.8 ^{+0.4} _{-0.3}$                & m/s     \\
 $\eta_2$                        & $\left(9 ^{+4} _{-3}\right) \times 10^{1}$ & $\left(1.0 ^{+0.5} _{-0.4}\right) \times 10^{2}$ & $\left(9 ^{+6} _{-3}\right) \times 10^{1}$ & $\left(1.2 ^{+0.5} _{-0.5}\right) \times 10^{2}$ & $\left(6 ^{+4} _{-2}\right) \times 10^{1}$ & $\left(5 ^{+4} _{-1}\right) \times 10^{1}$ & $\left(5 ^{+4} _{-1}\right) \times 10^{1}$ & $\left(6 ^{+6} _{-2}\right) \times 10^{1}$ & $\left(1.0 ^{+0.4} _{-0.4}\right) \times 10^{2}$ & $88 ^{+34} _{-28}$                   & $\left(7 ^{+5} _{-3}\right) \times 10^{1}$ & $96 ^{+31} _{-28}$                   & days    \\
 $\eta_3$                        & $41.8 ^{+2.3} _{-20.7}$                    & $42.4 ^{+1.0} _{-0.8}$                           & $42.5 ^{+1.5} _{-1.1}$                     & $41.9 ^{+1.3} _{-10.1}$                          & $38.6 ^{+3.3} _{-2.6}$                     & $39 ^{+4} _{-3}$                           & $39 ^{+4} _{-3}$                           & $38 ^{+4} _{-3}$                           & $32.0 ^{+0.9} _{-0.9}$                           & $31.5 ^{+0.7} _{-0.6}$               & $32.1 ^{+1.7} _{-1.2}$                     & $31.2 ^{+0.6} _{-0.7}$               & days    \\
 $\eta_4$                        & $1.5 ^{+2.0} _{-0.8}$                      & $0.45 ^{+0.28} _{-0.16}$                         & $0.5 ^{+0.4} _{-0.2}$                      & $0.6 ^{+0.7} _{-0.2}$                            & $1.5 ^{+1.4} _{-0.7}$                      & $1.5 ^{+1.5} _{-0.7}$                      & $1.5 ^{+1.5} _{-0.8}$                      & $1.6 ^{+1.6} _{-0.9}$                      & $0.8 ^{+0.4} _{-0.2}$                            & $0.82 ^{+0.25} _{-0.18}$             & $0.80 ^{+0.34} _{-0.21}$                   & $0.93 ^{+0.30} _{-0.22}$             &         \\
 RV slope                        & $0.28 ^{+0.20} _{-0.19}$                   & $0.29 ^{+0.11} _{-0.11}$                         & $0.36 ^{+0.11} _{-0.13}$                   & $0.07 ^{+0.12} _{-0.12}$                         & $0.43 ^{+0.32} _{-0.34}$                   & $0.4 ^{+0.3} _{-0.4}$                      & $0.6 ^{+0.4} _{-0.4}$                      & $0.09 ^{+0.27} _{-0.30}$                   & $-0.7 ^{+0.5} _{-0.5}$                           & $-0.5 ^{+0.5} _{-0.4}$               & $-0.5 ^{+0.5} _{-0.5}$                     & $-0.6 ^{+0.5} _{-0.5}$               & m/s/yr  \\
 RV quadr                        & $-0.13 ^{+0.16} _{-0.15}$                  & $-0.15 ^{+0.10} _{-0.09}$                        & $-0.13 ^{+0.10} _{-0.09}$                  & $-0.18 ^{+0.10} _{-0.09}$                        & $-0.03 ^{+0.29} _{-0.32}$                  & $-0.02 ^{+0.28} _{-0.31}$                  & $-0.05 ^{+0.30} _{-0.33}$                  & $0.11 ^{+0.26} _{-0.26}$                   & $0.4 ^{+0.4} _{-0.4}$                            & $0.3 ^{+0.4} _{-0.3}$                & $0.5 ^{+0.4} _{-0.4}$                      & $0.2 ^{+0.4} _{-0.4}$                & m/s/yr² \\[8pt]
 
\multicolumn{14}{c}{\large{\textit{joint GP on RVs and FWHM with quasi-periodic kernel}}} \\[8pt]

 $j^{RV}_{\rm \texttt{ESPR18}}$         & $0.76 ^{+0.14} _{-0.11}$             & $0.77 ^{+0.14} _{-0.13}$                   & $0.74 ^{+0.15} _{-0.13}$                   & $0.86 ^{+0.16} _{-0.13}$                   & $0.55 ^{+0.21} _{-0.15}$                   & $0.60 ^{+0.22} _{-0.18}$                   & $0.69 ^{+0.25} _{-0.21}$                   & $0.59 ^{+0.19} _{-0.15}$                   & $1.7 ^{+0.8} _{-0.7}$                & $1.3 ^{+0.8} _{-0.8}$                & $2.2 ^{+0.8} _{-0.7}$                & $1.4 ^{+0.6} _{-0.5}$                & m/s     \\
 $j^{RV}_{\rm \texttt{ESPR19}}$         & $0.51 ^{+0.06} _{-0.07}$             & $0.44 ^{+0.07} _{-0.07}$                   & $0.46 ^{+0.08} _{-0.08}$                   & $0.48 ^{+0.07} _{-0.06}$                   & $1.06 ^{+0.19} _{-0.18}$                   & $1.08 ^{+0.20} _{-0.20}$                   & $1.13 ^{+0.20} _{-0.20}$                   & $1.01 ^{+0.17} _{-0.17}$                   & $0.95 ^{+0.21} _{-0.19}$             & $0.67 ^{+0.15} _{-0.14}$             & $0.95 ^{+0.22} _{-0.19}$             & $0.53 ^{+0.14} _{-0.14}$             & m/s     \\
 $j^{FWHM}_{\rm \texttt{ESPR18}}$       & $2.5 ^{+0.4} _{-0.3}$                & $2.4 ^{+0.4} _{-0.3}$                      & $2.4 ^{+0.4} _{-0.3}$                      & $2.5 ^{+0.4} _{-0.4}$                      & $1.9 ^{+0.5} _{-0.4}$                      & $1.9 ^{+0.5} _{-0.4}$                      & $1.9 ^{+0.5} _{-0.4}$                      & $1.9 ^{+0.5} _{-0.4}$                      & $20 ^{+7} _{-5}$                     & $20 ^{+7} _{-5}$                     & $20 ^{+7} _{-5}$                     & $21 ^{+7} _{-5}$                     & m/s     \\
 $j^{FWHM}_{\rm \texttt{ESPR19}}$       & $2.26 ^{+0.20} _{-0.17}$             & $2.25 ^{+0.19} _{-0.18}$                   & $2.23 ^{+0.19} _{-0.17}$                   & $2.25 ^{+0.19} _{-0.17}$                   & $2.2 ^{+0.4} _{-0.3}$                      & $2.21 ^{+0.34} _{-0.31}$                   & $2.20 ^{+0.35} _{-0.31}$                   & $2.20 ^{+0.35} _{-0.31}$                   & $3.8 ^{+0.7} _{-0.5}$                & $3.8 ^{+0.6} _{-0.5}$                & $3.7 ^{+0.7} _{-0.5}$                & $3.8 ^{+0.6} _{-0.5}$                & m/s     \\
 RV offset                         & $0.7 ^{+1.4} _{-0.9}$                & $1.0 ^{+1.5} _{-1.6}$                      & $0.9 ^{+1.6} _{-1.1}$                      & $0.7 ^{+1.7} _{-0.9}$                      & $4.2 ^{+1.3} _{-2.7}$                      & $2.7 ^{+2.1} _{-2.1}$                      & $4.2 ^{+2.9} _{-3.4}$                      & $1.4 ^{+1.5} _{-2.6}$                      & $-4 ^{+4} _{-3}$                     & $-3 ^{+4} _{-3}$                     & $-4 ^{+4} _{-3}$                     & $-2.7 ^{+3.2} _{-2.8}$               & m/s     \\
 FWHM offset & $-0.6 ^{+0.7} _{-2.5}$               & $-0.09 ^{+0.34} _{-1.74}$                  & $-0.0 ^{+0.3} _{-0.5}$                     & $-0.1 ^{+0.3} _{-0.7}$                     & $3.9 ^{+1.6} _{-3.2}$                      & $3 ^{+4} _{-3}$                            & $4 ^{+3} _{-4}$                            & $1.5 ^{+3.5} _{-1.6}$                      & $-0.4 ^{+1.9} _{-2.6}$               & $-0.3 ^{+1.6} _{-2.4}$               & $-0 ^{+2} _{-4}$                     & $-0.3 ^{+1.5} _{-2.2}$               & m/s     \\
 %$f_{\rm sys}$                     & $5844.6 ^{+0.9} _{-0.8}$             & $5844.8 ^{+1.1} _{-1.4}$                   & $5844.5 ^{+1.1} _{-1.1}$                   & $5844.5 ^{+1.1} _{-1.4}$                   & $-0.7 ^{+1.0} _{-1.5}$                     & $-1.1 ^{+1.7} _{-1.0}$                     & $-0.3 ^{+1.6} _{-1.3}$                     & $-0.7 ^{+1.6} _{-1.5}$                     & $11 ^{+10} _{-9}$                    & $12 ^{+10} _{-10}$                   & $10 ^{+10} _{-11}$                   & $12 ^{+10} _{-10}$                   & m/s     \\
 $\eta_1$                        & $0.58 ^{+0.21} _{-0.13}$             & $0.47 ^{+0.11} _{-0.10}$                   & $0.50 ^{+0.13} _{-0.11}$                   & $0.44 ^{+0.13} _{-0.11}$                   & $1.6 ^{+0.8} _{-0.4}$                      & $1.6 ^{+0.8} _{-0.5}$                      & $1.7 ^{+1.1} _{-0.5}$                      & $1.2 ^{+0.7} _{-0.4}$                      & $2.1 ^{+0.5} _{-0.4}$                & $1.9 ^{+0.4} _{-0.3}$                & $2.0 ^{+0.5} _{-0.4}$                & $1.9 ^{+0.4} _{-0.3}$                & m/s     \\
 $\eta_1^{FWHM}$                     & $1.8 ^{+0.7} _{-0.7}$                & $2.0 ^{+0.7} _{-0.6}$                      & $2.0 ^{+0.7} _{-0.5}$                      & $2.1 ^{+0.7} _{-0.6}$                      & $2.2 ^{+0.9} _{-0.7}$                      & $2.4 ^{+0.9} _{-0.8}$                      & $2.2 ^{+0.9} _{-0.8}$                      & $2.6 ^{+0.9} _{-0.7}$                      & $19 ^{+6} _{-4}$                     & $19 ^{+6} _{-4}$                     & $18 ^{+6} _{-4}$                     & $19 ^{+6} _{-4}$                     & m/s     \\
 $\eta_2$                          & $80 ^{+32} _{-25}$                   & $\left(8 ^{+4} _{-3}\right) \times 10^{1}$ & $\left(7 ^{+4} _{-2}\right) \times 10^{1}$ & $\left(8 ^{+4} _{-3}\right) \times 10^{1}$ & $\left(7 ^{+4} _{-2}\right) \times 10^{1}$ & $\left(7 ^{+8} _{-3}\right) \times 10^{1}$ & $\left(8 ^{+8} _{-4}\right) \times 10^{1}$ & $\left(7 ^{+7} _{-3}\right) \times 10^{1}$ & $78 ^{+14} _{-12}$                   & $79 ^{+14} _{-12}$                   & $77 ^{+15} _{-13}$                   & $81 ^{+14} _{-12}$                   & days    \\
 $\eta_3$                          & $40.8 ^{+2.6} _{-19.5}$              & $42.2 ^{+1.1} _{-18.9}$                    & $42.3 ^{+1.4} _{-19.9}$                    & $40.9 ^{+1.7} _{-19.2}$                    & $39.6 ^{+3.4} _{-3.0}$                     & $40 ^{+3} _{-4}$                           & $40 ^{+3} _{-4}$                           & $39 ^{+4} _{-4}$                           & $30.8 ^{+0.6} _{-0.6}$               & $30.8 ^{+0.5} _{-0.5}$               & $30.6 ^{+0.6} _{-0.5}$               & $30.5 ^{+0.5} _{-0.5}$               & days    \\
 $\eta_4$                          & $1.6 ^{+1.9} _{-0.9}$                & $0.5 ^{+0.6} _{-0.2}$                      & $0.55 ^{+1.14} _{-0.24}$                   & $0.61 ^{+1.19} _{-0.23}$                   & $2.3 ^{+1.6} _{-1.2}$                      & $2.5 ^{+1.6} _{-1.4}$                      & $2.7 ^{+1.6} _{-1.4}$                      & $2.4 ^{+1.6} _{-1.4}$                      & $0.91 ^{+0.31} _{-0.22}$             & $0.84 ^{+0.25} _{-0.19}$             & $0.83 ^{+0.31} _{-0.22}$             & $0.91 ^{+0.27} _{-0.22}$             &         \\
 $\eta_4^{FWHM}$                      & $3.4 ^{+1.1} _{-1.3}$                & $3.6 ^{+1.0} _{-1.2}$                      & $3.5 ^{+1.0} _{-1.2}$                      & $3.6 ^{+1.0} _{-1.2}$                      & $3.1 ^{+1.3} _{-1.3}$                      & $3.2 ^{+1.2} _{-1.3}$                      & $3.1 ^{+1.3} _{-1.3}$                      & $3.3 ^{+1.1} _{-1.3}$                      & $2.5 ^{+1.0} _{-0.7}$                & $2.6 ^{+1.0} _{-0.7}$                & $2.5 ^{+1.0} _{-0.7}$                & $2.6 ^{+1.0} _{-0.7}$                & m/s     \\
 RV slope                          & $0.30 ^{+0.14} _{-0.26}$             & $0.28 ^{+0.11} _{-0.12}$                   & $0.35 ^{+0.13} _{-0.16}$                   & $0.08 ^{+0.12} _{-0.12}$                   & $0.5 ^{+0.5} _{-0.7}$                      & $0.3 ^{+0.5} _{-0.8}$                      & $0.4 ^{+0.6} _{-0.9}$                      & $0.0 ^{+0.4} _{-0.4}$                      & $-0.7 ^{+0.6} _{-0.6}$               & $-0.5 ^{+0.5} _{-0.5}$               & $-0.5 ^{+0.5} _{-0.4}$               & $-0.6 ^{+0.5} _{-0.5}$               & m/s/yr  \\
 RV quadr                          & $-0.14 ^{+0.19} _{-0.12}$            & $-0.12 ^{+0.13} _{-0.09}$                  & $-0.11 ^{+0.15} _{-0.11}$                  & $-0.18 ^{+0.11} _{-0.09}$                  & $-0.3 ^{+0.5} _{-0.5}$                     & $-0.1 ^{+0.8} _{-0.6}$                     & $-0.3 ^{+0.9} _{-0.6}$                     & $0.0 ^{+0.3} _{-0.4}$                      & $0.4 ^{+0.5} _{-0.4}$                & $0.3 ^{+0.4} _{-0.4}$                & $0.6 ^{+0.4} _{-0.4}$                & $0.3 ^{+0.4} _{-0.4}$                & m/s/yr² \\[3pt]

\hline

\end{tabular}
}

\setlength{\extrarowheight}{2pt}
\end{sidewaystable*}

\begin{figure*}[h]
\center
\includegraphics[width=0.6\textwidth]{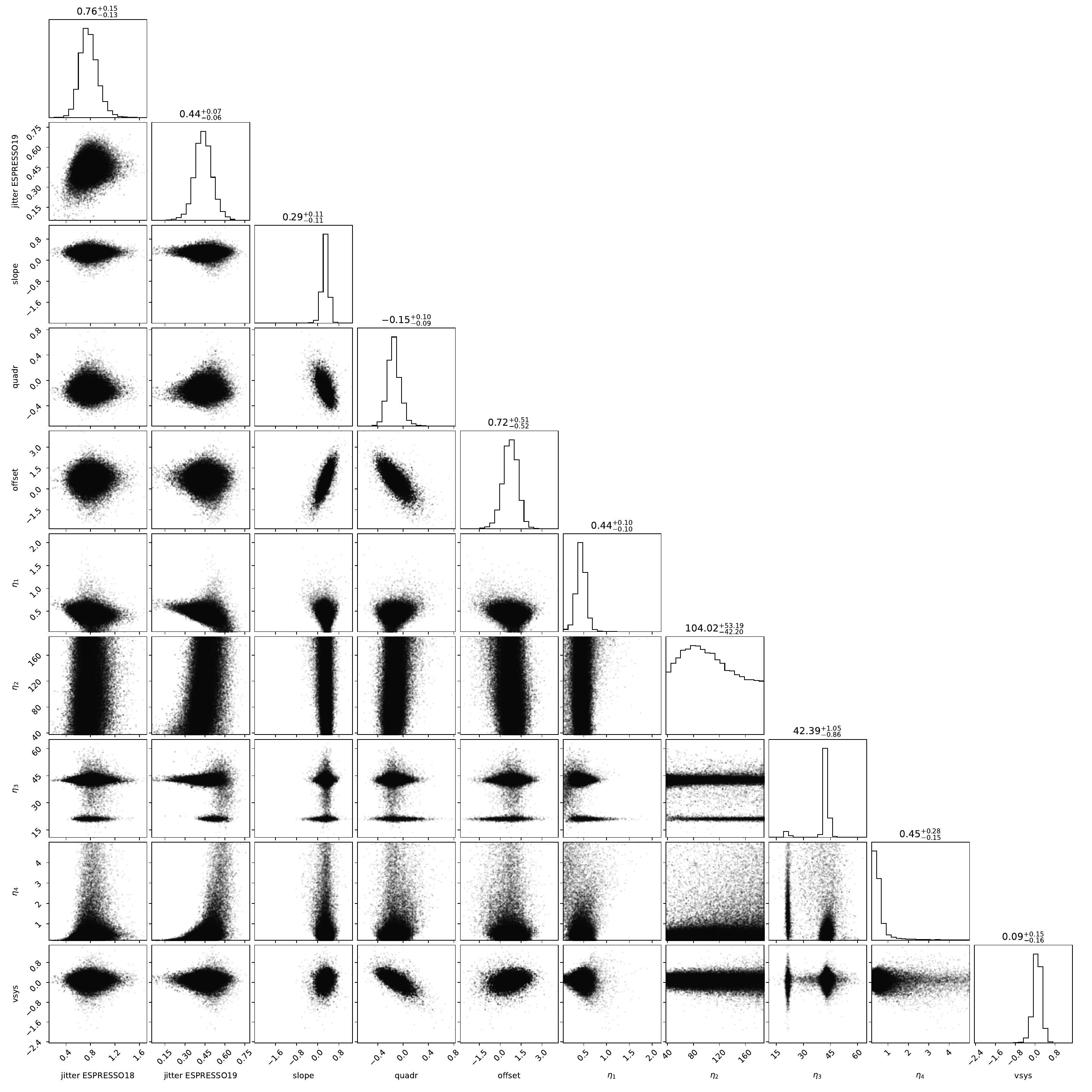}
\caption{Corner plot for the no-planet GP analysis done with \texttt{kima} for HD\,10700 \texttt{TM} RVs.}\label{Fig:cornerHD10700GPDRS}
\centering
\end{figure*}

\begin{figure*}[h]
\center
\includegraphics[width=0.6\textwidth]{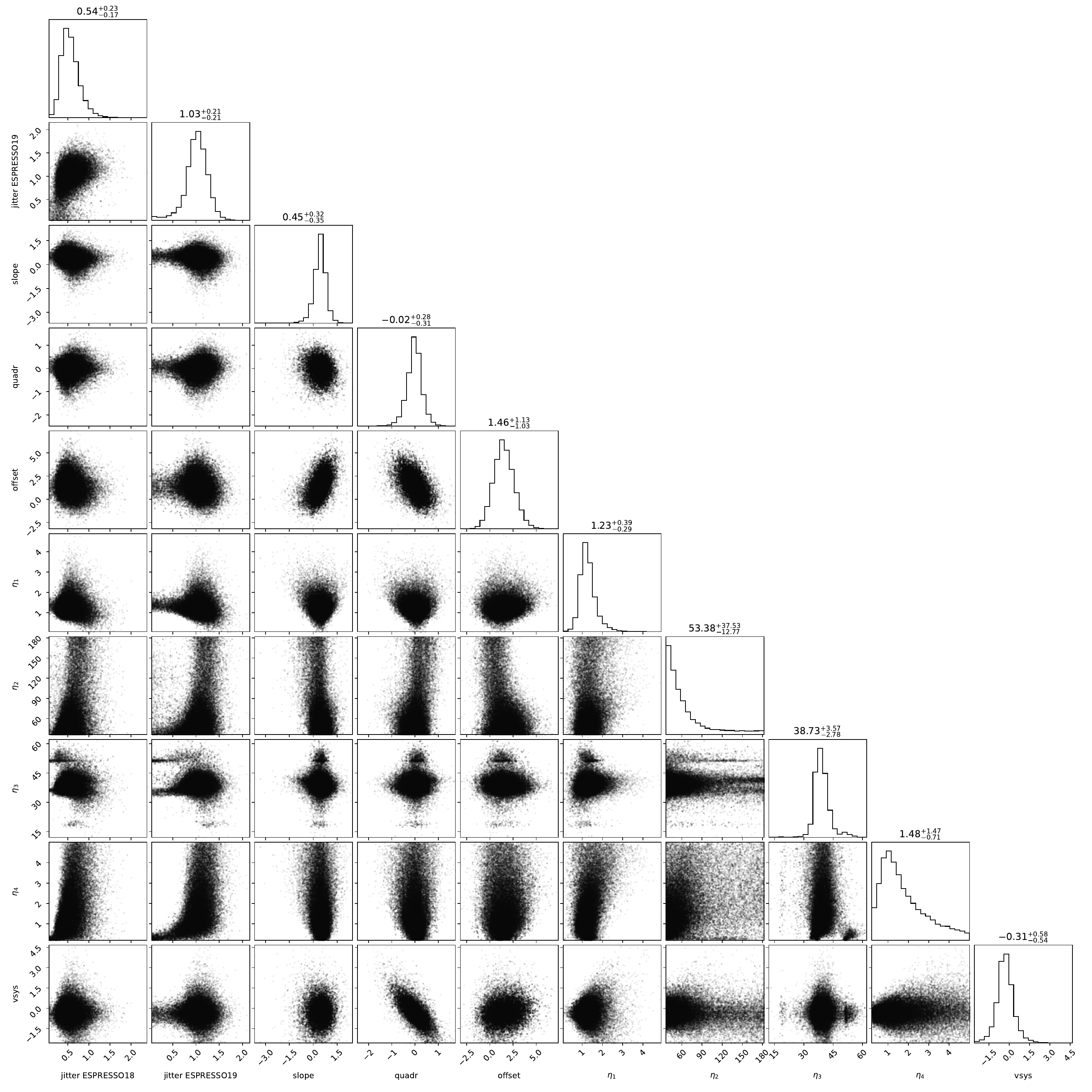}
\caption{Corner plot for the no-planet GP analysis done with \texttt{kima} for HD\,102365 \texttt{TM} RVs.}\label{Fig:cornerHD102365GPTM}
\centering
\end{figure*}

\begin{figure*}[h]
\center
\includegraphics[width=0.6\textwidth]{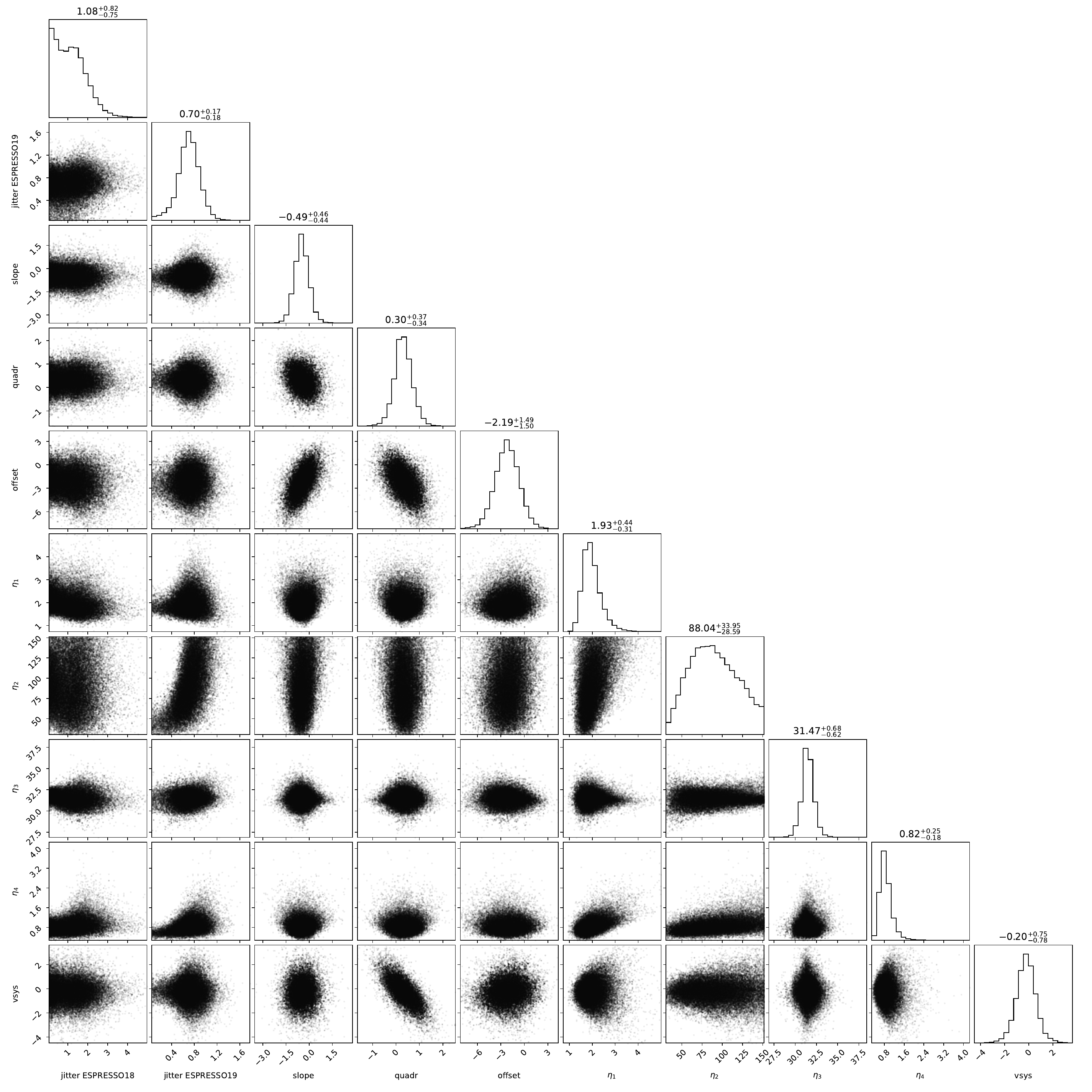}
\caption{Corner plot for the no-planet GP analysis done with \texttt{kima} for HD\,304636 \texttt{TM} RVs.}\label{Fig:cornerHD304636GPTM}
\centering
\end{figure*}

%--------------------------------------------------------------------

\setlength{\extrarowheight}{6pt}

\begin{table*}[h]
\centering

\caption{Median and confidence interval posterior values for the \texttt{kima} analysis for HD\,10700 RVs calculated on telluric-corrected spectra.\label{Tab:paramsGPstc}}

\resizebox{1.2\columnwidth}{!}{
\begin{tabular}{c|c c c c|c}
\hline \hline

    Parameter     & \texttt{DRS}$_{\mathrm{tc}}$               & \texttt{TM}$_{\mathrm{tc}}$                      & \texttt{TMb}$_{\mathrm{tc}}$               & \texttt{TMr}$_{\mathrm{tc}}$   &  Units       \\ 
    
    \hline

 $j^{RV}_{\rm \texttt{ESPR18}}$    & $0.78 ^{+0.14} _{-0.11}$                         & $0.75 ^{+0.15} _{-0.13}$                         & $0.75 ^{+0.15} _{-0.13}$                   & $0.79 ^{+0.15} _{-0.13}$                         & m/s     \\
 $j^{RV}_{\rm \texttt{ESPR19}}$    & $0.51 ^{+0.06} _{-0.07}$                         & $0.41 ^{+0.06} _{-0.06}$                         & $0.45 ^{+0.08} _{-0.08}$                   & $0.43 ^{+0.06} _{-0.06}$                         & m/s     \\
 RV offset                       & $0.9 ^{+0.8} _{-0.8}$                            & $0.7 ^{+0.5} _{-0.5}$                            & $0.9 ^{+0.5} _{-0.6}$                      & $0.2 ^{+0.5} _{-0.5}$                            & m/s     \\
 $\eta_1$                        & $0.56 ^{+0.21} _{-0.14}$                         & $0.45 ^{+0.09} _{-0.09}$                         & $0.45 ^{+0.11} _{-0.11}$                   & $0.43 ^{+0.09} _{-0.09}$                         & m/s     \\
 $\eta_2$                        & $\left(1.0 ^{+0.4} _{-0.3}\right) \times 10^{2}$ & $\left(1.0 ^{+0.6} _{-0.4}\right) \times 10^{2}$ & $\left(8 ^{+6} _{-3}\right) \times 10^{1}$ & $\left(1.1 ^{+0.5} _{-0.5}\right) \times 10^{2}$ & days    \\
 $\eta_3$                        & $40 ^{+4} _{-19}$                                & $42.4 ^{+0.9} _{-0.7}$                           & $42.5 ^{+1.4} _{-1.1}$                     & $42.2 ^{+0.9} _{-1.0}$                           & days    \\
 $\eta_4$                        & $1.8 ^{+1.8} _{-1.1}$                            & $0.42 ^{+0.23} _{-0.14}$                         & $0.5 ^{+0.4} _{-0.2}$                      & $0.47 ^{+0.22} _{-0.16}$                         &         \\
 RV slope                        & $0.28 ^{+0.20} _{-0.20}$                         & $0.25 ^{+0.10} _{-0.11}$                         & $0.34 ^{+0.11} _{-0.12}$                   & $0.07 ^{+0.10} _{-0.10}$                         & m/s/yr  \\
 RV quadr                        & $-0.15 ^{+0.17} _{-0.14}$                        & $-0.14 ^{+0.09} _{-0.09}$                        & $-0.14 ^{+0.10} _{-0.08}$                  & $-0.14 ^{+0.08} _{-0.08}$                        & m/s/yr² \\[3pt] \hline
\end{tabular}
}
\end{table*}

\setlength{\extrarowheight}{2pt}

\end{appendix} 

\end{document}